\documentclass[a4paper,11pt]{article}

\usepackage[utf8]{inputenc}
\usepackage[T1]{fontenc}

\usepackage{jcapmod}
\usepackage{caption}

\usepackage{amsmath,amsfonts,amssymb}
\usepackage{mathtools,physics}
\usepackage{textalpha}
\usepackage{cancel}
\usepackage{hyperref}
\usepackage{graphicx}
\usepackage{multirow}

\usepackage[a4paper,margin=2.5cm]{geometry}

\numberwithin{equation}{section}

\def\sqrtb{\mathpalette\DHLhksqrt}
\def\DHLhksqrt#1#2{%
\setbox0=\hbox{$#1\sqrt{#2\,}$}\dimen0=\ht0
\advance\dimen0-0.2\ht0
\setbox2=\hbox{\vrule height\ht0 depth -\dimen0}%
{\box0\lower0.4pt\box2}}

\begin{document}

\pagenumbering{roman}

\begin{titlepage}

\baselineskip=15.5pt \thispagestyle{empty}

\begin{center}
    {\fontsize{20.74}{24}\selectfont \bfseries Cosmological consequences of a principle of finite amplitudes}
\end{center}

\vspace{0.1cm}

\begin{center}
    {\fontsize{12}{18}\selectfont Caroline Jonas,$^{1}$ Jean-Luc Lehners,$^{1}$ and  Jerome Quintin$^{1}$}
\end{center}

\begin{center}
    \vskip8pt
    \textsl{$^1$ Max Planck Institute for Gravitational Physics (Albert Einstein Institute),\\
    Am M\"uhlenberg 1, D-14476 Potsdam, Germany}
\end{center}

\vspace{1.2cm}

\hrule
\vspace{0.3cm}
\noindent {\bf Abstract}\\[0.1cm]
Over 30 years ago, Barrow and Tipler proposed the principle according to which the action integrated over the entire four-manifold describing the universe should be finite. Here we explore the cosmological consequences of a related  criterion, namely, that semiclassical transition amplitudes from the early universe up to \emph{current} field values should be well defined. On a classical level, our criterion is weaker than the Barrow-Tipler principle, but it has the advantage of being sensitive to quantum effects. We find significant consequences for early universe models, in particular, eternal inflation and strictly cyclic universes are ruled out. Within general relativity, the first phase of evolution cannot be inflationary, and it can be ekpyrotic only if the scalar field potential is trustworthy over an infinite field range. Quadratic gravity eliminates all nonaccelerating backgrounds near a putative big bang (thus imposing favorable initial conditions for inflation), while the expected infinite series of higher-curvature quantum corrections eliminates Lorentzian big bang spacetimes altogether. The scenarios that work best with the principle of finite amplitudes are the no-boundary proposal, which gives finite amplitudes in all dynamical theories that we have studied, and string-inspired loitering phases. We also comment on the relationship of our proposal to the swampland conjectures.
\vskip10pt
\hrule
\vskip10pt

\end{titlepage}

\thispagestyle{empty}

\setcounter{page}{2}
\tableofcontents

\newpage
\pagenumbering{arabic}
\clearpage



\section{Introduction}

Cosmology has made enormous progress in the last century, with our understanding of the universe evolving from a static, infinite universe to a dynamical one in which space and time themselves have a history. In fact progress in physics has repeatedly been associated with getting rid of infinities: in this case the infinite arena of never changing space and time, which had led not only to puzzles regarding the gravitational stability of the universe but also to Olbers' paradox (i.e.~the question of why the night sky is dark when there should be stars in all directions). Physics abounds with examples where infinities were removed, including the appreciation of causality when the speed of light was found to be finite, the quantum resolution of the classically diverging attraction between an electron and the atomic nucleus at short distances, or the quantum taming of the ultraviolet catastrophe. In the present work we want to explore the consequences of requiring cosmological amplitudes to be well defined and thus finite.

We will work in a semiclassical setting, so that amplitudes may be thought of as path integrals of the form
\begin{equation}
 \mathcal{A}(\Phi_\mathrm{i}\to\Phi_\mathrm{f})=\int_{\Phi_\mathrm{i}}^{\Phi_\mathrm{f}}\mathcal{D}\Phi\,e^{\frac{i}{\hbar}S[\Phi]}\simeq \sum \mathcal{N}\,e^{\frac{i}{\hbar}S_\mathrm{cl}[\Phi_\mathrm{i}\to\Phi_\mathrm{f}]}\,.\label{eq:finAmpPrin}
\end{equation}
Such path integrals interpolate between specified field values (or their derivatives), denoted here collectively by $\Phi_\mathrm{i}$ and $\Phi_\mathrm{f}$, with each path being weighted by a phase proportional to the action~$S$. Several features are of importance. The first is that in the gravitational context the spacetime geometry is not fixed, and in particular the time elapsed between configurations is not specified at the outset. Rather, the amplitude selects the geometry. Therefore, when talking about the early history of the universe, we must specify field values that correspond to a possible "initial" configuration, such as a zero spatial volume of the universe, or perhaps a zero expansion rate when thinking about the so-called emergent or loitering scenarios. In some cases such configurations will effectively be in the infinite past, while in other cases the beginning will have occurred a finite time ago. Second, amplitudes such as \eqref{eq:finAmpPrin} can be written in terms of an expansion in Planck's constant $\hbar$ as sums over one or more saddle point solutions, plus corrections involving higher powers in $\hbar$. The saddle points are extrema of the action $S$, i.e., they are (in general complex) solutions of the classical equations of motion. The crucial point for our work is that such saddle points must necessarily have finite action $S_\mathrm{cl}[\Phi_\mathrm{i}\to\Phi_\mathrm{f}]$ if we are to obtain a meaningful expansion and a well-defined transition rate.\footnote{In Eq.~\eqref{eq:finAmpPrin} $\mathcal{N}$ denotes a normalisation factor, which will be unimportant for our present considerations. We also note that away from the saddle points we expect most field configurations to have infinite action. This is a feature and not a bug: in quantum mechanics, for instance, off-shell nowhere-differentiable paths may be viewed as the cause of the uncertainty principle, see e.g.~\cite{Skinner}.\label{footnote:offshellaction}} As we will see, this requirement alone already leads to surprisingly strong constraints on cosmological models. Finally, since our focus is on the amplitude, and not just on the action, we can go beyond the classical realm and explore some quantum aspects. In particular, there can be a competition between saddle point solutions and the associated quantum interference effects can lead to instabilities -- we will discuss such an example in the context of inflation. Closely associated to this last point, we may further desire that the amplitudes describe stable configurations -- we will include such information in our analysis when it is available.

Put simply, we are asking what we can find out about the history of the universe just from demanding semiclassical consistency. This is a very minimal assumption, though we should immediately point out one caveat: there is the possibility that the degrees of freedom used to describe the universe change entirely near the big bang, i.e.~that a nongeometric phase is reached where semiclassical gravity becomes inadequate. We will not be able to put any restrictions on such models, a recent example of which was presented in \cite{Agrawal:2020xek}. With that caveat in mind, let us point out that a principle of finite amplitudes can be very powerful: a specific example in the context of quadratic gravity was recently discussed in \cite{Lehners:2019ibe}, where it was concluded that only accelerating spacetimes are allowed near the big bang (we will improve the analysis of this case in this work). Thus, quadratic gravity may automatically impose initial conditions that are favorable to inflation. It remains unresolved whether or not quadratic gravity provides a viable theory of quantum gravity (see, e.g., \cite{Donoghue:2019clr,Bonanno:2020bil} and references therein). Nevertheless, this example illustrates that semiclassical physics can act as a strong selection principle on the initial conditions for the universe. Since everything follows from such initial conditions, it clearly seems worthwhile exploring the principle in much more generality. In the present work we take some steps in that direction. 

Before embarking on our study of specific examples, we should mention that Barrow and Tipler (BT) proposed a principle of finite action already in 1988 \cite{Barrow1988} (and recently updated in \cite{Barrow:2019gzc}), where they demanded that the action evaluated over the entire (past and future) history of the universe should be finite. Their motivation came from the observation that the action constitutes a fundamental physical quantity, respecting all known symmetries and being the basic quantity in deriving the dynamics of most systems; thus they wanted the action of the entire four-manifold $\mathcal{M}$ describing the universe to be finite. Clearly, our proposal is closely related to that of BT. However, since we cannot possibly observe the future (and since we do not want to suppose that the far future will affect the past history of our universe), we restrict the finiteness criterion to amplitudes interpolating to \emph{current} field configurations. In fact, in light of what we just argued, the restrictions that we will find will typically stem from the primordial/creation phase of the universe and not from the subsequent, hot big bang evolution. On classical solutions our proposal is thus weaker than BT, but phenomenologically that may be seen as a benefit, as the BT proposal seems to overconstrain the universe in some cases of interest. However, since we analyze the semiclassical amplitude and not just the action, our proposal is sensitive to quantum transitions and quantum interference effects that sometimes offer crucial additional insight. 

Since we will discuss many topics, it may be useful to provide a brief overview of the paper. 
We will study the consequences of this principle of finite amplitudes first for standard cosmological models and then progressively move on to more speculative models extending general relativity. In Sec.~\ref{sec:fluids}, we start by studying simple examples involving perfect fluids and scalar fields minimally coupled to ordinary general relativity. Here we recover the usual mixmaster dynamics, which is compatible with finite amplitudes, though typically sends any scalars that are present to infinite distances in field space. Then in Sec.~\ref{sec:early}, we turn to well-known theories of the very early universe, especially inflation and eternal inflation, but also bouncing models. Inflation comes out as being allowed only as a transient phase, while eternal inflation leads to ill-defined amplitudes and is thus ruled out. Ekpyrotic models lead to finite amplitudes and are thus compatible with the principle, but require that one trusts the potential over an infinite range. Cyclic universes cannot be strictly periodic and rather generically require a beginning. Loitering models, in which the universe starts out with zero expansion rate but nonzero volume, are discussed in \ref{sec:loitering} with specific technical implementations being discussed in the later parts of Sec.~\ref{sec:qgcorrections}. The no-boundary proposal provides a framework that quite generally renders cosmological amplitudes finite. This will be the subject of Sec.~\ref{sec:noboundary}. In Sec.~\ref{sec:qgcorrections}, we move on to discuss the implications of higher-curvature corrections to general relativity, which provide very strong constraints in comparison to ordinary general relativity. Specific examples of effective theories inspired by quantum gravity expectations are discussed in Secs.~\ref{sec:limiting}--\ref{sec:stringcosmo}, with an emphasis on genesis and loitering scenarios. We present our conclusions in Sec.~\ref{sec:discussion}.
We summarize in Table \ref{table1} the main results obtained in this work, classified by order of appearance in the text.
\begin{center}
\begin{tabular}{ |p{2.3cm}|p{3.8cm}|p{0.75cm}|p{7cm}| } 
		\hline
		\textbf{Scenario}& \textbf{Specifics} & \textbf{Sec.} & \textbf{Results and conclusions} \\ 
		\hline
		\multirow{4}{2.3cm}{General Relativity}& \multirow{2}{3.8cm}{with a scalar field} &\multirow{2}{0.75cm}{\ref{sec:fluids}} & Yes if $\forall\phi$, $|V(\phi)|<\infty$, four-volume $<\infty$.\\ 
		  & & &In FLRW/B9: Yes if $a^3V\stackrel{t\to 0}{\sim} t^{3s-2}$, $s>\frac{1}{3}$. \\
		  \cline{2-4}
		  &With a perfect fluid & \ref{sec:fluids} & EOS $p=w\rho\,$: Yes if $-1<w<+1$.\\ 
		  \cline{2-4}
		  &Black holes & \ref{sec:fluids} & Yes: singularity always in the future.\\ 
		\hline
		\multirow{5}{2.3cm}{Inflation} & \multirow{2}{3.8cm}{as the first phase of evolution} & \multirow{2}{0.75cm}{\ref{sec:inflation}} &No: $S_{\text{on-shell}}^\mathrm{(classical)}<\infty$ but interferences \\
		& &&$\Rightarrow$ unstable semiclassical amplitudes.\\
		\cline{2-4}
		& Eternal phase&\ref{sec:eternalinflation}&No: $S_{\text{on-shell}}\to\infty$.\\
		\cline{2-4}
		& Transient and \newline not eternal phase&\multirow{2}{0.75cm}{\ref{sec:eternalinflation}}&\multirow{2}{7cm}{Yes.}\\
		\hline
		\multirow{2}{2.3cm}{Bounce} & \multirow{2}{3.8cm}{with a contracting phase with $p=w\rho$} & \multirow{2}{0.75cm}{\ref{sec:bounceandcyclic}} & Yes if and only if $w>1$ \newline $\Leftrightarrow$ pure ekpyrotic phase.\\ 
		\hline
		\multirow{5}{2.3cm}{Cyclic universe\newline (EH action in FLRW)} & Exactly cyclic:  & \multirow{2}{0.75cm}{\ref{sec:bounceandcyclic}}&No: $S_{\text{on-shell}}\to\infty$, and moreover, \\
		& $\exists\,P\mbox{ s.t.}~a(t)=a(t+P)$ & &it violates the second law of thermodynamic.\\
		\cline{2-4}
		& $a$ larger at each cycle, but cyclic in local quantities, e.g., $H$ &\multirow{3}{0.75cm}{\ref{sec:bounceandcyclic}}&\multirow{3}{7cm}{Yes: $a$ reaches $0$ in finite proper time, but universe becomes geodesically incomplete.} \\
		\hline
		\multirow{3}{2.3cm}{No-boundary\newline proposal} &Defined with initial momentum/regularity condition \cite{DiTucci:2019bui,DiTucci:2019dji} &\multirow{3}{0.75cm}{\ref{sec:noboundary}} & \multirow{3}{7cm}{Yes: $S_{\text{on-shell}}<\infty$ and stable\newline semiclassical amplitudes.}\\
		\hline
		\multirow{6}{2.3cm}{$(\mathrm{Riem})^n$ higher-curv.~theories} & Quadratic gravity \newline ($n=2$) &\multirow{2}{0.75cm}{\ref{subsec:higherorder}} & \multirow{2}{7cm}{Yes: $S_{\text{on-shell}}<\infty$ but only for accelerating backgrounds.}\\
		\cline{2-4}
		&\multirow{3}{3.8cm}{$n\geq 3$ on Lorentzian FLRW bckg} &\multirow{3}{0.75cm}{\ref{subsec:higherorder}} & $k=0$: Yes if $n$ is bounded and $\exists$\newline sufficiently accelerated expansion.\newline $k\neq0$: No, $S_{\text{on-shell}}\to\infty$. \\
		\cline{2-4}
		& All $n$ included&\ref{subsec:higherorder}&No for all FLRW solutions.\\
		\hline
		\multirow{2}{2.3cm}{Limiting curvature} & \multirow{2}{3.8cm}{See \cite{Sakakihara:2020rdy}: nonsingular anisotropic universe} &\multirow{2}{0.75cm}{\ref{sec:limiting}} & Yes: $S_{\text{on-shell}}<\infty$ (but null geodesics are past incomplete). \\
		\hline
		\multirow{7}{2.3cm}{Emergent and loitering universes} & $k$-essence theory $\textrm{\cancel{NEC}}$, e.g., ghost condensate &\multirow{2}{0.75cm}{\ref{subsec:kessence}} & Yes, but other issues (e.g.,~unitarity violation of fluctuations). \\
		\cline{2-4}
		& (beyond-)Horndeski& \ref{subsec:Horndeski}&Yes for some models, at least at bckg level.\\
		\cline{2-4}
		& Loitering phase \newline in EH action &\multirow{2}{0.75cm}{\ref{sec:loitering}} & Yes, also with a scalar field and a perfect fluid action.\\
		\cline{2-4}
		& \multirow{2}{3.8cm}{Loitering phase in string cosmology} & \multirow{2}{0.75cm}{\ref{sec:stringcosmo}}&Dilaton gravity: No, $S_{\text{on-shell}}\to\infty$.\newline Non-pert.~curvature corr.~in $\alpha'$ \cite{Bernardo:2020nol}: Yes.\\
		\hline
\end{tabular}
\captionof{table}{Summary of the results of this work. In the results and conclusions column, ``yes'' and ``no'' refer to whether the principle of finite amplitudes is satisfied or not.}\label{table1}
\end{center}


\section{Implications for canonical scalar fields and perfect fluids} \label{sec:fluids}

Let us start by considering a simple example consisting of general relativity and a canonical scalar field, with action\footnote{The reduced Planck mass is defined in terms of Newton's gravitational constant as $M_\mathrm{Pl}^{-2}\equiv 8\pi G_\mathrm{N}$, while the speed of light is set to unity. The reduced Planck constant $\hbar$ is also set to unity, though it is sometimes written down explicitly when it serves an illustrative purpose.}
\begin{equation}
 S=\int\dd^4x\,\sqrt{-g}\left(\frac{M_\mathrm{Pl}^2}{2}R-\frac{1}{2}\partial_\mu\phi\partial^\mu\phi-V(\phi)\right)\,.\label{eq:SGRscalar}
\end{equation}
Variation with respect to the metric yields the Einstein field equation
\begin{equation}
 R_{\mu\nu}-\frac{1}{2}Rg_{\mu\nu}=\frac{1}{M_\mathrm{Pl}^2}T_{\mu\nu}\,,\label{eq:EEs}
\end{equation}
where the energy-momentum tensor of the canonical scalar field is given by
\begin{equation}
 T_{\mu\nu}=\partial_\mu\phi\partial_\nu\phi+g_{\mu\nu}\left(-\frac{1}{2}\partial_\alpha\phi\partial^\alpha\phi-V(\phi)\right)\,.\label{eq:Tmunuscalarfield}
\end{equation}
A saddle point of the amplitude \eqref{eq:finAmpPrin} involves the on-shell action, i.e.~the action evaluated on a solution of the classical equations of motion (EOMs). We are thus principally interested in whether or not the on-shell action blows up, and we may substitute the field equations into the action to simplify its expression.\footnote{Note that we thus allow for cancellations between gravity and matter terms. This is because in quantum gravity we do not expect there to be much distinction between gravity and matter.
For example, in extended supersymmetry all fields can be part of the same multiplet. In string theory, everything is made of strings.} Such a simplification is obtained by using the trace of \eqref{eq:EEs} and \eqref{eq:Tmunuscalarfield},
\begin{equation}
 M_\mathrm{Pl}^2R=-T=\partial_\mu\phi\partial^\mu\phi+4V(\phi)\,,
\end{equation}
which upon substitution into the action \eqref{eq:SGRscalar} yields the simple on-shell action
\begin{equation}
 S_\textrm{on-shell}=\int\dd^4x\,\sqrt{-g}\,V(\phi)\,. \label{eq:scalaronshell}
\end{equation}
This already bears interesting consequences in view of the principle of finite amplitudes. The field values we might be interested in today ($t=t_0$) take some finite values, say $\phi(t_0)=\phi_\mathrm{f}$ and $g_{ij}(t_0)=h_{ij}$. The action will then certainly be finite if the scalar field potential is bounded, i.e., $|V(\phi)|<\infty$ for any scalar field value, and if the four-volume of the universe up to today is finite, i.e.~if
\begin{equation}
 {\it{Vol}}=\iint^{t_0}\dd^3x\,\dd t\,\sqrt{-g} < \infty\,.
\end{equation}
Another possibility, which we will explore in Sec.~\ref{sec:loitering}, is that in the past the scalar potential decreases sufficiently fast as to render the above integral \eqref{eq:scalaronshell} finite even when integrating back to the infinite past. In the present section we will assume that this does not occur and that consequently the four-volume $\it{Vol}$ must be finite. The interesting aspect is that under these circumstances the action will be finite regardless of the behavior of the scalar field and of the detailed shape of the geometry. There are two immediate consequences of requiring the four-volume $\it{Vol}$ to be finite: the first is that the universe must have had a beginning, i.e., a big bang of some sort must have occurred. The second is that the spatial volume of the universe must be finite too. This is automatically the case in a closed universe, and it can be achieved in flat or open universes with suitable topological identifications (e.g.~a torus for flat space). Hence the spatial curvature can take any values (positive, negative or null), as long as the topology is nontrivial. In the following, we will typically set the spatial reference volume to unity, so as to avoid clutter in equations.

Note that the curvature of the universe may blow up in the approach to the big bang without causing a divergence in the action \cite{Barrow1988}. For instance, the four-volume is insensitive to anisotropies of Bianchi-IX form where the metric is written as
\begin{align}
 \dd s^2_\mathrm{IX}=-\dd t^2+a^2&\left[e^{2(\beta_++\sqrt{3}\beta_-)}(\sin\psi\,\dd\theta-\cos\psi\sin\theta\,\dd\phi)^2\right. \nonumber\\
 &\ \left.+\,e^{2(\beta_+-\sqrt{3}\beta_-)}(\cos\psi\,\dd\theta+\sin\psi\sin\theta\,\dd\phi)^2+e^{-4\beta_+}(\dd\psi+\cos\theta\,\dd\phi)^2\right]\,,\label{eq:B9}
\end{align}
with $a(t)$ being the scale factor and $\beta_\pm(t)$ the anisotropies. Up to angular terms the determinant of the metric is simply $\sqrt{-g} = a^3$. Thus, even though it is known that a typical approach to the big bang involves mixmaster behavior in which the anisotropies grow fast and then chaotically switch to other spatial directions (in which the metric can once again be approximated by Bianchi IX \cite{Misner:1969hg,Belinsky:1970ew}) the overall volume shrinks as $a^3 \sim t$ and thus the four-volume will be finite.

In fact, it turns out that requiring the potential to be bounded everywhere can be softened because of this property. In Bianchi IX or for a flat Friedmann-Lema\^itre-Robertson-Walker (FLRW) background metric
\begin{equation}
 \dd s^2_\mathrm{FLRW}=-\dd t^2+a(t)^2\dd\mathbf{x}^2\,,
\end{equation}
the on-shell action goes as
\begin{equation}
 S_\textrm{on-shell}\propto\int_0^{t_0}\dd t\,a^3V(\phi)\,.\label{eq:scalaronshell2}
\end{equation}
Above we have chosen the origin of the time coordinate such that it coincides with the big bang. It is clear from the above that it is the quantity $a^3V(\phi)$ which must not diverge faster than (or at the same rate as) $1/t$ as $t\rightarrow 0$ and which must remain bounded everywhere else. In particular, if $a^3\sim t^{3s}$ with $3s>1$ as $t\rightarrow 0$, the potential may well diverge as $1/t^2$, and the on-shell action shall nevertheless remain finite. As a specific example, an exponential potential of the form $V(\phi)=V_0e^{\sqrtb{2/s}\phi}$ with $s>1/3$ yields the exact solution $a(t)\propto t^s$ and $\phi(t)=-\sqrt{2s}\ln(t\sqrt{V_0/[s(3s-1)]})$ in FLRW.\footnote{Requiring $s>1/3$ is equivalent to restraining the ratio of the pressure to the energy density to be in the range $(-1,1)$, i.e.~essentially imposing the dominant energy condition. In such a range though, anisotropies would typically tend to dominate in the approach toward the big bang. In that sense, this exact FLRW solution is "special", but it is nevertheless informative. The more generic argument with anisotropies can be made as before, though no simple exact solutions are known in that case.}
Correspondingly, $a^3V(\phi)\propto t^{3s-2}$, and the on-shell action is finite even though the potential actually diverges as $V(\phi)\propto 1/t^2$ at the big bang. From the semiclassical viewpoint this is thus perfectly fine. We note however that it is a strong assumption to trust the potential over an infinite field range--for instance in string compactifications, which contain information going beyond four-dimensional effective theory, it generically does not hold \cite{Ooguri:2006in}. 

It is interesting to observe that the on-shell action \eqref{eq:scalaronshell} vanishes entirely in the case of an identically vanishing potential, i.e.~the case of a massless scalar field.
In such a case, the principle of finite amplitudes would be satisfied regardless of the past four-volume of the universe (and as a corollary, regardless of the spatial geometry). However, such matter has a propagation speed equal to the speed of light, which seems to preclude any measurement of a transition amplitude. Indeed, any physical measurement apparatus would be massive and would alter the matter content in such a way that the total on-shell action would not exactly vanish anymore. The same argument holds in the case of vacuum.

Many matter types of cosmological relevance are well approximated as perfect fluids with energy density $\rho$ and pressure $p$. For such fluids, the trace of the energy-momentum tensor is given by $T=\rho(-1+3w)$, assuming an equation of state (EOS) $p=w\rho$ relating pressure and energy density. By the trace of the Einstein equations, $R=-T/M_\mathrm{Pl}^2$, the on-shell action for the gravity plus perfect fluid system \cite{Schutz:1970my,Brown:1992kc} becomes
\begin{equation}
 S_\textrm{on-shell}=\int\dd^4x\,\sqrt{-g}\,\frac{M_\mathrm{Pl}^2}{2}R+\int\dd^4x\,\sqrt{-g}\,p=\frac{1}{2}\int\dd^4x\,\sqrt{-g}\,(1-w)\rho\,.
\end{equation}
This vanishes in the case of a vacuum ($\rho\equiv 0$) or for $w=1$, i.e.~for a stiff fluid, in analogy with the case of a massless scalar field. For the reasons  stated above, such cases are not perfectly realistic. For matter with other equations of state, we can make progress by specializing to a flat FLRW universe. Then for $w>-1$, the EOMs imply the well-known relations 
\begin{equation}
 a \propto t^{\frac{2}{3(1+w)}}\,, \quad \rho \propto a^{-3(1+w)} \propto t^{-2}\,, \quad S_\textrm{on-shell} \propto\int_0^{t_0} \dd t \, t^{-\frac{2w}{1+w}}\propto t_0^{\frac{1-w}{1+w}}-\lim_{t\to 0}t^{\frac{1-w}{1+w}}\,. \label{eq:fluid}
\end{equation}
The time integral in the action then converges as long as $-1<w\leq+1$. The borderline case of $w=+1$ was just discussed, while that of $w=-1$ corresponds to an effective cosmological constant. In such a case, the background solution is that of de Sitter spacetime in the flat slicing. We will analyze inflation in detail in Secs.~\ref{sec:inflation} and \ref{sec:eternalinflation}. Here we see that any perfect fluid with an EOS between $-1$ and $+1$ leads to a convergent classical action (again assuming a finite spatial volume of the universe) and thus a finite transition amplitude at leading order in $\hbar$, despite the generic appearance of a curvature singularity (and the divergence of the matter energy density) at the associated big bang.

In our discussion so far we have neglected boundary terms, more specifically the Gibbons-Hawking-York (GHY) term 
\begin{equation}
 \pm \int_{\partial\mathcal{M}} \dd ^3 x \, \sqrt{h} \, K\,,\label{eq:GHY}
\end{equation}
where $h$ is the determinant of the metric on the boundary surface $\partial\mathcal{M}$ and $K$ denotes the trace of the extrinsic curvature there. Such a boundary term is necessary for making the variational principle well posed when we specify field values rather than momenta, i.e.~for the case where we have in mind Dirichlet boundary conditions on the metric. The amplitudes that we discuss have two boundaries, at both end points of the time integration of the action. (There is no issue with the spatial integrations as we had to assume finite nonsingular spatial sections.) On the final boundary (e.g., today when an observation is made), all field values and their derivatives are finite; thus, a boundary term will not cause any divergence and we do not need to investigate it. On the initial boundary a Neumann boundary condition can be of interest \cite{DiTucci:2020weq} (and we will discuss this condition in Sec.~\ref{sec:noboundary}), and in that case we do not require the GHY term. However, for now we want to specify field values on the boundary and hence we have to include the GHY term. For a FLRW metric, we have $K \propto \dot{a}/a$ and $\sqrt{h} = a^3$; thus, the boundary term becomes
\begin{equation}
 \pm \int_{\partial\mathcal{M}} \dd ^3 x \, \sqrt{h} \, K \ \propto\  a^2 \dot{a} \ \propto\  t^{\frac{1-w}{1+w}}\,.
\end{equation}
This is once again finite (actually zero) at $t=0$ for $-1<w<+1$, and thus the conditions for finiteness of the amplitude are unaltered whether or not the GHY term is included.

For the case of general relativity with a canonical scalar field, one should also consider the variation with respect to $\phi$. Taking the canonical action \eqref{eq:SGRscalar}, one has
\begin{equation}
 \delta_\phi S=\int_{t_\mathrm{i}}^{t_\mathrm{f}}\dd t\int_{\partial\mathcal{M}}\dd^3x\,\sqrt{-g}\left(\Box\phi-V_{,\phi}\right)\delta\phi-\left[\int_{\partial\mathcal{M}}\dd^3 x\,\sqrt{h}\,n^\mu\nabla_\mu\phi\,\delta\phi\right]_{t_\mathrm{i}}^{t_\mathrm{f}}\,,\label{eq:varwrtphi}
\end{equation}
where $\Box\equiv\nabla^\mu\nabla_\mu$ is the d'Alembertian, and $n^\mu$ is the unit normal vector with respect to the hypersurface $\partial\mathcal{M}$.
Typically, demanding $\delta S/\delta\phi=0$ yields the EOM $\Box\phi-V_{,\phi}=0$ when the second term on the right-hand side is set to zero, i.e., fixing the variation on the boundary, $\delta\phi(t_\mathrm{i})=\delta\phi(t_\mathrm{f})=0$.
However, this can only be done provided $\sqrt{h}\,n^\mu\nabla_\mu\phi$ is nondivergent on the boundary.
For instance, this boundary term is $a^3\dot\phi$ in FLRW, and a typical solution in the approach to the big bang ($t_\mathrm{i}=0$) is $a\sim t^{1/3}$ and $\phi\sim\ln(t)$, so $a^3\dot\phi\sim\mathrm{constant}$ and the variational principle is safe.
This is not always the case, however, and one should always check that the boundary term does not diverge.
The fact that the classical solution may diverge on the boundary, e.g.~$\phi\to-\infty$ as $t\to 0$, does not prevent one from fixing the functional variation\footnote{Indeed, as an example, if arbitrary paths in field space starting at $\phi(t=0)=-\infty$ and parametrized by $\varepsilon$ are written as $\phi_\varepsilon(t)=\phi(t)+\varepsilon\eta(t)$ with $\eta(t=0)=0$, then the functional derivative of $\phi$ is simply $\delta\phi\equiv(\partial\phi_\varepsilon/\partial\varepsilon)|_{\varepsilon=0}=\eta$, which vanishes at $t=0$.} $\delta\phi(t_\mathrm{i}=0)=0$.
However, in the path integral, starting at $\phi(t_\mathrm{i}=0)=-\infty$ and $a(t_\mathrm{i}=0)=0$ may be imprecise in the sense that one would really want to sum over all possible paths with these initial conditions, but there are infinitely many different rates at which such a singularity may be reached.
We know from the formulation of the principle of finite amplitudes that we are interested in testing the convergence of saddle points, not arbitrary paths (since as already stated those may well lead to divergences off-shell), and so one should ensure that the initial conditions are set such that one lies on the classical trajectory.
Therefore, in this context, the initial conditions in the path integral should be really thought of as limits; schematically, in minisuperspace, the integration is over
\begin{equation}
 \int_{\bar{\phi}(\bar{a})}\mathcal{D}\phi\int_{\bar{a}}\mathcal{D}a\,,
\end{equation}
where $(\bar\phi,\bar{a})$ represents a classical solution. As indicated, one may then think of $\bar\phi$ as being a function of $\bar{a}$ and take the limit as $\bar{a} \to 0$.
The reasoning may appear circular: we need the classical solution to set the initial conditions in the appropriate limit, to then perform the variational principle, check its validity and the applicability of the classical solution we started with.
This should rather be seen as a consistency test, which applies for any classical field theory, irrespectively of the principle of finite amplitudes.

Before moving on to a discussion of very early universe models, there is one loose end that we should address: a point of concern might be that our principle of finite amplitudes might be too restrictive in that it might forbid the existence of black holes in light of their enclosed spacetime singularities. But black holes (without charge) are vacuum solutions of the Einstein equations, with vanishing Ricci scalar. Even if we include charge, (classically) the singularities always reside to the future as is apparent from the associated Carter-Penrose diagrams, due to the fact that space and time get swapped at the horizon.\footnote{When rotation and/or charge are included, the singularities reside near the Cauchy horizon, which comes to represent future infinity inside the black hole \cite{Dafermos:2012np}.} Thus, since we consider amplitudes up to the current time, black holes can happily coincide with the principle. In this context we would also like to mention the interesting recent work of Borissova and Eichhorn \cite{Borissova:2020knn}, which takes a differing point of view by proposing to turn these arguments around: given that we expect quantum gravity to resolve singularities, the authors of \cite{Borissova:2020knn} proposed that one may use diverging black hole actions (where the action is evaluated over the entire spacetime) as a criterion for selecting candidate quantum gravity theories, since in such theories the singularities would be filtered out by the path integral. It will be interesting to study this approach, which is akin to the original BT proposal, further.


\section{Implications for early universe models} \label{sec:early}

It has been appreciated at least since the 1970s that our universe was and is in a very special state \cite{1979grec.conf..504D}, which cannot really be explained by known physics. In particular it is mysterious why the universe is observed to be spatially so flat on large-scales, when gravity makes things clump. Other puzzling features include the large scale spatial homogeneity and isotropy, the number of galaxies (why so many, if one would have been sufficient? why not more, if galaxies harbor life?), the distribution of galaxies, the entropy of the universe (why so high, given that the universe is filled with radiation? why so low, given that it is still many orders of magnitude below the conjectured de Sitter bound?), the recent detection of vacuum energy (why is it so small? yet why is it large enough that it got discovered already in the present epoch?), the absence of antimatter, the origin of primordial magnetic fields and the absence of topological defects.  In order to explain at least some of these features, extensions of the standard hot big bang cosmological model have been proposed. We will analyze a number of the most prominent ones here.

\subsection{Inflation} \label{sec:inflation}

Inflation is a conjectured phase of accelerated expansion in the very early universe that has been proposed as an explanation of the observed large-scale spatial flatness, homogeneity and isotropy of the universe, while in addition providing a mechanism for amplifying quantum fluctuations into density fluctuations such as those seen in the cosmic background radiation \cite{Guth:1980zm,Linde:1981mu,Albrecht:1982wi,Mukhanov:1981xt}. Although the initial conception was that inflation followed an earlier radiation dominated phase, later incarnations (e.g., \cite{Linde:1983gd}) implicitly or explicitly suggested that inflation could have been the first phase of evolution, extending all the way back to the big bang. Here we would like to see to what extent these ideas are compatible with the principle of finite amplitudes. In the present section we will review relevant aspects of what is known about the situation where inflation extends all the way back to the big bang, ignoring the effects of eternal inflation, while in the next section \ref{sec:eternalinflation} we will focus on inflation as a transient phase, including implications of eternal inflation.

In a FLRW universe with curvature parameter $k$, the Friedmann equation relates the expansion rate $H$ to the energy densities of radiation (subscript "r"), matter (subscript "m"), and a scalar field $\phi$ with EOS $w_\phi$ (assumed to be slowly varying here) according to
\begin{equation}
 1 = -\frac{k}{(aH)^2} + \frac{\rho_\mathrm{r}}{3H^2a^4} + \frac{\rho_\mathrm{m}}{3H^2a^3} + \frac{\rho_\phi}{3H^2a^{3(1+w_\phi)}}\,, \label{eq:Friedmann}
\end{equation}
where $\rho_{\mathrm{r},\mathrm{m},\phi}$ represent the respective energy densities at the reference scale $a=1$. The homogeneous curvature term $k/(aH)^2$ has not been measured (statistically) significantly different from zero to date, yet during the phase of radiation domination [where $a(t) \propto t^{1/2}$] it grew in proportion to $1/(aH)^2 \propto t \propto a^2$. From the electroweak scale ($t_\mathrm{ew} \sim 10^{-12}\,\mathrm{s}$) until today this term thus grew by a factor of about $e^{60}$, and by another factor of $e^{60}$ if we extrapolate back to a hypothetical grand unified scale. This is the flatness puzzle. Inflation resolves it by assuming a preradiation phase during which the curvature term was driven to extremely small values. For a constant EOS $w$, we have that $a(t) \propto t^{\frac{2}{3(1+w)}}$ and consequently the curvature term evolves as $(aH)^{-2} \propto t^{\frac{2}{3}\frac{1+3w}{1+w}}$, implying that in an expanding universe it shrinks only if $-1 < w < -1/3$. This is the regime of inflation. \footnote{The borderline value $w=-1$, obtained from a constant potential (i.e.~a cosmological constant), corresponds to de Sitter spacetime in the flat slicing with $a(t) \propto e^{Ht}$.
The "singularity" $a=0$ is now obtained for $t \to - \infty$, but this point is ill defined in flat FLRW coordinates. More appropriate coordinates \cite{Spradlin:2001pw} show that exact de Sitter spacetime is nonsingular and extends beyond this bounce point to include a mirror, exponentially contracting phase. However, \emph{exact} eternal de Sitter spacetime is not relevant, physically speaking, to the extent that it cannot explain our observed universe. Nevertheless, small deviations from de Sitter can be relevant, and models that are sufficiently close to de Sitter in the past ($a\sim e^{Ht}$ and $|\dot H|/a^2<\infty$ as $t\to-\infty$) can similarly be expanded beyond the apparent singularity at $a=0$ (see \cite{Yoshida:2018ndv}). Resulting scenarios that experience a bounce or that are cyclic shall be discussed at more length in Sec.~\ref{sec:bounceandcyclic}.}
Note that requiring $1/(aH)=1/\dot{a}$ to shrink is equivalent to requiring $\dot{a}$ to grow, i.e., it corresponds to accelerated expansion. This phase can be modeled as a scalar field evolving in a sufficiently flat potential $V(\phi)$. In the inflationary range $-1 < w_\phi < -1/3$ the scalar field then comes to dominate over the other terms in Eq.~\eqref{eq:Friedmann}, and if this phase lasts sufficiently long the flatness puzzle disappears.

But can we assume that inflation was the first phase of evolution of the universe? The Friedmann equation implies that $V(\phi) \simeq 3M_\mathrm{Pl}^2H^2 \propto t^{-2}$, so that from \eqref{eq:scalaronshell} we can evaluate the on-shell action  back to vanishing scale factor $a=0$, corresponding to $t=0$, obtaining
\begin{equation}
 S_\textrm{on-shell} = \int \dd^4 x \, \sqrt{-g} \, V(\phi) = \int_{0}^{t_\mathrm{f}} \dd t \, a^3 V(\phi) \propto \int_{0}^{t_\mathrm{f}} \dd t \ t^{\frac{2}{1+w_\phi} -2}\,,
\end{equation} 
which converges to a finite result. Thus the classical action for inflation is finite, even when integrating back to the cosmological singularity. Here we assumed a highly symmetric metric ansatz, but the theorem of \cite{Borde:2001nh} shows that inflation is generically past incomplete, implying that the action for inflation may quite generically be taken to be finite to the past.

\begin{figure}[ht]
    \begin{center}
        \includegraphics[width=0.4\textwidth]{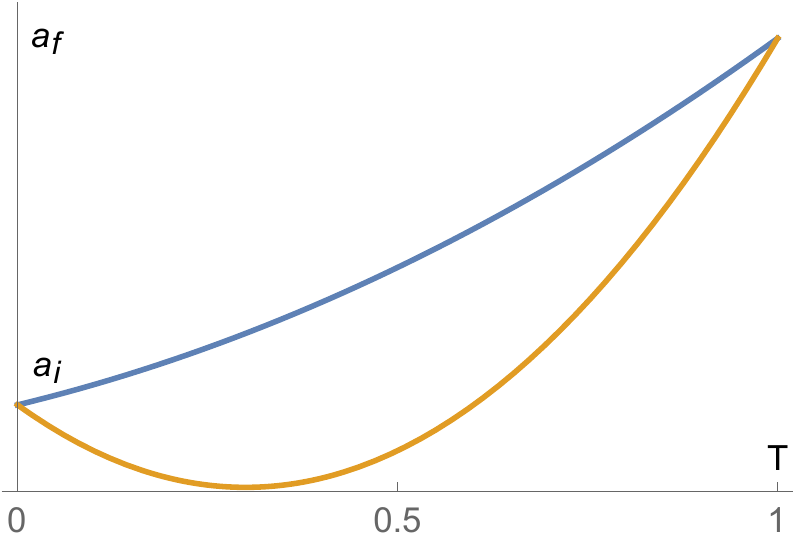}
    \end{center}
    \caption{When specifying the boundary conditions that the universe evolves from an initial size $a_\mathrm{i}$ to a final size $a_\mathrm{f}$, the semiclassical amplitude for inflation involves an interference between two background solutions. One solutions expands (blue), while the other bounces before reaching the desired final size (orange). The interference between the two solutions causes an instability in the fluctuations. The time coordinate in the above plot has been rescaled so as to run over the fixed range $0 \leq T \leq 1$.} \label{fig:infint}
\end{figure}

However, finiteness of the classical action is not necessarily sufficient to make sure that the semiclassical amplitude is well defined. In fact, as shown recently in \cite{DiTucci:2019xcr}, the quantum amplitude for inflation involves two classical solutions, and it is the interference of these two classical solutions that causes problems. We will now briefly review the salient conceptual features here -- for the detailed technical discussion we refer to \cite{DiTucci:2019xcr}. It is easiest to analyze the amplitude by evaluating it from a nonzero initial scale factor $a_\mathrm{i}$ to a final scale factor $a_\mathrm{f}$. For simplicity we may simply approximate the scalar field as being constant, assuming slow roll. If we just fix the initial and final field values, then the path integral will be a sum over all configurations that interpolate between the two values. The saddle points of the path integral will correspond to solutions of the classical EOMs that respect the boundary conditions. One may then easily see that there are two such solutions: the first corresponds to the standard expanding inflationary background, and the second corresponds to a solution that first contracts to zero size and then reexpands to reach the final size $a_\mathrm{f}$; see Fig.~\ref{fig:infint}. Both solutions contribute to the path integral with equal weight, leading to interference. In the limit that we take the initial size to zero, the two solutions merge. 

The problem arises when considering fluctuations, both gravitational waves and density perturbations, around these backgrounds. If one chooses the standard Bunch-Davies vacuum, then the perturbations around the expanding solution will obey a Gaussian distribution, as expected. But those around the bouncing background will obey an inverse Gaussian distribution, where large fluctuations are preferred (heuristically, one may note that the contracting solution is the time reverse of the expanding one--perturbations will effectively start out in the time reverse of the stable Bunch-Davies mode, which is the unstable mode). One may then show that in the limit of vanishing initial size the unstable perturbations remain dominant. Thus, when envisaging inflation to have been the first phase of the evolution of the universe, one finds that the semiclassical amplitude is in fact unstable and thus physically not meaningful. We note that the same conclusion was obtained by different means in \cite{Hofmann:2019dqu}.

There are at least three ways out of this negative result. The first is to assume that inflation did not occur. In this case some other mechanism must have been the reason for the early flatness and the cause of the primordial density fluctuations. We will review such alternative scenarios later in this section. The second possibility is that inflation occurred, but that it was not the first phase of evolution.\footnote{The initial conditions required for inflation to then start is an ongoing research topic of its own.} Indeed, as also shown in \cite{DiTucci:2019xcr}, when one assumes an initial state that is sufficiently large and in which the universe is expanding with sufficient certainty, then the bouncing solution becomes irrelevant to the path integral (technically, a Stokes phenomenon occurs, causing the integration contour for the lapse function not to pass through the bouncing saddle point anymore). This renders the amplitude finite and stable at the beginning of inflation, and in such a case inflation was a transient phase. As we will show in the next section, under certain circumstances this may not be quite enough yet to have a fully well-defined amplitude, as there can be a runaway effect associated to eternal inflation. A third possibility is that the boundary conditions at the big bang were such that, even though inflation was the first phase of evolution, the unstable branch did not contribute to the path integral. It turns out that this possibility requires truly quantum boundary conditions, not realizable in classical gravity. An explicit realization of this possibility is represented by the no-boundary wave function, to be discussed in Sec.~\ref{sec:noboundary}.

\subsection{Eternal inflation} \label{sec:eternalinflation}

\begin{figure}[ht]
\includegraphics[width=0.40\textwidth]{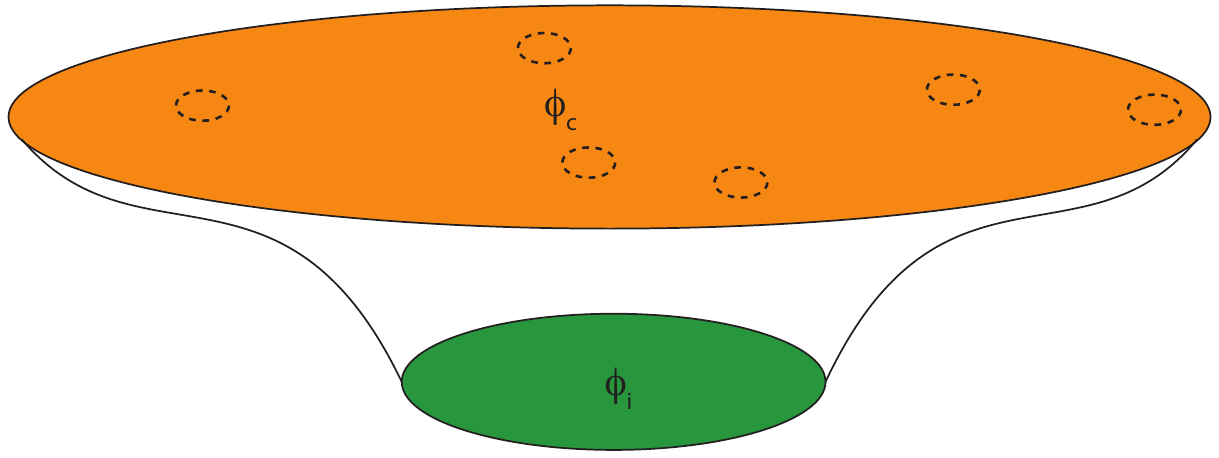}
\hspace{1.5cm}
\includegraphics[width=0.49\textwidth]{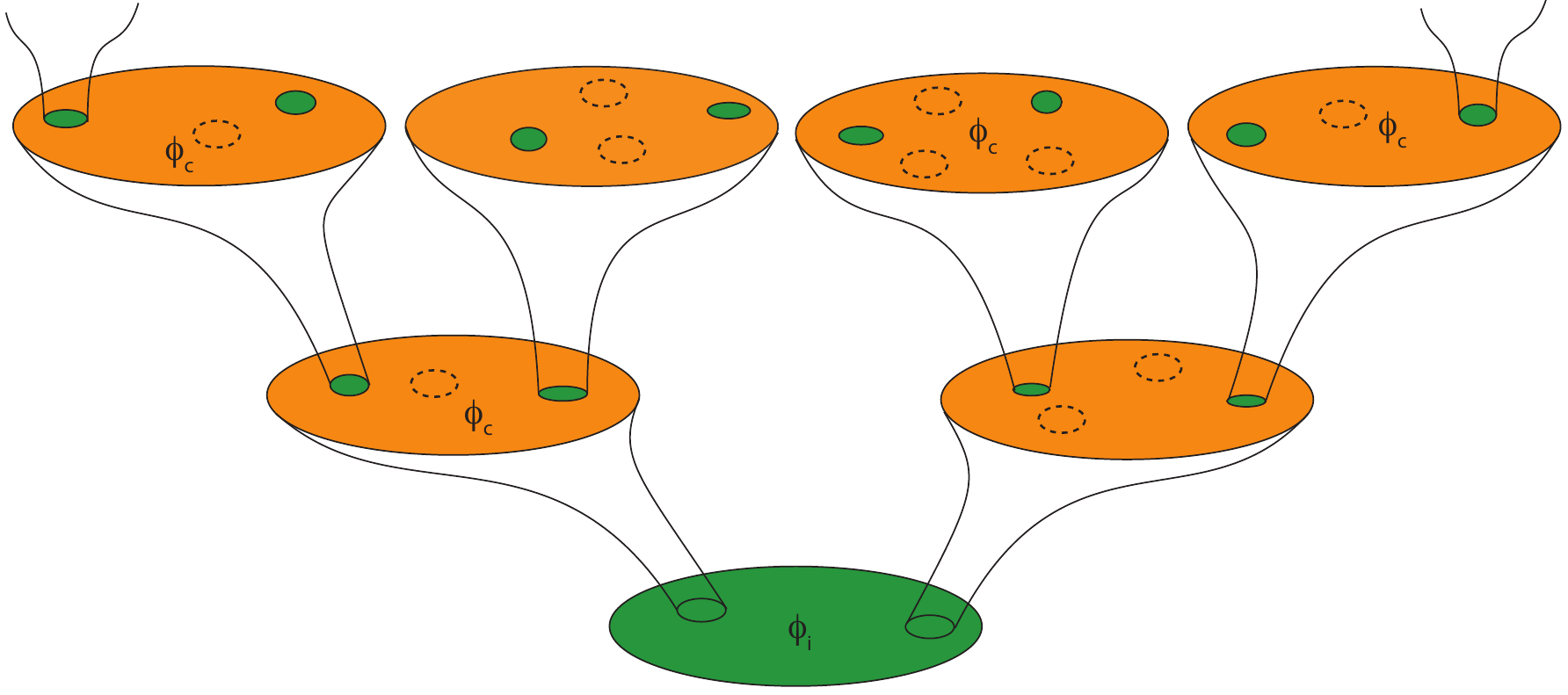}
\caption{The inflaton field evolves via a combination of classical rolling down the potential combined with random quantum jumps. When the classical rolling dominates, essentially the entire inflationary region ends inflation approximately simultaneously and then undergoes reheating followed by the ordinary hot big bang evolution. This is depicted in the left panel above, with green denoting inflation and orange the regions that have reheated. By contrast, if the quantum jumps are sufficiently important, then some regions will perpetuate inflation, as drawn in the right panel. The dashed ellipses indicate sample regions where inflation has just ended, and which (in our calculation) act as a proxy for our current state of the universe. The crucial point is that we have no way of distinguishing these sample regions at the end of inflation, and consequently we must include all of them.} \label{fig:cartoon}
\end{figure}

In the previous section we have seen that difficulties arise when inflation is conceived of as reaching back all the way to the big bang. As we mentioned, a potential resolution is to assume that inflation was a transient phase, following an earlier phase of noninflationary evolution. However, even in this case, there are circumstances that prevent inflationary amplitudes from being well defined. These have to do with the phenomenon of eternal inflation. When the inflationary potential is very flat, one may usefully think of the evolution of the inflaton field as a combination of two effects: a classical slow roll down the potential, accompanied by random jumps of the inflaton due to quantum fluctuations. The size of these quantum jumps is determined by the height of the potential, while the classical rolling is determined by the slope. Thus, the situation may arise where the potential is so flat that the quantum jumps become very significant. In a small region of the universe, it may happen that (during a Hubble time, say) the field jumps further up than it rolled down. Inflation will be prolonged in that region, which will consequently become larger than the other regions that did not undergo the jump up the potential. For sufficiently flat potentials, a runaway effect may develop where there always remain regions in the inflationary phase--inflation has become eternal; see also Fig.~\ref{fig:cartoon}. This situation has been much discussed, being alternatively seen as the greatest feature \cite{Linde:1986fc,Guth:2007ng,Bousso:2011up,Guth:2013sya} or the biggest problem \cite{Ijjas:2013vea,Ijjas:2014nta} of inflation. Here we will show that in such a case the corresponding amplitudes are ill defined.

It is a fact that we cannot directly measure the total expansion that our universe has undergone; rather, we can measure the current Hubble rate $H=\dot{a}/a$ and thus the corresponding energy density of the universe. In the context of inflation, the potential $V(\phi)$ is a good approximation to the total energy density, and thus we may use the scalar field value $\phi$ as a stand-in for the energy density (and the expansion rate). We are then interested in transitions from an initial field value $\phi_\mathrm{i}$, assumed to be in the inflationary region of the potential, to a final value $\phi_\mathrm{c}$ ("c" stands for critical). This final value will be taken to correspond to the end of inflation, where reheating occurs. For simplicity, we will assume that everywhere inflation ends, a standard hot big bang phase follows. For definiteness, we will consider a linear potential
\begin{equation}
    V(\phi) = V_0 - \alpha(\phi-\phi_\mathrm{i})\,,
\end{equation}
with $V_0,\,\alpha > 0$ and partly follow the recent exposition of \cite{Rudelius:2019cfh}. To describe inflation in this potential, it is useful to employ the stochastic formalism originally developed by Starobinsky \cite{Starobinsky:1986fx}. There he showed that on large scales, the inflaton may be described semiclassically by a Fokker-Planck equation of the form
\begin{equation}
 \partial_t P=\frac{1}{2}\left(\frac{H^3}{4\pi^2}\right)\partial_\phi^2 P+\frac{1}{3H}\partial_\phi \left(V_{,\phi} P\right)\,,
\end{equation}
a Schr\"odinger-like equation, whose solution $P[\phi,t]$ is interpreted as the probability that the scalar field has the value $\phi$ at time $t$. In this formalism, the background is approximated as being constant (which becomes a particularly good approximation when the potential is very flat), so that one can define a time coordinate everywhere. The solution to this heat-type equation is given by a Gaussian,
\begin{equation}
 P[\phi,t]=\frac{1}{\sqrt{2\pi}\sigma(t)}\mathrm{exp}\left[-\frac{(\phi-\mu(t))^2}{2\sigma(t)^2}\right]\,,
\end{equation}
with mean $\mu(t)$ and standard deviation $\sigma(t)$. Thus the scalar field may be thought of as undergoing Brownian motion on the potential.\footnote{The so-called "jumps" of the scalar field may also be analyzed directly in the path integral formalism; see \cite{Bramberger:2019zks}.} We can shift the time coordinate such that the scalar takes the value $\phi_\mathrm{i}$ at time $t=0$. Then, for the linear potential, we have \cite{Rudelius:2019cfh}
\begin{equation}
    \mu(t)=\phi_\mathrm{i} + \frac{\alpha}{3H}t\,,\qquad\sigma^2(t)=\frac{H^{3}}{4\pi^2}t\,.
\end{equation}
The mean describes the classical rolling, while the standard deviation characterizes the quantum fluctuations. We would like to calculate the action of those regions that interpolate between the initial value $\phi_\mathrm{i}$ and the final value $\phi_\mathrm{c}$ at which inflation ends. Explicitly $\phi_\mathrm{c}=\phi_\mathrm{i}+V_0/\alpha-M_\mathrm{Pl}/\sqrt{2}$ is the value at which the slow-roll parameter $\epsilon_V(\phi)=(M_\mathrm{Pl}^2/2)(V_{,\phi}/V)^2=(M_\mathrm{Pl}^2/2)(\alpha/V(\phi))^2$ reaches unity. The action is an integral over all the regions for which the scalar is in the inflationary region  $\phi < \phi_{\mathrm{c}}$. 			
With the interpretation that $P[\phi,t]$ represents the fraction of the universe with the scalar field value $\phi$ at time $t$, in conjunction with Eq.~\eqref{eq:scalaronshell}, this amounts to calculating
\begin{equation}
    S_{\phi_{\mathrm{c}}} = \int_{t=0}^{t=t_\mathrm{f}} \dd t\,a^3\,P[\phi < \phi_{\mathrm{c}}, t] V(\phi)\,,\label{eq:Seiev}
\end{equation}    
where the volume factor $a^3$ stems from the determinant of the metric and where we normalized the spatial integrations to yield unity. We have considered only those regions that have stayed in the inflationary phase throughout, hence the insertion of $P[\phi < \phi_{\mathrm{c}},t] = \int_{-\infty}^{\phi_{\mathrm{c}}}P[\phi,t]\,\dd\phi$. Note that the volume factor comes from the covariance of the action and not from a cosmological measure. In the expression above, the lower limit of integration $t=0$ corresponds to the assumed beginning of inflation and the upper limit of integration $t=t_\mathrm{f}$ corresponds to the end of inflation. An important point is that we must sum over all regions where inflation comes to an end, since there is no way of distinguishing these regions locally. In other words, if we assume that we live in a postinflationary period, then we have no way of knowing in which of the dashed ellipses of Fig.~\ref{fig:cartoon} we reside. Thus we must extend the upper limit of integration to $t_\mathrm{f}=+\infty$. Note however that this does not mean that we consider the infinite future of the evolution of the universe--anything occurring after reheating would have to be added on to the present calculation. The probability to still be inflating at time $t$ is
\begin{align}
    P[\phi < \phi_{\mathrm{c}},t] = \int_{-\infty}^{\phi_{\mathrm{c}}}P[\phi,t]\,\dd\phi &= \int_{-\infty}^{\phi_{\mathrm{c}}} \dd\phi\,\frac{\sqrt{2\pi}}{H^{3/2}\sqrt{t}} \exp[-\frac{2\pi^2}{H^3 t}\left(\phi - \phi_\mathrm{i} - \frac{\alpha}{3H}t \right)^2] \nonumber \\
    & = \frac{1}{2}\mathrm{erfc} \left[ \frac{\phi_\mathrm{i} - \phi_\mathrm{c} + \frac{\alpha}{3H}t }{\frac{H}{2\pi}\sqrt{2Ht}}\right] \nonumber \\
    & \simeq \frac{3H^{5/2}}{2\pi^{3/2}\alpha\sqrt{2t}} \exp[-\frac{2\pi^2\alpha^2}{9H^5}t]\, \qquad (t \gg H^5/\alpha^2)\,.
\end{align}
We can approximate the potential during inflation as $V(\phi) \approx V(\phi_\mathrm{c})$ (this will provide a lower bound on the action) and the scalar field evolution as $a(t)\simeq a_0 e^{Ht}$ with the slow-roll relation $3M_\mathrm{Pl}^2H^2 \simeq V(\phi_\mathrm{c})$. Putting everything together, we find that the action of the regions that have undergone inflation is approximated by
\begin{equation}
    S_{\phi_{\mathrm{c}}} \gtrsim \frac{a_0^3 V(\phi_\mathrm{c})^{9/4}}{3^{1/4}(2\pi)^{3/2}\alpha M_\mathrm{Pl}^{5/2}} \int^{t = + \infty} \dd t \ t^{-1/2}\exp[\left(3H-\frac{2\pi^2\alpha^2}{9H^5}\right)t]\,. \label{eq:actiondivergence}
\end{equation}
We do not specify a lower bound to the integral since we are using the late time approximation, which is the appropriate limit for determining the convergence or divergence of the integral.
The convergence of \eqref{eq:actiondivergence} will therefore only depend on the sign of the exponent. In particular, for 
\begin{equation}
    2\pi^2 \alpha^2 < (3H^2)^3\,, \label{eq:divergencecondition}
\end{equation}
we find that the action diverges, $S_{\phi_{\mathrm{c}}} \to \infty$, while it converges otherwise. The above
corresponds to the regime of eternal inflation, which may also be written as the condition 
\begin{equation}
    \frac{|V_{,\phi}|}{V^{3/2}} < \frac{1}{\sqrtb{2}\pi M_\mathrm{Pl}^3}\,. \label{eq:EI}
\end{equation}
Analogous calculations may be performed with other forms of the potential. In all cases, for sufficiently flat potentials obeying \eqref{eq:EI}, one finds that in the eternal inflation regime  the action of inflationary regions diverges, rendering the transition amplitudes ill defined. By contrast, for models that do not inflate eternally, the action converges and meaningful amplitudes may be obtained.

The more local point of view would be to look at individual worldlines and calculate the action for transitions from the beginning to the end of inflation. \textit{A priori} we do not know the range of the time integration in the action integral. Classically, one could just calculate the time it takes for the scalar field to roll down the potential from $\phi_\mathrm{i}$ to $\phi_\mathrm{c}$. But when quantum fluctuations are included, jumps up the potential may occur and this will prolong the time. Hence, one should calculate the expectation value of the time it takes for such a transition to occur. In the quantum context, one cannot simply think of single worldlines that one can follow from the beginning to the end of inflation. Indeed, a quantum transition can only occur with a certain probability (which is well approximated by the random walk $P[\phi,t]$), so effectively each time a transition can occur, the worldlines may be thought of as splitting into worldlines where no jump occurred and new worldlines where a jump did occur. The probability function $P[\phi,t]$ describes the fraction of the volume of the universe with the scalar field value $\phi$ at time $t$, and hence must be multiplied by the spatial volume $a^3$ to obtain an expression that is proportional to the total number of worldlines (assuming a suitable coarse graining) with these properties. Thus the expectation value for the time spent in the inflationary phase is given by
\begin{equation}
    \langle t_\mathrm{f}\rangle =\int_{0}^{+\infty} \dd t\ a(t)^3 P[\phi < \phi_\mathrm{end}, t] \, t\,.
\end{equation}
This expectation value blows up in the same manner as the action \eqref{eq:actiondivergence} when the potential is sufficiently flat, i.e.~when Eq.~\eqref{eq:EI} holds. This is another manifestation of the unphysical results one obtains in the eternal inflation regime. In noneternal inflation the expectation value is close to the classical rolling time.

In the stochastic approach, the quantum fluctuations are replaced by random jumps of the scalar field value, with the result that the fields are effectively treated classically. Thus, the action is the only meaningful quantity we can calculate. It is in that sense that we computed Eq.~\eqref{eq:Seiev}, which can be interpreted as the on-shell action evaluated over the expectation value of the scalar field potential. It would be very interesting to also evaluate a transition amplitude purely in the path integral approach. We will sketch the outline of such a calculation here. An important ingredient in the stochastic approach is that in between jumps of the scalar field, the evolution is classical. This may be justified by the fact that inflation is very efficient in decohering the universe \cite{Kiefer:1998qe}.
In the path integral formalism, one would then have to calculate a transition amplitude conditioned on having these almost-classical periods of expansion between the quantum transitions of the scalar field. This means that effectively there is a measurement of the scalar field value after each jump, and thus the amplitude may be represented as a sum over field configurations having undergone a specified number $n$ of scalar field jumps,
\begin{equation}
 \mathcal{A} \approx \sum_n \int \mathcal{D}g_{\mu\nu}\mathcal{D}\phi\ \exp[\frac{i}{\hbar}S]\approx \sum_n {\it{Vol}}_{\mathrm{inf},n} \, \exp[A(n)] \, \exp[\frac{i}{\hbar}S_\textrm{phase}(n)]\,.
\end{equation}
Here the amplitude $A(n)$ will be a negative real number representing the likelihood of having undergone $n$ up-jumps, while the volume factor ${\it{Vol}}_{\mathrm{inf},n}$ arises due to the fact that each up-jump creates new interpolating geometries in proportion to the volume produced by the inflationary expansion. For each $n$ the time integration in $S_\textrm{phase}$ will have to be adjusted such that the integration lasts until the scalar field has reached the critical value $\phi_\mathrm{c}$, i.e., the final integration time $t_\mathrm{f}(n)$ will be a function of $n$ and can be used as a substitute label for $n$. Then up to time $t_\mathrm{f}$ the number of interpolating geometries having undergone inflation throughout will display the same kind of divergence at large $n$ (for sufficiently flat potentials) as that implied by the stochastic framework,
\begin{equation}
 {\it{Vol}}_{\mathrm{inf},n} \, \exp[A(n)] \sim a^3\,P[\phi < \phi_{\mathrm{c}}, t_\mathrm{f}] \sim \exp[\left(3H-\frac{2\pi^2\alpha^2}{9H^5}\right)t_\mathrm{f}]\,.
\end{equation}
As we said, this is just an outline of the calculation in the path integral formalism, and it would be of interest to make it precise. The stochastic calculation that we presented above provides an effective description that is particularly trustworthy in flat potentials \cite{Creminelli:2008es}, i.e.~exactly for the cases of greatest interest.

Our conclusion is that at the semiclassical level models that lead to eternal inflation are ill defined. A strong interpretation of this result is to say that eternal inflation cannot occur and that the scalar potential is required to be such that it does not allow for eternal inflation to take place. As shown by Rudelius \cite{Rudelius:2019cfh}, the corresponding requirements on the potential are essentially identical to those of the de Sitter swampland conjectures \cite{Obied:2018sgi,Ooguri:2018wrx,Agrawal:2018own,Garg:2018reu}. Thus we can see that the de Sitter swampland conjectures might already be implied by semiclassical physics, without requiring input from string theory. Put differently, these results suggest that finding a de Sitter vacuum in string theory will require going beyond semiclassical physics, if such vacua exist at all. An independent approach to these questions was also pursued by Dvali \textit{et al.}~\cite{Dvali:2017eba}, in which an inflationary spacetime is described as a coherent state of gravitons. There it was found that the semiclassical description breaks down after a time scale proportional to the inverse of the Hubble rate cubed. Requiring this breakdown not to happen leads to conditions on the scalar potential that are very similar to the de Sitter swampland conjectures, and that also render eternal inflation impossible \cite{Dvali:2018jhn}. It is striking that these different approaches converge to very similar conclusions. If eternal inflation cannot exist, the associated paradoxes do not arise. In our view, this puts  inflation  on a stronger conceptual footing.

In summary, we have found that the semiclassical amplitude is ill defined if one assumes that inflation reaches all the way back to the big bang or if the potential is such that inflation becomes eternal. By contrast, a transient, noneternal inflationary phase is unproblematic and may play a significant role in addressing the puzzles of the very early universe.

\subsection{Nonsingular universes}\label{sec:nonsingular}

\subsubsection{Bouncing and cyclic universes}\label{sec:bounceandcyclic}

In this subsection, we shall explore very early universe scenarios that are nonsingular alternatives to inflationary cosmology.
Recalling the flatness problem that was used to motivate inflation in section \ref{sec:inflation}, one wants the curvature term in the Friedmann equation, which evolves as $(aH)^{-2}\propto t^{\frac{2}{3}\frac{1+3w}{1+w}}$, to shrink sufficiently.
While this is only possible if $-1<w<-1/3$ in an expanding universe, this is achieved for any EOS $w>-1/3$ in a contracting universe.
Therefore, alternatives to inflation often involve a prolonged phase of contraction followed by a bounce, i.e., a nonsingular transition from contraction to expansion, which is then followed by the standard hot big bang history (see, e.g., \cite{Brandenberger:2016vhg} for a review of various models).
Depending on the EOS in the contracting phase when vacuum fluctuations exit the Hubble radius, different primordial power spectra of curvature and tensor perturbations can be generated, with various contenders for the explanation of the cosmic microwave background.
For example, matter domination ($w=0$) \cite{Wands:1998yp,Finelli:2001sr,Brandenberger:2012zb} or an ekpyrotic field ($w>1$) \cite{Khoury:2001wf,Lehners:2007ac,Buchbinder:2007ad,Lehners:2008vx} can yield a near scale-invariant primordial scalar power spectrum, making them alternative testable scenarios for the very early universe.
Ensuring anisotropies do not grow to become dominant during contraction is typically only achievable with $w>1$ though.

In this context, there are two main possibilities for the past history of the universe prior to the big bounce: the universe could be contracting forever in the past or it could undergo a series of contraction and expansion cycles; in both cases the physical time runs all the way to $-\infty$.
Let us explore the consequences of the principle of finite amplitudes for these possibilities separately.

In the case of a universe that is forever contracting to the past and which undergoes a single bounce, it suffices to evaluate the contribution to the total on-shell action
\begin{equation}
 S_\textrm{on-shell}\propto\int_{-\infty}^{-t_\mathrm{b}}\dd t\,a^3\rho\,.\label{eq:Sonshellbounce}
\end{equation}
In the above, we assume the contracting universe is dominated by a perfect fluid in FLRW up to the time of the nonsingular bounce, $-t_\mathrm{b}$.
The total action receives additional contributions from the nonsingular bounce phase and the subsequent standard hot big bang expanding phases.
However, those necessarily yield finite contributions to the amplitude since the energy density in these respective phases is bounded (by virtue of being nonsingular\footnote{This statement holds only at the background level. If perturbations are included, then the perturbed action must also yield a finite contribution to the amplitude. Avoiding instabilities and divergences in the perturbed actions of nonsingular cosmologies is a subject of ongoing research. We shall explore a few examples later in Sec.~\ref{sec:qgcorrections}.}) and since they span finite time intervals.
Considering the universe to be dominated by a perfect fluid with EOS parameter $w$ in the far past, Eq.~\eqref{eq:Sonshellbounce} reduces to
\begin{equation}
 S_\textrm{on-shell}\propto\int_{-\infty}\dd t\,(-t)^{-\frac{2w}{1+w}}\,,
\end{equation}
which converges only if\footnote{As before, we ignore the obviously convergent limit of the integral and focus on the limit that could possibly diverge. Note that $w<-1$ appears convergent, but this case is not consistent with a contracting solution. Note also that $w=1$ is allowed only because, in that case, the on-shell action identically vanishes (recall Sec.~\ref{sec:fluids}), thus removing the would-be divergence.} $w\geq 1$.
For $-1\leq w<1$, it is straightforward to see that as the volume keeps growing ($a^3\to\infty$ as $t\to-\infty$), the energy density does not tend to zero sufficiently fast to ensure the convergence of the action.
Therefore, the only single-bounce cosmologies that can yield well-defined amplitudes are the ones where the matter content in the asymptotic past is very stiff ($w\geq 1$, e.g., an ekpyrotic scalar).
Recalling that the matter energy density scales as $\rho\propto a^{-3(1+w)}$, we note that any other matter content with EOS less stiff than $w\geq 1$ would dominate in the limit $a\rightarrow\infty$, thus leading to a divergent action.
In conclusion, the only single-bounce cosmology allowed by the principle of finite amplitudes is a "pure" ekpyrotic scalar field model with $w_\phi>1$, in which no other fields are present in the asymptotic past.
Matter and radiation would then only be produced at reheating.
The open question in this case is whether one can trust the effective description over the infinite field range of $\phi$ required to reach $t = - \infty$.
That of course is questionable (cf.~the distance conjecture in string compactifications \cite{Ooguri:2006in}), but at the level of the semiclassical theory there is nothing immediately wrong with that.

Moving on to the discussion of cyclic universes, let us simplify the treatment by considering only the Einstein-Hilbert (EH) term in the action (in FLRW),
\begin{equation}
 S_\mathrm{EH}\propto\int_{-\infty}^{t_0}\dd t\,a^3(2H^2+\dot H)\,.\label{eq:SEHFLRWc}
\end{equation}
Let us distinguish two ways in which the universe could undergo a series of bounces and turnovers in the limit $t\to -\infty$:
(i) the universe is exactly cyclic, in the sense that the scale factor is periodic, i.e., there exists a period $P$ from one cycle to another such that $a(t)=a(t+P)$;
(ii) the maximal size of the universe in each cycle [$a_\mathrm{max}\equiv a(t_\mathrm{turnover})$] gets smaller and smaller as we look back in time, and thus it is asymptotically zero, but there is nevertheless a periodicity in some physical quantities [e.g., in the Hubble parameter, $H(t)=H(t+P)$].
If we separate the integration as a sum over cycles, one can write \eqref{eq:SEHFLRWc} as
\begin{equation}
 S_\mathrm{EH}\propto\sum_{n=0}^\infty\int_{t_{n}}^{t_{n+1}}\dd t\,a^3(2H^2+\dot H)\,,\label{eq:SEHFLRWcs}
\end{equation}
where $P=t_n-t_{n+1}$ represents the period of the $n$th cycle, and where we explicitly integrate back in time, i.e., $t_1$ ($<t_0$) corresponds to the time of the first bounce back in time, $t_2$ ($<t_1$) is the bounce time of the previous cycle, etc., all the way to $t_\infty=-\infty$.
In case (i), one clearly sees that the integrand is always nonzero, so the infinite sum generally diverges.
Moreover, this case is not physically well motivated from the onset since the second law of thermodynamics implies that the maximal size of the universe from one cycle to another should grow as time evolves.
This brings us to case (ii), where the integrand in \eqref{eq:SEHFLRWcs} is asymptotically zero as we go far back in time ($n\to\infty$).
Therefore, the action may converge.
As an example, the average Hubble parameter could be approximately constant, such that the average volume of each cycle grows exponentially (for an example of such scenario, see \cite{Barrow:1995cfa,Steinhardt:2001st,Steinhardt:2004gk,Lehners:2008qe,Barrow:2017yqt,Barrow:2017zar,Ijjas:2019pyf}).
In other words, when averaging over cycles, the universe could be de Sitter like.
In such a case, we recover the result that, classically, the action is finite.\footnote{Whether or not a quantum instability arises--as discussed in Sec.~\ref{sec:inflation} for inflation--is not obvious though.
Also, in the case where $\langle H\rangle_P>0$ and $\langle a\rangle_P\to 0$ as $t\to-\infty$, where $\langle\rangle_P$ means the time average over the cycle period $P$, we note that the spacetime is necessarily geodesically incomplete \cite{Borde:2001nh}.}
In the same vein, a cyclic universe which starts from a zero-size singularity at some finite time in the past can also yield finite semiclassical amplitudes.
Indeed, such cyclic universes only undergo a finite number of cycles up to today, so the analysis of convergence reduces to computing the action in the approach to the big bang in the very first cycle.

In summary, bouncing and cyclic universes are typically highly constrained by a principle of finite amplitudes.
Universes that quickly become infinitely large in the infinite past and cyclic universes that do not reach zero size (at least asymptotically) are all excluded by this principle.
It thus appears very difficult for a universe that never had a beginning to produce well-defined transition amplitudes.
A possible counterexample shall be loitering universes, which is the topic of the next subsection.

\subsubsection{Loitering and emerging universes, aka genesis scenarios} \label{sec:loitering}

There exists another class of nonsingular cosmological models that differ from typical bouncing and cyclic universes.
If the universe is asymptotically Minkowski in the past, in the sense that $a\to\mathrm{constant}$ and $H\to 0$ (and $\dot H\to 0$) as $t\to-\infty$, then one says that the universe is "loitering" or "quasi-static" in that limit.
It is straightforward to see how such scenarios might satisfy the principle of finite amplitudes since the integrand of the Einstein-Hilbert action \eqref{eq:SEHFLRWc} in the limit $t\to-\infty$ would simply vanish.
For a perfect fluid or scalar field, it is also clear that the on-shell action remains finite since the loitering limit is equivalent to $\rho\to 0$ or $V(\phi)\to 0$, while $a\to\mathrm{constant}$ and $t\to-\infty$.

Some motivation for loitering phases appears in string cosmology, where one expects two kinds of stringy modes to be of particular relevance when the universe is small at early times: momentum modes and winding modes.
If the universe has a toroidal geometry, then the winding modes that wrap around various circular directions halt the expansion of the universe in these directions.
Meanwhile the momentum modes contribute to enhance the Hubble rate, i.e., they support the expansion of the universe.
The combination of the two effects can lead to a so-called loitering phase in which the universe is almost static for a period of time, hovering at the Hagedorn temperature, which can be interpreted as the maximal temperature of a gas of strings.
One idea is that such a phase could describe the very earliest moments of the universe.
The reason for this is T-duality, which implies that if the radius of a particular circular direction becomes smaller than the string length (as measured by momentum modes), then there exists a physically equivalent description in terms of winding modes propagating in a geometry for which the circle radius is inverted.
Thus, a universe smaller than the string length becomes equivalent to one larger than the string length.
It then becomes natural to envisage a beginning of the universe at the string length, with both momentum and winding modes being of equal importance.
As suggested by Brandenberger and Vafa \cite{Brandenberger:1988aj}, some winding modes could subsequently annihilate, effectively releasing some directions from the tug-of-war between momentum and winding modes and letting some spatial directions expand indefinitely.
They showed that one-dimensional strings would in fact tend to annihilate preferentially in three spatial directions, thus providing a mechanism for obtaining a universe with three large and six (or seven, depending on the theory) small dimensions.

The model described above is the idea behind string gas cosmology \cite{Brandenberger:1988aj} (see, e.g., \cite{Tseytlin:1991xk,Easther:2004sd,Battefeld:2005av,Brandenberger:2008nx,Gasperini:2007zz} and references therein for reviews).
String cosmology background solutions, which recover the expected behavior $a\to\mathrm{constant}$ as $t\to-\infty$, have been hard to find however.
Since this is beyond the scope of general relativity and "typical" matter, we postpone the discussion of possible models to the next section.
We will present a couple of string cosmology attempts and address the finiteness of amplitudes in such examples in Sec.~\ref{sec:stringcosmo}.

Loitering phases have also been explored from an effective field theory (EFT) perspective, i.e.~agnostic of what the correct ultraviolet-complete theory should be.
There, one postulates some theory of modified gravity and explores solutions that may satisfy $a\to\mathrm{constant}$ as $t\to-\infty$.
The loitering phase is terminated as the scale factor and the Hubble parameter starts growing, which is only possible if $\dot H>0$, effectively violating the null energy condition (NEC) (a necessary requirement to have a nonsingular cosmology in the first place).
For that reason, such cosmological models are also sometimes called "emerging" universes or "genesis" scenarios.
When the reheating phase is reached at high Hubble scale, the NEC is recovered and standard hot big bang cosmology begins.
See, e.g., \cite{Creminelli:2010ba,Creminelli:2012my,Nishi:2015pta,Kobayashi:2015gga,Nishi:2016ljg,Yoshida:2017swb,Mironov:2019qjt,Volkova:2019jlj,Ageeva:2020gti,Ilyas:2020zcb} for examples of such scenarios and their implementation in modified gravity.
In Sec.~\ref{sec:Horndeski} we will explore a few such examples and comment on their validity with respect to the principle of finite amplitudes.

\subsection{No-boundary proposal} \label{sec:noboundary}

There exists a rather different approach to early universe path integrals, known as the ``no-boundary proposal'' \cite{Hawking:1981gb,Hartle:1983ai}. Unlike the models described in the previous sections, the no-boundary proposal is not  
a new dynamical model of the early universe, rather it is a prescription for obtaining a well-defined transition amplitude \emph{given} a dynamical model of the universe. In that sense it is highly relevant to our present discussion.\footnote{We do not present new technical results regarding the no-boundary proposal here. We rather wish to relate its properties to our general analysis of cosmological amplitudes.} The main idea of the no-boundary proposal is that in early universe transition amplitudes, the initial boundary should be removed altogether, leaving only the final boundary. In the path integral, one should then sum over all geometries that do not have a boundary in the past--in other words, one would wish to sum over compact, regular four-manifolds having as their only boundary a specified final hypersurface. Such a sum would automatically render the amplitude finite. Since different final boundary configurations would obtain different amplitudes in general, one would thus obtain a theory of initial conditions. In other words, given a dynamical model of the early universe, the no-boundary proposal can in principle yield different (relative) probabilities for different late-time outcomes. 

Such an idea is evidently very attractive. Before asking whether the predictions (in the context of various dynamical models) might agree with what we observe in the universe, one first has to ask whether the required four-geometries actually exist. In fact, as long as the geometry is restricted to be Lorentzian, they do not--there simply are no compact and regular Lorentzian geometries. However, it is important to realize that the four-manifolds summed over in the path integral must \emph{not} be identified with the physical spacetime. The physical spacetime rather must be thought of as being given by a sequence of final boundary geometries. And just as an ordinary real integral may be well approximated by a complex saddle point, so it may occur that a gravitational path integral is best approximated via a complex saddle point. In such a case, compact and regular geometries can be found. The simplest example consists of an analytic continuation of de Sitter space. In the presence of a positive cosmological constant\footnote{The reduced Planck mass is set unity in this subsection.} $\Lambda = 3 H^2$, the de Sitter solution in the closed slicing is given by
\begin{equation}
 \dd s^2 = - \dd t^2 + \frac{1}{H^2}\cosh^2(Ht) \; \dd\Omega_{(3)}^2\,,
\end{equation}
where $\dd\Omega_{(3)}^2$ represents the line element on the 3-sphere. The de Sitter space may be thought of as a hyperboloid, with the waist of size $a=1/H$ occurring at $t=0$. At that location one may analytically continue the time coordinate into the Euclidean time direction, via the continuation
\begin{equation}
 t= \mp i \left(\tau - \frac{\pi}{2H}\right)\,, \qquad \frac{\pi}{2H} \geq \tau \geq 0\,.
\end{equation}
This is reminiscent of a Wick rotation in field theory, though for now we have kept both possible directions of rotation. The metric then becomes that of a four-sphere,
\begin{equation}
 \dd s^2 = + \dd\tau^2 + \frac{1}{H^2}\sin^2 (H\tau) \; \dd\Omega_{(3)}^2\,,
\end{equation}
and thus the spacetime (now really just a space) is indeed smoothly rounded off at $\tau=0$. Slightly deformed solutions of this type also exist for slow-roll inflation, effectively with the replacement $H \to \sqrt{V(\phi)/3}$ and the scalar field being essentially constant \cite{Lyons:1992ua}. In ekpyrotic models, the geometries look rather different, with a large Euclidean part followed by a transition to a contracting Lorentzian geometry \cite{Battarra:2014xoa}. In all cases, one may also add both scalar and tensor perturbations \cite{Halliwell:1984eu}. Their Fourier modes are given by spherical harmonics with wave number $\ell$, and for simplicity we will retain only a single tensor mode here, since this will be sufficient to illustrate the main consequences of the no-boundary proposal. 

For now, let us go back to the (possibly perturbed) slow-roll solution. If we specify a final scale factor value $\bar{a}$, scalar field value $\bar{\phi}$, and tensor perturbation amplitude $\bar\delta$, then the wave function\footnote{The technical distinction between a wave function and a transition amplitude depends on the detailed definition of the path integral; see \cite{Feldbrugge:2017kzv,DiazDorronsoro:2017hti}. In the absence of a precise definition, we will use the two terms interchangeably in this section.} is approximated by a sum over saddle point contributions $S_\sigma$,
\begin{align}
 \Psi(\bar{a},\bar{\phi},\bar{\delta}) & \approx  \sum_\mathrm{relevant} \exp[\frac{i}{\hbar}S_\sigma(\bar{a},\bar\phi,\bar\delta)] \nonumber \\ 
 &= \sum_\mathrm{relevant} \exp \Bigg[\pm \left(\frac{12\pi^2}{\hbar V(\bar\phi)} -\frac{\ell(\ell+1)(\ell+2)}{2\hbar V(\bar\phi)}\bar{\delta}^2\right) \nonumber \\ &
 \qquad \qquad \qquad \,\,  \pm \;i\;\left(\frac{4\pi^2}{\hbar}\sqrt{\frac{V(\bar\phi)}{3}}\left(\bar{a}^2 - \frac{3}{V(\bar\phi)}\right)^{3/2}+\frac{3\ell(\ell+2)\bar{a}}{2\hbar \sqrt{V(\bar\phi)}}\bar{\delta}^2 \right)\Bigg]\,. \label{nbwf}
\end{align}
Note that there are four different saddle points, depending on the choice of plus/minus signs. This is because to reach a given scale factor $\bar{a}$, we have two choices of Wick rotation and moreover we could let $t$ run in the positive or negative direction. This way we always get four saddle points related by complex conjugation and time reversal. The question then is: which saddle points are relevant and should be included in the sum? Before addressing this question, let us point out a few features of the above expression: if we neglect the perturbations, then we see that the wave function is weighted by $\mathrm{exp}[\pm 12\pi^2/\hbar V(\bar\phi)]$, and thus we really obtain different relative weightings (and thus probabilities) for different positions on the potential. Also, at large scale factor the phase changes in proportion to $\bar{a}^3$, thus showing that a classical universe is an (nontrivial) outcome of the proposal \cite{Hartle:2008ng}. As for the perturbations, we see the emergence of a scale-invariant spectrum, since the dispersion is proportional to $\ell^{-3}$. In other words, the no-boundary proposal implies the Bunch-Davies vacuum, which does not have to be put in as an additional assumption \cite{Halliwell:1984eu}. The last statement however only holds when we are able to choose the plus sign in the first factor in\footnote{We should note that this is the sign advocated by Hartle and Hawking \cite{Hartle:1983ai}, and it is opposite to the sign used by Vilenkin in the closely related tunnelling proposal \cite{Vilenkin:1982de}.} \eqref{nbwf}. Otherwise perturbations are actually enhanced, leading to unphysical predictions \cite{Feldbrugge:2017fcc}. This brings us back to the crucial question of which saddle points are to be included in the path integral.  

There is a conundrum here: on the one hand the no-boundary proposal suggests that we should not have a second boundary, yet on the other hand we must integrate from \emph{some} field value up to the values specified on the final boundary. In a sense, we are supposed to impose boundary conditions such that there is no boundary. The saddle points described above all have the property that $a(\tau=0)=0$, i.e., they start from zero size. A natural prescription is thus to integrate over all geometries that start at zero size, very much like what we have done in previous sections. This corresponds to imposing Dirichlet boundary conditions at the initial boundary \cite{Feldbrugge:2017kzv,DiazDorronsoro:2017hti}. It was shown in \cite{Feldbrugge:2017mbc} however that in this case, whenever one includes a saddle point with stable perturbations, one automatically has to include the complex conjugate saddle point, which has the opposite Wick rotation and unstable perturbations. Note that the Wick rotation is revealed by the derivative of the scale factor, $a_{,\tau}(\tau=0)=\pm 1$ or $\dot{a}(\tau=0)=\pm i$. But because of the uncertainty principle we cannot fix both $a$ and $\dot{a}$ simultaneously. This suggests an alternative approach however, developed in \cite{Louko:1988bk,DiTucci:2019dji,DiTucci:2019bui}: we may simply fix the initial expansion rate $\dot{a}=+i$ and then only the stable saddle points contribute.\footnote{It is interesting to note that this prescription also reproduces the canonical partition function in asymptotically anti-de Sitter space when the cosmological constant is negative \cite{DiTucci:2020weq}.} In this case the wave function reduces to the expression
\begin{align}
 \Psi_\mathrm{Neumann}(\bar{a},\bar{\phi},\bar{\delta}) \approx &\,\exp[+ \left(\frac{12\pi^2}{\hbar V(\bar\phi)} -\frac{\ell(\ell+1)(\ell+2)}{2\hbar V(\bar\phi)}\bar{\delta}^2\right)]\nonumber\\
 &\times\left\{\exp[+i\;\left(\frac{4\pi^2}{\hbar}\sqrt{\frac{V(\bar\phi)}{3}}\left(\bar{a}^2 - \frac{3}{V(\bar\phi)}\right)^{3/2}+\frac{3\ell(\ell+2)\bar{a}}{2\hbar \sqrt{V(\bar\phi)}}\bar{\delta}^2 \right)] + \textrm{c.c.} \right\}\,.
\end{align}
Now the perturbations are truly in the Bunch-Davies vacuum. We are however still left with two contributing saddle points (the second giving the complex conjugate contribution above), which correspond to universes that are the time reverse of each other. A prominent view is that these will quickly decohere as the universe grows and can be treated independently once the universe's size exceeds the Hubble radius \cite{Hartle:2008ng}. Note that the weighting for the background favors small values of the potential. This constitutes an unresolved aspect of the no-boundary proposal when applied to inflation: the no-boundary proposal prefers a very short inflationary phase at low values of the potential. This seems to be in some tension with observations, though one should bear in mind that we do not yet understand the structure of the scalar potential (as exemplified by recent swampland discussions \cite{Agrawal:2018own}) and only have a partial understanding of probabilities in quantum cosmology \cite{Vilenkin:1988yd}. For ekpyrotic cosmologies, the wave function takes an analogous form, but with the replacement $V(\bar\phi) \to |V(\bar\phi)|$ and with the understanding that $\bar\phi$ represents the value of the ekpyrotic scalar at the beginning of the ekpyrotic phase. This means that the predictions for ekpyrotic models are rather opposite to those of inflation: in this case the no-boundary proposal predicts a long ekpyrotic phase (quickly rendering the universe classical \cite{Lehners:2015sia}), starting at small magnitudes of the potential. The weighting is also preserved across effective nonsingular bounces \cite{Lehners:2015efa}. Thus, in a potential landscape, the no-boundary proposal favors ekpyrotic regions, but whether or not these lead to realistic universe still hinges on understanding the bounce and reheating in fundamental physics.

The no-boundary proposal provides a remarkable illustration of how cosmological amplitudes may be finite: it simultaneously avoids singularities in spacetime and in the saddle point action, and to this extent its implications remain fully in the realm of semiclassical physics. What is more, no-boundary solutions are stable to the addition of higher curvature corrections stemming from loop corrections to general relativity \cite{Jonas:2020pos} (see also \cite{Cano:2020oaa,Narain:2021bff}), thus indicating that no-boundary solutions may persist in full quantum gravity. From this vantage point one may be tempted to reformulate the subject of the present paper as asking whether there exist other prescriptions for semiclassical cosmological amplitudes that are equally sound. We will come back to this point in the discussion section.


\section{Examples involving physics beyond general relativity} \label{sec:qgcorrections}

\subsection{Approach to the big bang in higher-order derivative gravity\label{subsec:higherorder}}

So far we restricted our attention to models based on general relativity, i.e., up to here the action for gravity was given by the Einstein-Hilbert term
\begin{equation}
 S_\textrm{EH}=\int\dd^4x\,\sqrt{-g}\,\frac{M_\mathrm{Pl}^2}{2}R\,,
\end{equation}
possibly supplemented by a cosmological constant and boundary terms. However, general relativity is known to be nonrenormalizable \cite{Goroff:1985th}. Loop corrections at order $\hbar^L$ induce terms that involve the Riemann tensor to the power $L+1$, and such terms may very well play an important role in the approach to the big bang. For this reason we want to investigate theories of the form 
\begin{equation}
 S=\frac{1}{2}\int\dd^4x\,\sqrt{-g}\,f(R_{\mu\nu\rho\sigma})\,,\label{eq:fRmunurhosigma}
\end{equation} 
where $f(R_{\mu\nu\rho\sigma})$ is an \textit{a priori} arbitrary (scalar, polynomial) function of the Riemann tensor. An exception is played by quadratic gravity, which stands out by being a renormalizable theory of gravity in and of itself \cite{Stelle:1976gc,Stelle:1977ry}, given by the action
\begin{equation}
 S_\mathrm{quad}=\int\dd^4x\,\sqrt{-g}\left(\frac{M_\mathrm{Pl}^2}{2}R+\frac{\omega}{3\sigma}R^2-\frac{1}{2\sigma}C^2+\theta\mathcal{G}\right)\,,\label{eq:quadgrav}
\end{equation}
where $\omega$, $\sigma$ and $\theta$ are coupling constants, and
where the Weyl tensor squared $C^2$ and the Gauss-Bonnet combination $\mathcal{G}$ (which yields a total derivative) can be expressed in terms of the Riemann and Ricci tensors as
\begin{align}
 C^2&\equiv C_{\mu\nu\rho\sigma}C^{\mu\nu\rho\sigma}=R_{\mu\nu\rho\sigma}R^{\mu\nu\rho\sigma}-2R_{\mu\nu}R^{\mu\nu}+\frac{1}{3}R^2\,,\nonumber\\
 \mathcal{G}&=R_{\mu\nu\rho\sigma}R^{\mu\nu\rho\sigma}-4 R_{\mu\nu}R^{\mu\nu}+R^2\,.\label{eq:C2}
\end{align}
Quadratic gravity will play a dual role: we may treat it as a complete theory on its own, but also as an example of the impact of higher curvature terms. The main potential flaw of quadratic gravity is that it suffers from the presence of a ghost, and it is still a matter of debate how this should be interpreted \cite{Salvio:2019ewf,Donoghue:2019fcb}.\footnote{If quadratic gravity, viewed as a fundamental theory, does violate unitarity, then we expect our principle of finite amplitudes to be violated for any cosmology that includes perturbations about a given background. For this reason, we restrict ourselves to background solutions in this subsection and remain agnostic about the resolution of the ghost problem in such a case. If quadratic gravity is only an effective theory, in which case the ghost might be outside the regime of validity of the theory or be resolved thanks to higher-curvature quantum corrections, then we can still ask what are the implications of our principle of finite amplitudes.} Quadratic gravity nonetheless remains a popular theory, e.g., in the context of cosmology, Starobinsky's inflationary model makes use of it \cite{Starobinsky:1986fx} and is one of the favored inflationary mechanisms to explain the latest observations from the 2018 Planck survey \cite{Akrami:2018odb}. In more general $f(R_{\mu\nu\rho\sigma})$ theories, a ghost also appears if the theory stops at a given order in the curvature tensor, but for an infinite series the ghost may be resummed and disappear. Low-energy expansions of string theory are of this type and additionally contain terms with derivatives acting on the Riemann tensor; see, e.g., \cite{Metsaev:1987zx,Green:2010wi}.

If we want to investigate the approach to the big bang, then in the path integral we must be able to fix the metric on the initial boundary. This means that we must supplement the bulk action with the appropriate surface terms. For $f(R_{\mu\nu\rho\sigma})$ theories, this has been studied in \cite{Deruelle:2009zk}. Using the Arnowitt-Deser-Misner (ADM) decomposition of the metric
\begin{equation}
 \dd s^2=-N^2\dd t^2+h_{ij}(\dd x^i+N^i\dd t)(\dd x^j+N^j\dd t)\,,\label{eq:ADMdecomposition}
 \end{equation} 
the authors of \cite{Deruelle:2009zk} showed that the generalization of the GHY boundary term is given by
\begin{equation}
 S_\mathrm{boundary}=-\int_{\partial\mathcal{M}}\dd^3x\,\sqrt{h}\,\Psi^{ij}K_{ij}\,,\qquad\textrm{with}\quad \left.\delta\Psi^{ij}\right|_{\partial\mathcal{M}}=0\,,\label{eq:boundaryterm}
\end{equation}
where $K_{ij}$ is the extrinsic curvature on the boundary hypersurface $\partial\mathcal{M}$ and
\begin{equation}
 \Psi^{ij}[f]=-\frac{1}{2}h^{ik}h^{jl}n^{\mu}n^{\nu}\frac{\partial f}{\partial R^{\mu k \nu l}}\,,
\end{equation}
with $n^\mu=(1/N,-N^i/N)$ denoting the unit normal vector to the boundary hypersurface.
Adding the boundary term \eqref{eq:boundaryterm} trades the boundary condition on the extrinsic curvature $K_{ij}$ for a boundary condition on $\Psi^{ij}$. In general, this will also involve new degrees of freedom, i.e., fixing $\Psi^{ij}$ might not only fix the metric but also additional scalar/tensor degrees of freedom. 

It is instructive to look at a few examples. We start with the Ricci scalar and show that it reproduces the GHY result. Using the ADM decomposition we can write
\begin{equation}
 R=g^{\mu\rho}g^{\nu\sigma}R_{\mu\nu\rho\sigma}=h^{ik}h^{jl}R_{ijkl}-2n^\mu n^\rho h^{ij}R_{\mu i\rho j}\,,
\end{equation}
where we used the symmetries of the Riemann tensor to simplify the expression. Then we find
\begin{equation}
 \frac{\partial R}{\partial R_{\mu i\nu j}}=-2n^{\mu}n^{\nu}h^{ij}\quad\Rightarrow\quad \Psi^{ij}[R]=h^{ij}\,.
\end{equation}
Therefore $\Psi^{ij}K_{ij}=K$ and we indeed recover the result \eqref{eq:GHY}. Fixing $\Psi^{ij}$ thus fixes the three-metric on the boundary, as expected. For $f=R^n$, we have\footnote{More generally for $f=f(R)$, one recovers the known result (e.g., \cite{Dyer:2008hb}): $\Psi^{ij}[f(R)]=h^{ij}f_{,R}$.} $\Psi^{ij}[R^n]=nh^{ij}R^{n-1}$, so fixing $\Psi^{ij}$ requires in addition that $R$ is specified on the boundary, i.e., there is one new scalar degree of freedom. This is valid for all $f(R)$ theories and is consistent with the fact that all these theories are equivalent to general relativity plus a scalar field \cite{Barrow:1988xh}. Consider now the Ricci tensor squared $R^{\mu\nu}R_{\mu\nu}$. Its ADM decomposition is 
\begin{align}
 R^{\mu\nu}R_{\mu\nu}=&~R_{ij}R^{ij}-2n^\mu n^\nu h^{ij}R_{\mu i}R_{\nu j}+n^\mu n^\nu n^\alpha n^\beta R_{\mu\nu}R_{\alpha\beta}\nonumber\\
 =&~(R^k{}_{ikj}-n^\mu n^\nu R_{\mu i\nu j})R^{ij}-2n^\mu n^\nu h^{ij}(R^k{}_{\mu k i}-n^\alpha n^\beta R_{\alpha\mu\beta i})R_{\nu j}\nonumber\\
 &+n^\mu n^\nu n^\alpha n^\beta(R^k{}_{\mu k\nu}-n^{\rho}n^{\sigma}R_{\rho\mu\sigma\nu})R_{\alpha\beta}\,,
\end{align}
so
\begin{align}
 &\frac{\partial(R_{\alpha\beta}R^{\alpha\beta})}{\partial R_{\mu i\nu j}}=-2n^{\mu}n^{\nu}R^{ij}+2n^{\mu}n^{\nu}h^{ij}n^{\rho}n^{\sigma}R_{\rho\sigma}\nonumber\\
 \quad\Rightarrow&\quad \Psi^{ij}[R_{\mu\nu}R^{\mu\nu}]=-n_{\alpha}n_{\beta}(g^{\alpha\beta}R^{ij}+h^{ij}R^{\alpha\beta})\,.\label{eq:riccisquareboundary}
\end{align}
Similarly, for the Riemann tensor, we find
\begin{align}
 R^{\mu\nu\rho\sigma}R_{\mu\nu\rho\sigma}=&\ R^{ijkl}R_{ijkl}+n^\mu n^\nu n^\alpha n^\beta\left(4R_{\mu i\nu j}R_\alpha{}^i{}_\beta{}^j+2R_{ij\mu\nu}R^{ij}{}_{\alpha\beta}\right)\nonumber\\
 &+4n^\mu n^\nu n^\rho n^\alpha n^\beta n^\gamma R_{\mu\nu\rho i}R_{\alpha\beta\gamma}{}^{i}+4n^\mu n^\alpha R_{\mu ijk}R_{\alpha}{}^{ijk}\,,
\end{align}
so that
\begin{equation}
 \frac{\partial(R_{\alpha\beta\gamma\delta}R^{\alpha\beta\gamma\delta})}{\partial R_{\mu i\nu j}}=8n^\mu n^\nu n^\alpha n^\beta R_{\alpha}{}^i{}_\beta{}^j\quad\Rightarrow\quad \Psi^{ij}[R_{\mu\nu\rho\sigma}R^{\mu\nu\rho\sigma}]=-4n_\alpha n_\beta R^{\alpha i\beta j}\,.\label{eq:riemannboundaryterm}
\end{equation}
We can then deduce the boundary term of the Weyl tensor squared, defined above in Eq.~\eqref{eq:C2}, and we find 
\begin{align}
 &\Psi^{ij}[C^2]=-n_{\alpha}n_{\beta}\Big(4R^{\alpha i\beta j}-2(g^{\alpha\beta}R^{ij}+h^{ij}R^{\alpha\beta})-\frac{2}{3}R h^{ij}g^{\alpha\beta}\Big)=-4n_{\alpha}n_{\beta}C^{\alpha i\beta j}\,,\label{eq:C2boundary}
\end{align}
which matches the boundary term for $C^2$ found in \cite{Hawking:1984ph,Hohm:2010jc}. However, the boundary term for the Gauss-Bonnet combination $\mathcal{G}$ cannot be obtained through the method exposed in \cite{Deruelle:2009zk}, as it is a degenerate topological term in four-dimensional spacetime. Instead, the correct surface term for $\mathcal{G}$ (i.e.~enabling the imposition of Dirichlet boundary conditions $\left.\delta g_{\mu\nu}\right|_{\partial\mathcal{M}}=0$) is given by the Myers action \cite{Myers:1987yn,Deruelle:2017xel}
\begin{equation}
 \Psi^{ij}[\mathcal{G}]=2\left(JK^{ij}-2G^{ij}_\mathrm{b}\right)\,,
\end{equation}
where $G^\mathrm{b}_{ij}$ is the Einstein tensor constructed from the boundary-induced metric $h_{ij}$ and 
\begin{equation}
 J=-h^{ij}\left(-\frac{2}{3}K_{il}K^{lp}K_{pj}+\frac{2}{3}KK_{il}K^{l}{}_{j}+\frac{1}{3}K_{ij}(K_{lp}K^{lp}-K^2)\right)\,.
\end{equation}

We have everything in place now to investigate the finiteness of cosmological amplitudes of interest. We will start with the simplest backgrounds, namely, FLRW metrics with spatial curvature $k=-1,0,+1$ for open, flat, or closed spatial slices. On such backgrounds, the action of a $f(R_{\mu\nu\rho\sigma})$ theory simplifies considerably and may be written as \cite{Jonas:2020pos}
\begin{equation}
 S=\int\dd t\,a^3N\sum_{p_1,p_2\in\mathbb{N}}c_{p_1,p_2}\big(A_1^{(k)}\big)^{p_1}A_2^{p_2} \label{eq:A1A2action}
\end{equation}
for some coefficients $c_{p_1,p_2}$ and with
\begin{equation}
 A_1^{(k)}=\frac{\dot{a}^2+kN^2}{a^2N^2}\quad\textrm{and}\quad A_2=\frac{\ddot{a}N-\dot{a}\dot{N}}{aN^3}\,.
\end{equation}
For example $R=6\big(A_1^{(k)}+A_2\big)$ and $\mathcal{G}=24A_1^{(k)}A_2$.  The order in the Riemann tensor of the expression $f(R_{\mu\nu\rho\sigma})$ is $P=p_1+p_2$.
The constraint equation (a first integral of the equation of motion) for the action \eqref{eq:A1A2action} is given by \cite{Jonas:2020pos}
\begin{align}
 &\sum_{p_1,p_2}c_{p_1,p_2}\bigg[2p_1(p_2-1)\frac{a\dot{a}^2}{N^2}A_2^{p_2}\big(A_1^{(k)}\big)^{p_1-1}+(1-p_2)a^3A_2^{p_2}\big(A_1^{(k)}\big)^{p_1}\nonumber\\
 &\quad\quad+p_2(p_2-1)\frac{a\dot{a}\dddot{a}}{N^4}A_2^{p_2-2}\big(A_1^{(k)}\big)^{p_1}-p_2(2p_1+p_2-3)\frac{a\dot{a}^2}{N^2}A_2^{p_2-1}\big(A_1^{(k)}\big)^{p_1}\bigg]=0 \label{eq:friedmannhigherorder}
\end{align}
upon setting $\dot N=0$.
We may use \eqref{eq:friedmannhigherorder} to simplify the action \eqref{eq:A1A2action} on-shell.

We illustrate this method by applying it first to quadratic gravity. Since the Weyl tensor vanishes for FLRW metrics, the action \eqref{eq:quadgrav} reduces to
\begin{align}
 S&=\int\dd^4x\,\sqrt{-g}\,\left(\frac{M_\mathrm{Pl}^2}{2}R+\frac{\omega}{3\sigma} R^2+\theta\mathcal{G}\right)\nonumber\\
 &=\int\dd t\,\bigg[3M_\mathrm{Pl}^2\Big(\frac{a\dot{a}^2}{N}+aNk+\frac{a^2\ddot{a}}{N}-\frac{a^2\dot{a}\dot{N}}{N^2}\Big)+\frac{12\omega}{\sigma}\Big(\frac{\dot{a}^4}{aN^3}+\frac{2k\dot{a}^2}{aN}+\frac{Nk^2}{a}+\frac{2\dot{a}^2\ddot{a}}{N^3}\nonumber\\
 &\qquad+\frac{2k\ddot{a}}{N}-\frac{2\dot{a}^3\dot{N}}{N^4}-\frac{2k\dot{a}\dot{N}}{N^2}+\frac{a\ddot{a}^2}{N^3}-\frac{2a\dot{a}\ddot{a}\dot{N}}{N^4}\Big)+24\theta\Big(\frac{\dot{a}^2\ddot{a}}{N^3}+\frac{k\ddot{a}}{N}-\frac{\dot{a}^3\dot{N}}{N^4}-\frac{k\dot{a}\dot{N}}{N^2}\Big)\bigg].\label{eq:SquadFLRW}
\end{align}
The constraint \eqref{eq:friedmannhigherorder} for this action reads
\begin{equation}
 0=3M_\mathrm{Pl}^2\Big(\frac{a\dot{a}^2}{N^2}+ak\Big)+\frac{12\omega}{\sigma}\bigg[-\frac{3\dot{a}^4}{aN^4}-\frac{2k\dot{a}^2}{aN^2}+\frac{k^2}{a}-\frac{a\ddot{a}^2}{N^4}+\frac{2\dot{a}^2\ddot{a}}{N^4}+\frac{2a\dot{a}\dddot{a}}{N^4}\bigg]\,.
\end{equation}
We may use it to replace the term proportional to $k^2/a$ in the action, which will change the prefactors of the other terms in the on-shell action. If we now assume an approach to the big bang of the form $a(t) \sim t^s$, then the on-shell action will only be made up of terms scaling as\footnote{We are using an ansatz $a(t) \sim t^s$ which is generic in the presence of standard matter. An exception is the case of a pure cosmological constant, in which case the Riemann tensor is finite and constant everywhere and will thus not diverge in the approach to the big bang. In such a case we however encounter the "incompleteness" problem described in Sec.~\ref{sec:inflation}, namely that the two backgrounds then interfere and fail to reproduce an appropriate ground state for the perturbations \cite{DiTucci:2019xcr}.}
\begin{equation}
 t^{3s}\,,\quad t^{s+1}\,,\quad t^{3s-1}\,,\quad t^{s-1}\,,\quad t^{3s-3}\,.
\end{equation}
The action converges when $t \to 0$ as long as $s>1$, i.e.~as long as the background expands in an accelerated fashion.  As previously discussed, we also need to make sure that the boundary terms are well behaved. For the action \eqref{eq:SquadFLRW}, they read
\begin{align}
 &-M_\mathrm{Pl}^2\int\dd^3x\,\sqrt{h}\,K-\frac{2\omega}{3\sigma}\int\dd^3x\,\sqrt{h}\,RK-4\theta\int\dd^3x\,\sqrt{h}\left(J-2G^\mathrm{b}_{ij}K^{ij}\right)\,.
\end{align}
On a FLRW background, we have
\begin{equation}
 K=K_{ij}h^{ij}=\frac{\dot{a}}{aN}h_{ij}h^{ij}=\frac{\dot{a}}{aN}\,,\quad J=\frac{2\dot{a}^3}{N^3a^3}\,,\quad G_{ij}^\mathrm{b}K^{ij}=-\frac{3k\dot{a}}{Na^3}\,,
\end{equation}
so using the ansatz $a\sim t^s$ the boundary terms yield the exponents $t^{3s-1}$ (GHY term), $t^{3s-3}$, and $t^{s-1}$, which are once again finite (zero) when $s>1$ at $t=0$. Thus we recover the result that in an inflationary background (which may arise directly as a solution of pure quadratic gravity or from a coupling to a scalar field with a sufficiently flat potential), quadratic gravity leads to a finite action in the approach to the big bang \cite{Lehners:2019ibe}.

Now we can extend this method to the general action \eqref{eq:A1A2action}.
Using the ansatz $a\sim t^{s}$, we find that
\begin{equation}
 A_1^{(k)}\sim t^{-2}+k\cdot t^{-2s}\,,\qquad A_2\sim t^{-2}\,,
\end{equation}
so the behavior will be very different for $k$ equal or not equal to $0$.

In the spatially flat case $k=0$, the action \eqref{eq:A1A2action} has the exponents $t^{1+3s-2P}$, so it is finite for $s>(2P_\mathrm{max}-1)/3$, where $P_\mathrm{max}$ is the highest order of $P=p_1+p_2$ that appears in the action (generally $P_\mathrm{max}\geq 3$ for a theory that is at least cubic in curvature). All the terms of the constraint equation go like $t^{3s-2P}$, so we can in general get rid of the highest order in $t$ and keep as condition for finiteness $s>(2P_\mathrm{max}-3)/3 \geq 1$.
However this will lead to a divergence in the action unless the quantity $P_\mathrm{max}$ is bounded. Thus, on flat FLRW, generic $f(R_{\mu\nu\rho\sigma})$ theories have finite on-shell action only if they have a bounded order in the Riemann tensor and if the scale factor undergoes sufficiently accelerated expansion. \footnote{Here we assumed no symmetry induced cancellations between different on-shell terms in the action, as our emphasis was on generic theories. A prominent exception are $f(R)$ theories, where in some sense the higher-curvature terms are spurious, as they may be replaced by a scalar field.}

When the spatial sections are not flat, $k\neq 0$, it becomes impossible to keep the action finite. There are $p_1+1$ different orders in $t$ for the action, 
\begin{equation}
 t^{1+3s-2P+2(1-s)\cdot j}\quad\mbox{with}\quad j\in\lbrace 0,\dots,p_1\rbrace\,.
\end{equation}
The $j=1$ and $j=2$ terms yield the conditions
\begin{equation}
 s>2P-3\quad\textrm{and}\quad s<5-2P\,.\label{eq:contradiction}
\end{equation}
Since $P=p_1+p_2\geq 2$, these conditions imply $s>1$ for the first and $s<1$ for the second, which are mutually exclusive. We may still hope to be able to use the constraint equation to ease the tension. Plugging $a\sim t^s$ into \eqref{eq:friedmannhigherorder}, we find that the orders in $t$ appearing in the constraint equation are 
\begin{equation}
 t^{3s-2P+2(1-s)\cdot j}\quad\mbox{with}\quad j\in\lbrace 0,\dots,p_1\rbrace\,.
\end{equation}
Thus, the constraint equation will generally enable us to replace the order $t^{3s-2P+2(1-s)p_1}$ by a combination of the $p_1$ other orders in $t$, but for $p_1>2$ the contradiction \eqref{eq:contradiction} cannot be resolved. Whenever the order in the Riemann tensor is higher than 2, using the constraint equation is therefore not sufficient (again for generic theories, see the previous footnote) to obtain a finite on-shell action. 

Hence a principle of finite amplitudes is seen to be highly constraining regarding higher-curvature theories of gravity: if the highest power in the Riemann tensor $P_\mathrm{max}$ is larger than $2$, then Lorentzian FLRW backgrounds with nonzero spatial curvature are eliminated. Spatially flat backgrounds only lead to a finite action if the scale factor $a \sim t^s$ is highly accelerating with $s>(2P_\mathrm{max}-3)/3$. If there are infinitely many higher-curvature terms, as expected from loop corrections, then all Lorentzian FLRW backgrounds with standard matter contributions are eliminated. As we already mentioned in Sec.~\ref{sec:noboundary}, the no-boundary proposal shows that if one is prepared to depart from Lorentzian metrics by enlarging the four-manifolds that are summed in the path integral to the complex domain, then a finite amplitude may be obtained, even for an infinite sum of higher order terms containing the Riemann tensor \cite{Jonas:2020pos}. But a Lorentzian big bang becomes impossible. 

From our considerations so far, it has emerged that quadratic gravity plays a somewhat special role, as it is constraining enough to require accelerated backgrounds, but not so constraining as to eliminate Lorentzian big bang spacetimes altogether. Thus it is worthwhile examining this theory in a little more detail, by also looking at what happens once anisotropies are included. For ordinary general relativity, the generic solution when $t\to 0$ is the Belinski-Khalatnikov-Lifshitz (BKL) approach to singularities \cite{Belinsky:1970ew}, where $a\sim t^{1/3}$. It is also known that this type of solution continues to exist when one adds higher-order curvature terms. However, we just saw that in order for quadratic gravity to yield finite action, we need to approach the singularity with an accelerating solution, i.e., the scale factor must grow faster than $a\sim t$. This means that for quadratic gravity, the generic Kasner solutions are ruled out by the principle of finite amplitudes. Then, nongeneric, accelerating solutions become important because they are the only ones left. The question is now whether such solutions exist, so that they may play the role of saddle points of the path integral.

To answer this, we first consider the action $R+R^2+\textrm{matter}$, which can be transformed by going to the Einstein frame into the action $R+\textrm{inflationary scalar field}+\textrm{matter}$. To analyze anisotropies, we can, e.g., consider a Bianchi-I metric 
\begin{equation}
 \dd s^2_\mathrm{I}=-\dd t^2+a^2\left(e^{2(\beta_++\sqrt{3}\beta_-)}\dd x^2+e^{2(\beta_+-\sqrt{3}\beta_-)}\dd y^2+e^{-4\beta_+}\dd z^2\right)\,, \label{eq:B1}
\end{equation}
which leads to the Friedmann equation $3M_\mathrm{Pl}^2H^2=3(\dot{\beta}_+^2+\dot{\beta}_-^2)+\rho_{\textrm{infl}}+\rho_{w}$. Here the matter is taken to be a perfect fluid $p=w\rho$, so that using the continuity equation $\dot{\rho}+3H(\rho+p)=0$, we find $\rho_w\propto a^{-3(1+w)}$, and by integrating the equation of motion for the anisotropy $\ddot{\beta}_\pm+3H\dot{\beta}_\pm=0$, one gets $\dot{\beta}_\pm\propto a^{-3}$. Therefore, the Friedmann equation becomes
\begin{equation}
 3M_\mathrm{Pl}^2H^2=\frac{\sigma_0^2}{a^6}+\rho_{\text{infl}}+\frac{\rho_0}{a^{3(1+w)}}\,,
\end{equation}
where $\sigma_0^2\equiv(\dot\beta_+^2+\dot\beta_-^2)|_{a=1}$ and $\rho_0\equiv\rho_w(a=1)$ are integration constants that represent the energy density in anisotropies and matter at the reference scale $a=1$, respectively.
On an accelerating solution $\rho_{\textrm{infl}} \approx \textrm{constant}$, and hence (for $w\leq 1$) the anisotropic part proportional to $1/a^6$ eventually dominates the dynamics in the approach to $a\to 0$. The solution will then morph into a standard Kasner solution, leaving us with the conclusion that in the presence of anisotropies, there are no dynamical solutions to the $R+R^2$ action that accelerate all the way back to the singularity. 

Next, let us add to this action the Weyl squared term, so that we recover full quadratic gravity.
Remarkably, exact anisotropic inflating solutions to quadratic gravity \eqref{eq:quadgrav} supplemented by a positive cosmological constant $\Lambda$ have been found in \cite{Barrow:2006xb}. For a metric of Bianchi-I type, denoting $a(t)=e^{Ht}$, $\beta_+(t)=\sigma_+t$ and $\beta_-(t)=\sigma_-t$ for constant $H$ and $\sigma_\pm=\dot\beta_\pm$, the solutions are given by
\begin{equation}
 \left\lbrace
\begin{aligned}
 &H^2=\frac{4}{27}\frac{\Lambda}{M_\mathrm{Pl}^2}(2-\omega) - \frac{M_\mathrm{Pl}^2}{36} \sigma\,,\\
 &\Sigma^2\equiv\dot\beta_+^2+\dot\beta_-^2=\sigma_+^2+\sigma_-^2=\frac{2}{27}\frac{\Lambda}{M_\mathrm{Pl}^2}(1+4\omega) + \frac{M_\mathrm{Pl}^2}{18} \sigma\ .
\end{aligned}
 \right.\label{eq:exactaniinfsol}
\end{equation}
Solutions exist provided the coupling constants $\omega,\,\sigma$ take values such that the right-hand sides above are positive. Note that these solutions are intrinsic to quadratic gravity, as they do not have a well-defined limit when $1/\sigma\to 0$. (The isotropic de Sitter solution is recovered with $\omega=-1/4$ and $\sigma\to 0$.) Closely related solutions also exist for other types of Bianchi metrics (II, IV, VIh, and VIIh; see \cite{Barrow:2005qv,Barrow:2006xb,Barrow:2009gx,Middleton:2010bv}).
Interestingly, the above solution is nonsingular, in the sense that the four-curvature remains bounded and time runs all the way to $t=-\infty$. Whether this represents the full spacetime (or whether it may be extended as in the global de Sitter patch) is not known. A similar solution to the above can be obtained from a purposely built limiting-curvature theory, which will be the subject of the next subsection. More comments about spacetime extendibility will be given then. Computing the on-shell action of quadratic gravity on this solution, one finds
\begin{align}
 \frac{\omega}{3\sigma}\int\dd^4x\,\sqrt{-g}\,R^2= &~\frac{12\omega}{\sigma}\int_{-\infty}^{t_0}\dd t\, a^3\left[H^2+ \frac{\ddot{a}}{a}+(\dot{\beta}_+^2 + \dot\beta_-^2)\right]^2
 =~\frac{4 \omega}{\sigma}\frac{e^{3Ht_0}}{H}\left(2H^2+\Sigma^2\right)^2\,,\label{R2bulk}
\end{align}
\begin{align}
 -\frac{1}{2\sigma}\int\dd^4x\,\sqrt{-g}\,C^2 = &~-\frac{6}{\sigma} \int_{-\infty}^{t_0} \dd t \, a^3 \left[ (\ddot{\beta}_+^2 + \ddot{\beta}_-^2) +2 H (\dot{\beta}_- \ddot{\beta}_-+\dot{\beta}_+ \ddot{\beta}_+-2  \dot{\beta}_+^3 + 6 \dot{\beta}_-^2 \dot{\beta}_+) \right. \nonumber \\
 & \qquad\qquad\left. +H^2 (\dot{\beta}_+^2 + \dot\beta_-^2) +4 (\dot{\beta}_+^2+\dot\beta_-^2)^2 +4 (\dot{\beta}_-^2 \ddot{\beta}_+ -\dot{\beta}_+^2 \ddot{\beta}_+  +2  \dot{\beta}_- \ddot{\beta}_- \dot{\beta}_+) \right]\nonumber\\
 =&~-\frac{2}{\sigma}\frac{e^{3Ht_0}}{H}\left(4H\sigma_+(3\sigma_-^2-\sigma_+^2)+H^2\Sigma^2+4\Sigma^4\right)\,,
\end{align}
\begin{align}
 \theta\int\dd^4x\,\sqrt{-g}\,\mathcal{G}= &~24\theta\int_{-\infty}^{t_0}\dd t\,a^3\bigg[H^2\left( \frac{\ddot{a}}{a}-2(\dot{\beta}_+^2 + \dot\beta_-^2)\right)+2
 \left(\dot{\beta}_-^2 \ddot{\beta}_+ +2 \dot{\beta}_- \ddot{\beta}_-
 \dot{\beta}_+-\dot{\beta}_+^2 \ddot{\beta}_+\right)\nonumber\\
 &\qquad\qquad\quad-\frac{\ddot{a}}{a}
 \left(\dot{\beta}_-^2+\dot{\beta}_+^2\right)+H
 \left(6 \dot{\beta}_-^2 \dot{\beta}_+ -2(\dot{\beta}_- \ddot{\beta}_-+\dot{\beta}_+\ddot{\beta}_++\dot{\beta}_+^3)\right)\bigg]\nonumber\\
 =&~8\theta\frac{e^{3Ht_0}}{H}\left(H^4-3H^2\Sigma^2+2H\sigma_+(3\sigma_-^2-\sigma_+^2)\right)\,,\label{eq:GBterm}
\end{align}
while the associated boundary terms are [note that the factor 2 in front comes from the $1/2$ in the definition of \eqref{eq:fRmunurhosigma}]
\begin{align}
 &-2\cdot\frac{\omega}{3\sigma}\int\dd^3x\,\sqrt{h}\,\Psi^{ij}[R^2]K_{ij}=-\frac{4\omega}{3\sigma}\left[a^3Rh^{ij}K_{ij}\right]_{-\infty}^{t_0}\nonumber\\
 &\qquad\qquad\qquad\qquad\qquad\qquad\quad =-\frac{24\omega}{\sigma}\left[a^3H \left(\frac{\ddot{a}}{a}+H^2+(\dot{\beta}_+^2 + \dot\beta_-^2)\right)\right]_{-\infty}^{t_0}\nonumber\\
 &\qquad\qquad\qquad\qquad\qquad\qquad\quad=-\frac{24\omega}{\sigma}e^{3Ht_0}H \left[2H^2+\Sigma^2\right]\,,
\end{align}
\begin{align}
 &-2\cdot\left(-\frac{1}{2\sigma}\right)\int\dd^3x\,\sqrt{h}\,\Psi^{ij}[C^2]K_{ij}=-\frac{4}{\sigma}\left[a^3 n_{0}n_{0}C^{0 i 0 j}K_{ij}\right]_{-\infty}^{t_0}\nonumber\\
 &\qquad\qquad\qquad\qquad\qquad\qquad\quad\,=\frac{12}{\sigma}\left[a^3H \big(\dot{\beta}_-^2+\dot{\beta}_+^2\big)+a^3\big(6
 \dot{\beta}_-^2 \dot{\beta}_+-2
 \dot{\beta}_+^3+\dot{\beta}_- \ddot{\beta}_-+\dot{\beta}_+ \ddot{\beta}_+\big)\right]_{-\infty}^{t_0}\nonumber\\
 &\qquad\qquad\qquad\qquad\qquad\qquad\quad\,=\frac{12}{\sigma} e^{3Ht_0}\left[H\Sigma^2+2\sigma_+(3
 \sigma_-^2-
 \sigma_+^2)\right]\,,
\end{align}
\begin{align}
 &-2\cdot\theta\int\dd^3x\,\sqrt{h}\,\Psi^{ij}[\mathcal{G}]K_{ij}=-4\theta\left[J-2G^\mathrm{b}_{ij}K^{ij}\right]_{-\infty}^{t_0}\nonumber\\
 &\qquad\qquad\qquad\qquad\qquad\quad\ \, =-8\theta\left[a^3\left(H-2\dot{\beta}_+\right) \left(2H
 \dot{\beta}_++H^2+(\dot{\beta}_+^2-3
 \dot{\beta}_-^2)\right)\right]_{-\infty}^{t_0}\nonumber\\
 &\qquad\qquad\qquad\qquad\qquad\quad\ \, =-8\theta e^{3Ht_0}\left(H^3-3H\Sigma^2-2\sigma_+(\sigma_+^2-3\sigma_-^2)\right)\,.\label{GBboundary}
\end{align}
All these contributions are finite (with the Gauss-Bonnet contributions summing to zero); hence, the solution \eqref{eq:exactaniinfsol} satisfies the principle of finite amplitudes. 

\begin{figure}[ht!]
	\centering
	\includegraphics[width=9cm]{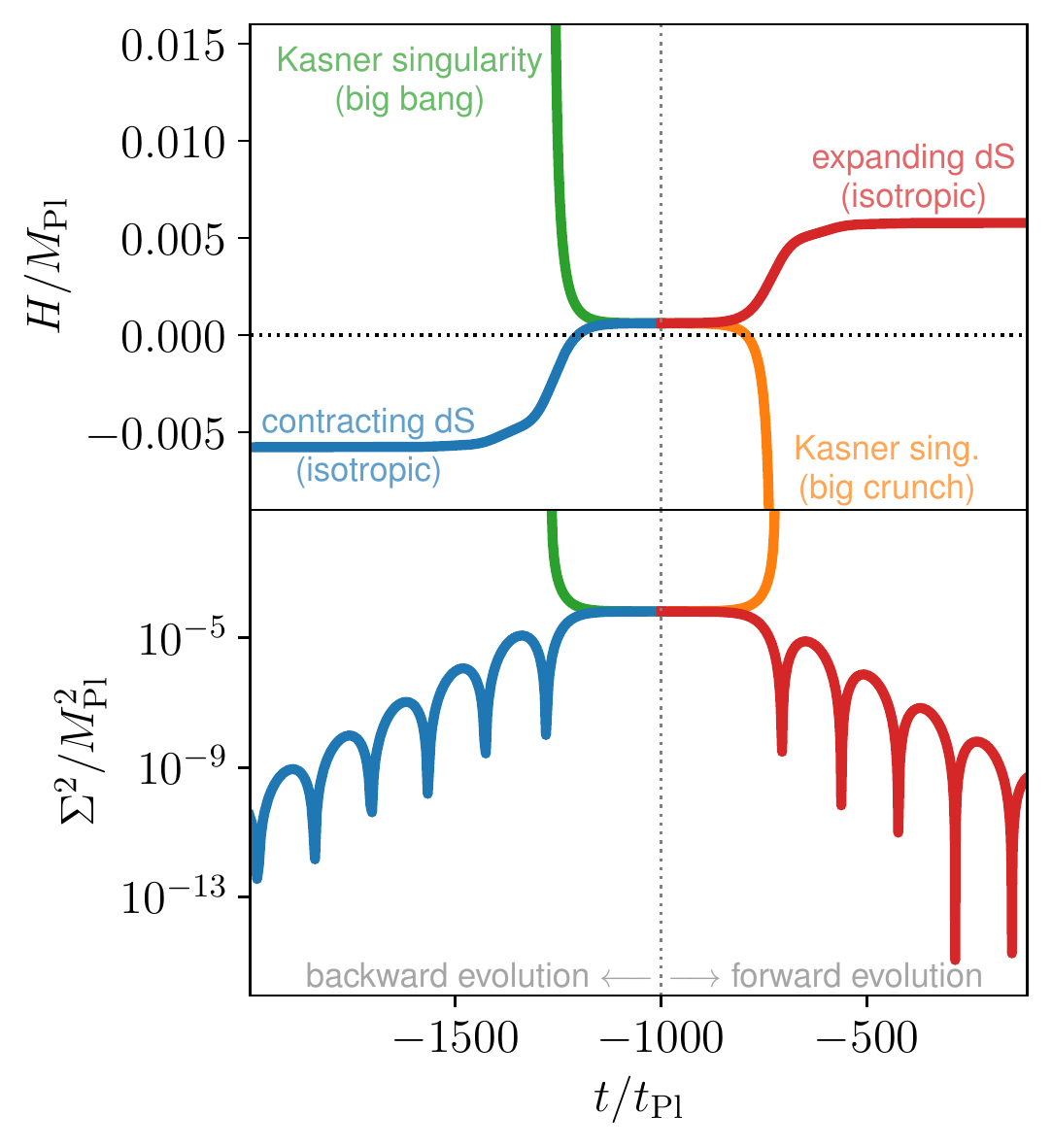}\caption{Time evolution of the anisotropies $\Sigma^2\equiv\dot\beta_+^2+\dot\beta_-^2$ and of the Hubble expansion rate $H$ in quadratic gravity with a cosmological constant $\Lambda$, starting from the exact anisotropically inflating solution \eqref{eq:exactaniinfsol} at $t_\mathrm{ini}=-1000\,t_\mathrm{Pl}$ and adding small perturbations about $\ddot\beta_\pm=0$ and $\dot H=0$. The following numerical values are used: $\Lambda=10^{-4}M_\mathrm{Pl}^4$, $\sigma=10^{-3}$ and $\omega=1/10$. Without loss of generality, we set $\dot\beta_-\equiv 0$, so $\Sigma^2$ is fully determined by $\dot\beta_+^2$. Small perturbations in the initial conditions are implemented in this example with $\ddot\beta_+(t_\mathrm{ini})=\pm 10^{-5}\sigma_+M_\mathrm{Pl}$, and $\dot H(t_\mathrm{ini})$ is set such that the constraint equation is satisfied initially.}\label{fig:quadgravaniinfpert}
\end{figure}

However, this solution is unstable under perturbations, and in reality one expects quantum fluctuations (in the anisotropies for example) to occur, so we have to look at the time evolution when small perturbations are added. The result of a numerical analysis is shown in Fig.~\ref{fig:quadgravaniinfpert}.
The exact anisotropic and inflating solution (constant $\Sigma^2$ and $H$) is initialized at $t=-1000\,t_\mathrm{Pl}$. Toward the future, either the anisotropies grow and the universe re-collapses (orange curves), or the anisotropies shrink to zero and the universe reaches an isotropic de Sitter expansion phase (red curves). To the past, either the anisotropies blow up and the universe collapses with a Kasner-type metric (green curves), or it bounces from a contracting de Sitter phase, while anisotropies are small (blue curves). We conclude that this solution is not a past attractor,\footnote{In fact, the only known past attractors in quadratic gravity appear to be a Kasner-like anisotropic solution and a radiation-like isotropic solution \cite{Barrow:2006xb,Muller:2017nxg}, both of which are not accelerating and hence are ruled out according to our principle of finite amplitudes. Moreover, there does not appear to be any consistent (even approximate) power-law solution in the scale factor and anisotropies near the big bang with the accelerating condition $s>1$ in quadratic gravity.} but if it is selected by the principle of finite amplitudes, it can evolve into an isotropic inflationary phase. A similar analysis was carried on in \cite{Barrow:2009gx}, where they show that in this scenario traces of the primordial anisotropic inflation could potentially be observable through anisotropic statistical effects in the cosmic microwave background. From our point of view this seems unlikely, as for these solutions the big bang is infinitely far in the past and thus anisotropies had to decay for the universe to survive.

Thus we may conclude that in the context of quadratic gravity a principle of finite amplitudes indeed severely restricts potential histories since it only allows accelerated expansion near the big bang. However, the geometry of spacetime may be anisotropic; in this sense our analysis refines that presented in \cite{Lehners:2019ibe}.\footnote{It was shown in \cite{Lehners:2019ibe} that the principle of finite amplitudes only allows for isotropic and homogeneous beginnings in the context of quadratic gravity. However, the arguments were making use of the divergence of the off-shell action, which by itself does not imply an ill-defined amplitude. Here we have rectified this analysis by using the on-shell action. In any case the result we get is very similar to the one obtained previously, so the conclusions are overall preserved.} Nevertheless, one might say that such anisotropic beginnings are allowed only under the most favorable conditions, as accelerated expansion is efficient in suppressing anisotropies dynamically. Quadratic gravity may thus be argued to naturally lead to initial conditions that are propitious for inflation. Whether or not the incompleteness problem discussed in Sec.~\ref{sec:inflation} also applies to quadratic gravity is thus worth studying in future work.

\subsection{Nonsingular cosmology with limiting curvature} \label{sec:limiting}

One might have the expectation that a full theory of quantum gravity avoids singularities altogether. At present it is not clear whether this expectation will really be borne out, but in the spirit of the present work one may wonder whether an effective theory that prevents curvature invariants from blowing up might automatically lead to finite amplitudes. In particular, if one has a theory in which anisotropies cannot diverge in the approach to small volumes, then one may expect the on-shell action to remain finite as we have seen in the previous subsection.
However, the resulting nonsingular cosmology may well extend in time to $-\infty$, which can become an issue regarding finite amplitudes.
Limiting curvature theories are a natural test bed to address this question since by definition they are constructed such that specific curvature invariants remain bounded at all times.
The idea is to add a series of terms to the Lagrangian of the form $\chi\mathcal{I}-V(\chi)$, where $\chi$ is an auxiliary scalar field that acts as a Lagrange multiplier, and $\mathcal{I}$ is a curvature-invariant function.
Upon variation with respect to $\chi$, the resulting constraint equation, $\mathcal{I}=V_{,\chi}$, indicates that the curvature-invariant function is naturally bounded provided the potential is set up such that it is itself bounded, i.e., $|V_{,\chi}|<\infty$.

Original implementations (e.g., \cite{Mukhanov:1991zn,Brandenberger:1993ef}) built curvature-invariant functions using quadratic spacetime curvature invariants such as $R^2$, $R_{\mu\nu}R^{\mu\nu}$, etc., but such theories often came with their difficulties, such as instabilities \cite{Yoshida:2017swb}.
A recent proposal \cite{Sakakihara:2020rdy} rather suggests to build extrinsic curvature invariants, i.e.~using curvature scalars such as $K^2$, $K^\mu{}_\nu K^\nu{}_\mu$, etc., where $K_{\mu\nu}$ is the extrinsic curvature tensor of a hypersurface defined by some new field in the theory, e.g.~a timelike, unit vector field\footnote{The foliation could also be defined with respect to a scalar field $\phi$ with constant field hypersurfaces. The normal vector would then be $\nabla^\mu\phi$, and the theory would be a generalization of mimetic gravity (see \cite{Sakakihara:2020rdy} for details). Defining the theory that way, however, usually comes with instabilities.} $u^\mu$, which is normal to the hypersurface and which defines the foliation.
On shell (when the constraint $u^\mu u_\mu=-1$ is satisfied), curvature invariants defined as $\mathcal{I}_1\equiv K^2$ and $\mathcal{I}_2\equiv K^\mu{}_\nu K^\nu{}_\mu-K^2/3$ become $(\nabla_\mu u^\mu)^2$ and $\nabla_\mu u^\nu \nabla_\nu u^\mu-(\nabla_\mu u^\mu)^2/3$, respectively.
These particular functions are interesting since in a Bianchi-I background metric of the form \eqref{eq:B1} they reduce to $K^2=9H^2$ and $K^\mu{}_\nu K^\nu{}_\mu-K^2/3=6\Sigma^2$, where as before $\Sigma^2\equiv\dot\beta_+^2+\dot\beta_-^2$ is the parameter characterizing the shear anisotropies.
Thus, one can see that limiting the curvature invariants $\mathcal{I}_1$ and $\mathcal{I}_2$ implies bounding the (averaged) Hubble parameter and the shear.
Covariantly the on-shell action (i.e., assuming again the constraint $u^\mu u_\mu=-1$ is satisfied) in such a case looks as follows,
\begin{equation}
 S_\textrm{on-shell}=M_\mathrm{Pl}^2\int\dd^4x\,\sqrt{-g}\left[\frac{R}{2}+\chi_1(\nabla_\mu u^\mu)^2+\chi_2\left(\nabla_\mu u^\nu \nabla_\nu u^\mu-\frac{1}{3}(\nabla_\mu u^\mu)^2\right)-M_\mathrm{L}^2V(\chi_1,\chi_2)\right]\,,
\end{equation}
where $M_\mathrm{L}$ represents some new mass scale associated with the limiting curvature scale (this way, the potential is meant to be a dimensionless function).
The potential is set such that the curvature invariants tend to constants at high scales, while general relativity is recovered at low scales.
The reader is invited to see \cite{Sakakihara:2020rdy} for the details, but it is sufficient to state here that a given potential yields the following Bianchi-I asymptotic solution at early times,
\begin{equation}
 a(t)\simeq a_0e^{M_\mathrm{L} t/3}\,,\qquad\beta_+(t)\simeq\pm\frac{M_\mathrm{L} t}{\sqrt{6}}\,,\qquad (t\rightarrow-\infty)
\end{equation}
where $\beta_-$ was set to zero for concreteness.
Accordingly, the Hubble parameter and the shear tend to constants (analogously to the anisotropic inflation solution in quadratic gravity discussed in the previous subsection),
\begin{equation}
 H(t)\simeq\frac{M_\mathrm{L}}{3}\,,\qquad\Sigma(t)\simeq\frac{M_\mathrm{L}}{\sqrt{6}}\,,\qquad (t\rightarrow-\infty)
\end{equation}
while the auxiliary fields and the potential scale as $\chi_1\simeq-\chi_2-1/12$ and $V(\chi_1,\chi_2)\simeq-1/12$.
Substituting the asymptotic solution in the on-shell action yields
\begin{align}
 S_\textrm{on-shell}&=M_\mathrm{Pl}^2\int_{-\infty}\dd t\,a^3\left[6H^2+3\dot H+3\Sigma^2+9\chi_1H^2+6\chi_2\Sigma^2-M_\mathrm{L}^2V(\chi_1,\chi_2)\right]\nonumber\\
 &\simeq\frac{7}{6}a_0^3M_\mathrm{Pl}^2M_\mathrm{L}^2\int_{-\infty}\dd t\,e^{M_\mathrm{L} t}\,,
\end{align}
to leading order as $t\to-\infty$. As before, we only emphasize the lower integration limit since it is the one determining the convergence of the on-shell action in this context.
From the above, $S_\textrm{on-shell}$ is clearly a finite number since the (averaged) scale factor goes to zero as $t\to-\infty$.
Therefore, it appears the effective limiting curvature theory can yield finite amplitudes even though the Bianchi-I background\footnote{The statement certainly only holds at the background level. When inhomogeneities are included, one has to check that no instabilities form, which could potentially lead to divergences in the perturbed action. The preliminary analysis of \cite{Sakakihara:2020rdy} showed most instabilities are avoided in the model above, but this might not hold for arbitrary initial conditions or with the addition of matter.} cosmology is nonsingular.

Let us point out that for the above solution, comoving timelike geodesics are past complete since the cosmological time $t$ (i.e., the proper time for comoving observers) is defined all the way to $t \rightarrow -\infty$.
However, null geodesics are past incomplete as in flat de Sitter space (or any inflationary primordial universe \cite{Borde:2001nh}).
To see this, one can compute the affine length $\lambda$ of a null geodesic that goes all the way to $t\rightarrow -\infty$ with (following, e.g., \cite{Fleury:2016htl})
\begin{equation}
 \lambda[-\infty,t_\mathrm{f}]=\int_{-\infty}^{t_\mathrm{f}}\frac{\dd t}{a}\left(k_x^2e^{-2(\beta_++\sqrt{3}\beta_-)}+k_y^2e^{-2(\beta_+-\sqrt{3}\beta_-)}+k_z^2e^{4\beta_+}\right)^{-1/2}\,,
\end{equation}
where the $k_i$'s
are the spatial components of a four-dimensional wave vector of any light ray, which are constants of motion \cite{Fleury:2014rea}.
For the constant-$H$ and constant-$\Sigma$ solution, one can check that the integral gives, for arbitrary $k_i$'s, a function whose $t\rightarrow-\infty$ limit vanishes.
Consequently, $\lambda[-\infty,t_\mathrm{f}]$ is a finite length at any finite time $t_\mathrm{f}$; hence, null geodesics are incomplete.
Correspondingly, $t=-\infty$ is a past spacetime boundary.
Whether this boundary represents the end of the spacetime or whether the spacetime can be extended beyond is not obvious though.
In an isotropic spacetime, this depends on the rate at which de Sitter-like evolution is reached asymptotically \cite{Yoshida:2018ndv}.
One might expect something similar here, but there is probably an additional requirement on the rate at which shear anisotropies tend to a constant.
One may conjecture that if it is reached sufficiently fast (or in the case of an exact solution as in quadratic gravity), the boundary could be extended and then $t=-\infty$ would not be the "beginning of time"; rather, there would possibly be a bounce and the action would presumably diverge in the true asymptotic past.
If the constant-anisotropy phase is not reached fast enough though, the $t=-\infty$ boundary could actually be a (parallelly propagated) curvature singularity (even though spacetime scalar curvature invariants do not blow up), and hence the on-shell action would be finite as $a\rightarrow 0$ ($t\to-\infty$).

\subsection{Emergent universe in (beyond-)Horndeski theory and subclasses thereof}\label{sec:Horndeski}

As discussed in Sec.~\ref{sec:loitering}, if one wants to consider a nonsingular loitering phase in the very early universe, then one has to consider some theory of modified gravity or a proper quantum gravity motivated theory such as string theory. The latter will be explored in the next subsection, but for now let us consider the effective approach in which one simply postulates a modified gravity theory and tests its theoretical consistency, in a similar fashion to the previous subsection on limiting curvature.

\subsubsection{Example in k-essence theory\label{subsec:kessence}}

A simple extension of general relativity is to introduce a scalar field that enters the action as a generic function of itself and its kinetic term, $X\equiv-\partial_\mu\phi\partial^\mu\phi/2$, i.e.,
\begin{equation}
 S=\int\dd^4x\,\sqrt{-g}\left(\frac{M_\mathrm{Pl}^2}{2}R+P(X,\phi)\right)\,.\label{eq:Sk}
\end{equation}
This theory is known as $k$-essence \cite{ArmendarizPicon:2000dh,ArmendarizPicon:2000ah}.
Varying with respect to the metric yields the Einstein field equations with energy-momentum tensor given by
\begin{equation}
 T_{\mu\nu}=Pg_{\mu\nu}+P_{,X}\partial_\mu\phi\partial_\nu\phi\,.
\end{equation}
The trace of the Einstein field equations then gives the relation
\begin{equation}
 -M_\mathrm{Pl}^2R=T=4P-2XP_{,X}\,,
\end{equation}
which upon substitution into \eqref{eq:Sk} yields the on-shell action
\begin{equation}
 S_\textrm{on-shell}=\int\dd^4x\,\sqrt{-g}(XP_{,X}-P)\,.\label{eq:Sosk}
\end{equation}
A nice property of $k$-essence theory is that it can implement the idea of a ghost condensate \cite{ArkaniHamed:2003uy}, from which one can construct cosmological backgrounds that violate the NEC for some time interval.

Here is an example of how this can be done, following, e.g., \cite{Lin:2010pf}. Let us specify to the case where the theory can be split into a kinetic and a potential function as $P(X,\phi)=K(X)-V(\phi)$. For our purpose, we consider
\begin{equation}
 K(X)=\frac{(2X-M^4)^2}{8\Lambda}\,,
\end{equation}
where $\Lambda^{1/4}$ and $M$ are new mass scales and $X=M^4/2$ is where the ghost condensate forms. Then, in a FLRW background, the field's energy density and sum of energy density and pressure are given by
\begin{align}
 \rho&=2XP_{,X}-P=\frac{1}{8\Lambda}(12X^2-4M^4X-M^8)+V(\phi)\,, \nonumber \\
 \rho+p&=2XP_{,X}=\frac{X}{\Lambda}(2X-M^4)\,.
\end{align}
About the ghost condensate $X=M^4/2$, we note that $\rho=V$ and $\rho+p=0$. Accordingly, a loitering phase where $H\rightarrow 0$ and $\dot H\rightarrow 0$ is achievable if $V(\phi)\rightarrow 0$ as $t\rightarrow -\infty$.

Let $\phi(t)=M^2t$ correspond to the ghost condensate. Then, let us add small fluctuations, $\phi(t)=M^2t+\delta\phi(t)$, so that up to order $\dot{\delta\phi}^2/M^4$ the Friedmann equations are
\begin{align}
 3M_\mathrm{Pl}^2H^2=\rho&\simeq\frac{M^6}{\Lambda}\dot{\delta\phi}+V(\phi)\,,\nonumber\\
 -2M_\mathrm{Pl}^2\dot H=\rho+p&\simeq\frac{M^6}{\Lambda}\dot{\delta\phi}\,,
\end{align}
while for the $\phi$ EOM,
\begin{equation}
 (P_{,X}+2XP_{,XX})\ddot\phi+3HP_{,X}\dot\phi+P_{,X\phi}\dot\phi^2-P_{,\phi}=0\,,
\end{equation}
becomes
\begin{equation}
 \frac{M^4}{\Lambda}(\ddot{\delta\phi}+3H\dot{\delta\phi})=-V_{,\phi}\,.
\end{equation}
Taking $V(\phi)\propto\phi^{-2\alpha}$ for some positive integer $\alpha$ as an ansatz, so that $V(\phi)\rightarrow 0$ and $V_{,\phi}\rightarrow 0$ as $\phi\rightarrow-\infty$, we get $\dot{\delta\phi}\rightarrow 0$ as $t\rightarrow-\infty$, and we see that the desired loitering phase is achieved asymptotically.
Violation of the NEC is subsequently realized.
To demonstrate this, let us consider the $\delta\phi$ EOM to leading order as $H\rightarrow 0$ and $|\delta\phi|\ll|M^2t|$ ($t\rightarrow-\infty$) with, e.g., $V(\phi)=V_0\Lambda^{1/2}\phi^{-2}$.
One then obtains
\begin{equation}
 \dot{\delta\phi}\simeq-\frac{V_0\Lambda^{3/2}}{M^{10}t^2}\,,\label{eq:deltaphidotsol}
\end{equation}
where we set the initial condition $\dot{\delta\phi}(t\rightarrow-\infty)=0$.
Therefore, $\dot{\delta\phi}<0$ and $\dot H\propto -\dot{\delta\phi}>0$ as time evolves, indicating NEC violation.
When the universe reaches a certain energy scale, it would then reheat, with the newly formed particles ending the NEC-violating phase and starting the standard hot big bang cosmology evolution.

Let us now consider the on-shell action for such a model. From \eqref{eq:Sosk} we have
\begin{equation}
 S_\textrm{on-shell}=\int\mathrm{d}^4x\,\sqrt{-g}\left[\frac{4X^2-M^8}{8\Lambda}+V(\phi)\right]\,.
\end{equation}
For the FLRW background solution described earlier, this reduces to (to leading order as $t\to-\infty$)
\begin{equation}
 S_\textrm{on-shell}\simeq\int_{-\infty}\mathrm{d}t\,a_0^3\left[\frac{M^6}{2\Lambda}\dot{\delta\phi}+V(\phi)\right]\,,
\end{equation}
where $a_0$ is the constant corresponding to $H\rightarrow 0$.
By construction, $\dot{\delta\phi}$ and $V(\phi)$ both tend to $0$ as $t\rightarrow-\infty$.
Therefore, the overall contribution to the action vanishes in the limit $t\rightarrow-\infty$.
This can be checked explicitly, e.g., for $V(\phi)\propto\phi^{-2}$ and with the solution \eqref{eq:deltaphidotsol}: the leading-order term in the integrand goes as $1/t^2$, which then scales as $1/t\to 0$ upon integration.
Correspondingly, the action evaluated at any finite time does not diverge.

Let us mention that a "crude" ghost condensate as above usually comes with other issues though, such as gradient instabilities, strong coupling, or unitarity violation (see e.g.~\cite{deRham:2017aoj}); furthermore, the effective potential would have to hold over an infinite field range.
Somewhat more successful scenarios can be constructed in Horndeski theories and beyond, which we will mention below. It remains unclear whether NEC-violating models can be constructed which are devoid of any inconsistencies, since typically the relevant issues are only investigated in a given background, whereas the theory should remain consistent in all possible deformations of the background that may be reached dynamically. Evidently this is a difficult check, but if it can be realized then our toy model example already illustrates how loitering phases can be in agreement with the principle of finite amplitudes.

\subsubsection{Some comments about Horndeski theory and beyond\label{subsec:Horndeski}}

There exist various models of genesis scenarios in Horndeski theory and beyond in the literature that could be analyzed,
including many that claim to have stable linear perturbations across time.
In every case, one can take the background solution, evaluate the corresponding on-shell action and confirm the finiteness as $t\to-\infty$.
For instance in \cite{Mironov:2019qjt,Mironov:2019haz} where beyond-Horndeski theory is used, the background solution and on-shell action in the asymptotic past reduce to the early model of \cite{Creminelli:2010ba}, which we find is consequent with the principle of finite amplitudes.
Regarding linear perturbations, if it is confirmed that no instability arises, then this is also in agreement with the principle of finite amplitudes.
Indeed, any fixed-wavelength perturbation that evolved back to $t\to-\infty$ should reach its well-defined Minkowski vacuum, and the properly renormalized on-shell action for such perturbations (e.g., with momentum cutoffs) shall remain finite across time if no ghost or gradient instability arises.

Such theories should really be viewed as EFTs that can describe physics only up to some cutoff scale and which are assumed to arise from integrating out heavy degrees of freedom from a more ultraviolet-complete theory.
The background energy scale should thus always remain below the EFT's cutoff scale; otherwise, one would enter a strongly coupled regime where the EFT breaks down.
This may be satisfied in certain Horndeski models (e.g., \cite{Ageeva:2018lko,Ageeva:2020gti,Ageeva:2020buc}), but those also indicate that the strong coupling scale shrinks as we look far back in time, thus making the regime over which the EFT is valid smaller and smaller.
Furthermore, even before strong coupling is reached one may have unitary violation in such theories violating the NEC \cite{deRham:2017aoj}.
Unitarity violation is somewhat equivalent to violating the principle of finite amplitudes, though the formulation is different. In particular, the former is more applied to the context of particle physics, whereas we are interested in the whole cosmology.
Nevertheless, if we perform perturbation theory and separate the background and the fluctuations, unitarity violation of the fluctuations would necessarily imply the violation of the principle of finite amplitudes in our context.

In summary, effective models of loitering cosmology in the very early universe using Horndeski theory and beyond do not indicate an immediate violation of the principle of finite amplitudes (it can certainly be satisfied at the background level), but those models have their limitations (by virtue of being EFTs), and more analysis is needed to demonstrate their proper robustness.

\subsection{Loitering phase in string cosmology}\label{sec:stringcosmo}

We analyzed a few examples of genesis models in the previous subsection within effective theories of modified gravity.
However, let us recall from Sec.~\ref{sec:loitering} that strong motivation for such emerging early universe scenarios came from string cosmology.
Therefore, it is worth exploring effective models of string theory and their cosmological applications in the context of early universe loitering phases.

\subsubsection{Example in dilaton gravity}

As a first example, we may consider the low-energy, tree-level effective theory of string theory known as dilaton gravity in $d+1$ spacetime dimensions,
\begin{equation}
 S=\frac{1}{2\ell_\mathrm{s}^{d-1}}\int\dd^{d+1}x\,\sqrt{-g}\,e^{-2\phi}\left(R+4\partial_\mu\phi\partial^\mu\phi\right)+S_\mathrm{m}\,,
\end{equation}
where $\ell_\mathrm{s}$ is the string length, $\phi$ is the dilaton field, and $S_\mathrm{m}$ represents the action of additional matter content.
The above ignores any possible potential for the dilaton field and sets the string theory Neveu-Schwarz--Neveu-Schwarz (NS-NS) two-form to zero.
Let us now consider an isotropic cosmological background for simplicity, although a more general analysis could allow for different scale factors in different directions.
Introducing the shifted dilaton $\Phi=2\phi-d\ln a$ and writing the scale factor as $a=e^\lambda$, the action reduces to
\begin{equation}
 S=\frac{1}{2\ell_\mathrm{s}^{d-1}}\int\dd t\,e^{-\Phi}\left(d\dot\lambda^2-\dot\Phi^2\right)+S_\mathrm{m}\label{eq:Sdg}
\end{equation}
after integration by parts.
For the matter action, we shall assume a gas of strings with momentum and winding modes contributing equally, with respective EOS $p=\rho/d$ and $p=-\rho/d$, so that the total pressure vanishes.
Given this matter content, the EOMs that follow from \eqref{eq:Sdg} admit a loitering solution of the form (see, e.g., \cite{Tseytlin:1991xk})
\begin{equation}
 \dot\lambda\sim\frac{1}{t^2}\,,\qquad\Phi\sim-2\ln(-t)\,.\qquad(t\to-\infty)
\end{equation}
The above shows that the shifted dilaton kinetic term $\dot\Phi^2\sim 4/t^2$ and the Hubble parameter $\dot\lambda$ tend to zero in the asymptotic past, and a loitering phase is indeed achieved with $a\to\mathrm{constant}$.
This also implies that the energy density tends to a constant as $\rho\propto a^{-d}$.
Upon substitution in \eqref{eq:Sdg}, the on-shell action for this solution to leading order as $t\to-\infty$ reduces to
\begin{equation}
 S_\textrm{on-shell}\sim\int_{-\infty}\dd t\,\left(-4+\rho\right)\,,\label{eq:Sdgos}
\end{equation}
which diverges.
This may be surprising since the vanishing of the Hubble parameter and kinetic term usually implies the vanishing of the integrand.
However, here the nonminimal coupling to the dilaton with $e^{-\Phi}\sim t^2$ actually yields a constant value for $e^{-\Phi}\dot\Phi^2$, which blows up upon integration to the infinite past.
Thus this picture of string cosmology is in conflict with the principle of finite amplitudes.

\subsubsection{Example with nonperturbative curvature corrections}

The previous example was limited to tree-level interactions (no string loops) and to zeroth order in curvature corrections.
Going beyond this is usually very difficult (especially in string theory), but recent developments indicate that curvature corrections to all orders in $\alpha'\sim\ell_\mathrm{s}^2$ may be expressed in a simple form when restricted to homogeneous backgrounds \cite{Hohm:2019jgu} (thanks to $\mathrm{O}(d,d)$ symmetry \cite{MEISSNER199133,Meissner:1996sa}).
Specifically, the FLRW action with no NS-NS two-form field is claimed to be of the form 
\begin{equation}
 S=\frac{1}{2\ell_\mathrm{s}^{d-1}}\int\dd t\,a^de^{-2\phi}\left[-4\dot\phi^2+4d\dot\phi H-d^2H^2-F(H)\right]+S_\mathrm{m}\,,\label{eq:Salphap}
\end{equation}
where the function $F(H)$ encapsulates the $\alpha'$ corrections,
\begin{equation}
 F(H)=2d\sum_{k=1}^\infty(-\alpha')^{k-1}c_k2^{2k}H^{2k}\,.\label{eq:FofH}
\end{equation}
The coefficients $c_k$ are to be determined from full string theory calculations, order by order.
To zeroth order, it is known that $c_1=-1/8$, but it is already a difficult task to determine $c_2$ to first order in $\alpha'$.
Thus, at present one may rather keep the $c_k$'s arbitrary, require convergence of the infinite sum, and check for consistency of the theory given the desired symmetries. In particular, if the function $F(H)$ takes the appropriate form, it was shown that string-frame de Sitter-like solutions (i.e.~$H=\mathrm{const.}$) were admissible, both in vacuum \cite{Hohm:2019jgu,Hohm:2019ccp} and with matter \cite{Bernardo:2019bkz,Bernardo:2020zlc} (see also \cite{Nunez:2020hxx} for other string- and Einstein-frame de Sitter-like solutions and Einstein-frame loitering solutions as will be discussed below).

This was recently applied to a very early universe loitering phase in the context of string gas cosmology \cite{Bernardo:2020nol}.
Exploring the regime where the string gas is dominated by winding modes with EOS $p=-\rho/d$, the appropriate convergent $F(H)$ function (i.e., nonperturbative in $\alpha'$) permits a solution of the form
\begin{equation}
 \phi(t)=\frac{d-1}{2}Ht\,,\qquad H=\mathrm{const.}\,,\qquad a(t)\propto e^{Ht}\,.\label{eq:SFsol}
\end{equation}
From this, the square bracket in \eqref{eq:Salphap} is simply a constant, and so the gravi-dilaton sector of the on-shell action becomes
\begin{equation}
 S_\textrm{on-shell}\propto\int_{-\infty}\dd t\,a^de^{-2\phi}=\int_{-\infty}\dd t\,e^{dHt}e^{-(d-1)Ht}=\int_{-\infty}\dd t\,e^{Ht}\,,
\end{equation}
and hence the integral converges in the string frame.

The loitering phase in this context appears when looking at the conformally equivalent Einstein frame (EF), through the transformation
\begin{equation}
 a_\mathrm{E}=e^{-2\phi/(d-1)}a\,,\qquad\dd t_\mathrm{E}=e^{-2\phi/(d-1)}\dd t\,,\qquad\phi_\mathrm{E}=\frac{2}{\kappa}\frac{1}{\sqrt{d-1}}\phi\,,
\end{equation}
where $\kappa^2\equiv M_\mathrm{Pl}^{1-d}$. Physically speaking, we are assuming that additional matter is coupled in the Einstein frame and that rulers and clocks are made of this matter.
From Eq.~\eqref{eq:SFsol} and writing $a(t)=a_0e^{Ht}$, this tells us that
\begin{equation}
 a_\mathrm{E}=a_0=\mathrm{const.}\,,
\end{equation}
confirming that $H_\mathrm{E}\equiv\dd a_\mathrm{E}/\dd t_\mathrm{E}=0$, i.e., the theory admits a loitering phase in the EF.
The time transformation tells us that
\begin{equation}
 t_\mathrm{E}=-\frac{e^{-Ht}}{H}\,,
\end{equation}
where we set the lower integration limits such that both the string-frame and EF times run from $-\infty$.
Then in the EF the dilaton evolves as
\begin{equation}
 \phi_\mathrm{E}(t_\mathrm{E})=-\frac{\sqrt{d-1}}{\kappa}\ln\left[H(-t_\mathrm{E})\right]\,,\label{eq:phiEFevo}
\end{equation}
and so a time derivative gives
\begin{equation}
 \frac{\dd\phi_\mathrm{E}}{\dd t_\mathrm{E}}=\frac{\sqrt{d-1}}{\kappa(-t_\mathrm{E})}\,,
\end{equation}
which tends to zero as $t_\mathrm{E}\rightarrow-\infty$.
Thus, looking at the gravi-dilaton sector of the EF action,
\begin{equation}
 S^\mathrm{(EF)}=\int_{-\infty}\dd t_\mathrm{E}\,a_\mathrm{E}^d\left[\frac{3}{\kappa^2}\left(2H_\mathrm{E}^2+\frac{\dd H_\mathrm{E}}{\dd t_\mathrm{E}}\right)+\frac{1}{2}\left(\frac{\dd\phi_\mathrm{E}}{\dd t_\mathrm{E}}\right)^2+\mathcal{O}(\alpha')\right]\,,\label{eq:EFS}
\end{equation}
the only possibly nonzero terms in the on-shell action as $t_\mathrm{E}\rightarrow-\infty$ are the $\mathcal{O}(\alpha')$ terms.
We claim, however, that those terms vanish on-shell as well, which would confirm the finiteness of the action.
In the string frame, the $\mathcal{O}(\alpha')$ terms entered the action in the form
\begin{equation}
 S\supset\int\dd t\,a^de^{-2\phi}A(H)\,,\label{eq:alphapSF}
\end{equation}
where $A(H)$ is essentially the function $F(H)$ of Eq.~\eqref{eq:FofH} starting at $k=2$ (order $\alpha'$).
The above expression is then transformed to the EF as
\begin{equation}
 \int\dd t\,a^de^{-2\phi}A(H)=\int\dd t_\mathrm{E}\,a_\mathrm{E}^de^{4\phi/(d-1)}A(H)\,,
\end{equation}
where $\phi$ and $H$ can in turn be written in terms of EF variables, after which $e^{4\phi/(d-1)}A(H)$ represents the $\mathcal{O}(\alpha')$ terms in \eqref{eq:EFS}.
Since $H$ is a constant and $\phi\rightarrow-\infty$ in the asymptotic past, that quantity vanishes.
Indeed, one can generally write
\begin{equation}
 H_\mathrm{E}=e^{\frac{2\phi}{d-1}}\left(H-\frac{2}{d-1}\dot\phi\right)\,,\qquad\frac{\dd\phi_\mathrm{E}}{\dd t_\mathrm{E}}=e^{\frac{2\phi}{d-1}}\dot\phi\,,
\end{equation}
so for $H_\mathrm{E}=0$, we have
\begin{equation}
 H=\frac{2}{d-1}\frac{\dd\phi_\mathrm{E}}{\dd t_\mathrm{E}}e^{-\frac{\kappa\phi_\mathrm{E}}{\sqrt{d-1}}}\,,
\end{equation}
which is confirmed to be a constant from \eqref{eq:phiEFevo}.
Thus, an arbitrary polynomial function of $H$ is still a constant once transformed to the EF, while the factor $e^{4\phi/(d-1)}=e^{2\kappa\phi_\mathrm{E}/\sqrt{d-1}}$ goes to $0$ as $\phi_\mathrm{E}\rightarrow-\infty$. Hence, the $\mathcal{O}(\alpha')$ terms in the EF action vanish on-shell.
In summary, it appears the above model provides an example of EF loitering phase in string cosmology, which is consistent with the principle of finite amplitudes.


\section{Discussion and conclusions} \label{sec:discussion}

In this paper, we have investigated the cosmological implications that arise from requiring transition amplitudes from the far past to the present to be finite and well defined. A closely related principle was proposed by Barrow and Tipler some thirty years ago \cite{Barrow1988} and updated in \cite{Barrow:2019gzc}. In their work, they proposed that the action of the entire universe, integrated over its entire past and future "history", should be finite. The two principles are closely related, in that they both force the universe to have finite spatial volume and that they eliminate histories in which the big bang is approached in such a manner that the action diverges. This is not very constraining for ordinary general relativity, where it remains the case that the typical approach to the big bang consists of mixmaster behavior. However, in semiclassical theories containing higher-curvature terms (which should be relevant in the approach to the big bang), the consequences are very severe: quadratic gravity allows inflationary initial conditions only, i.e., it eliminates all nonaccelerating backgrounds. The implications for theories that have an infinite series of higher-curvature corrections are even stricter: no Lorentzian big bang solution is allowed at all. In fact, the only currently known way in which the amplitude may be made finite in this case, while remaining in the semiclassical regime, is by extending metrics to the complex domain and implementing the no-boundary proposal.

In contrast to the principle of finite action, requiring finite amplitudes does not restrict the future evolution of the universe. Thus we do not eliminate universes that might expand forever. However, well-defined amplitudes are sensitive to additional information besides the action: the semiclassical path integral may, e.g., be described by two or more saddle points that may interfere. Such a case occurs for (Lorentzian) inflationary models, if one assumes that inflation goes back all the way to the big bang. Then two saddle points may interfere in such a way as to generate the wrong vacuum for perturbations. This is a problem of quantum initial condition for inflation, which needs to be better understood. Interestingly, eternal inflation is not allowed by the principle of finite amplitudes, as we find that the action already diverges when calculating a transition to present day field values. When applied to bouncing cosmologies, we find that models in which the first bounce is preceded by an infinitely long, pure ekpyrotic phase are allowed (though they require an infinite field displacement), while cyclic models need to have a singular beginning--thus \emph{strictly} cyclic cosmologies are excluded. These conclusions are independent of the nature of the high-curvature bounces (i.e.~how one violates the NEC to obtain nonsingular bounces) and rather only rely on robust low-energy semiclassical physics in the infinite past.

Barrow and Tipler noted that demanding finite action leads to an interesting give-or-take: eliminating divergences in the action typically leads to spacetime singularities, while eliminating spacetime singularities typically leads to divergences in the action. The one overarching exception to this dilemma is the no-boundary proposal, which manages to eliminate the initial singularity while also entailing finite action. If semiclassically consistent models allowing for an effective violation of the NEC exist, then there is the additional possibility of an initial loitering phase, in which the universe starts out at a finite scale factor. Loitering phases are motivated by string theory, and it is probably in this context (taking into account the effects of the extended nature of strings and of winding modes) that consistent models will be found. 

Our work also makes contact with Penrose's much older Weyl curvature hypothesis \cite{Penrose:1900mp}. Penrose noted that the singularities in the big bang and at the centers of black holes (as well as in big crunches) differ significantly from each other: the Weyl curvature was very small in the initial singularity, but it diverges in final singularities. From this observation Penrose conjectured that the Weyl curvature is related to gravitational entropy, and that the increase in the Weyl curvature is connected to the second law of thermodynamics, suitably generalized to include gravitational entropy. We would like to emphasize that our results go quite some way toward explaining the initial smallness of the Weyl curvature. In no-boundary models, highly symmetric, isotropic spacetimes have the highest probability. In quadratic gravity, the only allowed big bang evolutions also lead to isotropic, and moreover accelerating, geometries. Meanwhile loitering phases can only succeed if the anisotropies, which have the potential to cause the universe to collapse, are kept small. Thus the early universe scenarios that satisfy the principle of finite amplitudes automatically lead to small Weyl curvatures at early times.

Our principle may simply be seen as requiring a consistent semiclassical description of early universe physics.\footnote{Once again, addressing the possibility of a non-geometric early universe history that does not fit into the realm of semiclassical gravity is beyond the scope of our principle.} In this sense it is closely related to recent investigations of the string theory swampland, where consistency conditions arising from the interplay of quantum gravity and field theory are also investigated, mostly with input from string theory. In both cases, eternal inflation does not seem to be allowed; see, e.g., \cite{Agrawal:2018own}. Also, one could very well include the trans-Planckian censorship conjecture as a semiclassical consistency requirement \cite{Bedroya:2019snp}, and then further swampland conditions automatically follow \cite{Bedroya:2020rmd}. One aspect of the swampland conditions for which we do not see any counterpart here is the distance conjecture \cite{Ooguri:2006in}, which rests on the assumption that large field displacements reveal towers of new light states in the theory. If this conjecture is combined with our principle of finite amplitudes, very few semiclassical theories of the early universe might survive.

The most interesting part of the principle of finite amplitudes is its restrictive power, which changes what one considers as typical and as special. Here we have seen that dynamical evolutions, which from the classical viewpoint appear to be typical, such as BKL/mixmaster behavior, are eliminated outright in effective quantum gravity. On a more quantum note, eternal inflation is also eliminated. This provides a sharpening of the question of initial conditions and a fresh perspective on the big bang. At the time of writing, the gold standard in terms of amplitudes is provided by the no-boundary proposal, since it leads to finite amplitudes, eliminates spacetime singularities, yet does not require speculative physics such as NEC-violating models.\footnote{Yet some important questions remain also here, since at least naively the no-boundary proposal predicts a Hubble rate that is too small.} Quadratic gravity and loitering phases may provide worthy contenders and other appealing approaches may be discovered in the future. The consistency of semiclassical gravity is a powerful sieve, motivating us to take seriously the few models that it lets through.

\vskip23pt
\subsection*{Acknowledgments}
We would like to thank the late John Barrow for inspiring correspondence on the topics of this paper. 
We gratefully acknowledge the support of the European Research Council (ERC) in the form of the ERC Consolidator Grant CoG 772295 ``Qosmology.''
J.\,Q.~further acknowledges financial support in part from the \textit{Fond de recherche du Qu\'ebec---Nature et technologies} postdoctoral research scholarship and the Natural Sciences and Engineering Research Council of Canada Postdoctoral Fellowship.
\vskip23pt


\newpage 
\phantomsection
\addcontentsline{toc}{section}{References}

\let\oldbibliography\thebibliography
\renewcommand{\thebibliography}[1]{
  \oldbibliography{#1}
  \setlength{\itemsep}{0pt}
  \footnotesize 
}

\bibliographystyle{JHEP2}
\bibliography{FiniteAmplitudes}

\providecommand{\url}[1]{#1}\providecommand{\href}[2]{#2}\begingroup\raggedright\begin{thebibliography}{100}

\bibitem{Skinner}
D.~Skinner, \emph{Quantum field theory ii},
  \url{https://www.damtp.cam.ac.uk/user/dbs26/AQFT.html}, 2018.

\bibitem{Agrawal:2020xek}
P.~Agrawal, S.~Gukov, G.~Obied and C.~Vafa, \emph{{Topological Gravity as the
  Early Phase of Our Universe}},
  \href{https://arxiv.org/abs/2009.10077}{{\ttfamily arXiv:2009.10077}}.

\bibitem{Lehners:2019ibe}
J.L.~Lehners and K.~Stelle, \emph{{A Safe Beginning for the Universe?}},
  \href{https://doi.org/10.1103/PhysRevD.100.083540}{\emph{Phys. Rev. D}
  {\bfseries 100} (2019) 083540}
  [\href{https://arxiv.org/abs/1909.01169}{{\ttfamily arXiv:1909.01169}}].

\bibitem{Donoghue:2019clr}
J.F.~Donoghue, \emph{{A Critique of the Asymptotic Safety Program}},
  \href{https://doi.org/10.3389/fphy.2020.00056}{\emph{Front. in Phys.}
  {\bfseries 8} (2020) 56} [\href{https://arxiv.org/abs/1911.02967}{{\ttfamily
  arXiv:1911.02967}}].

\bibitem{Bonanno:2020bil}
A.~Bonanno, A.~Eichhorn, H.~Gies, J.M.~Pawlowski, R.~Percacci, M.~Reuter
  et~al., \emph{{Critical reflections on asymptotically safe gravity}},
  \href{https://doi.org/10.3389/fphy.2020.00269}{\emph{Front. in Phys.}
  {\bfseries 8} (2020) 269} [\href{https://arxiv.org/abs/2004.06810}{{\ttfamily
  arXiv:2004.06810}}].

\bibitem{Barrow1988}
J.D.~Barrow and F.J.~Tipler, \emph{Action principles in nature},
  \href{https://doi.org/10.1038/331031a0}{\emph{Nature} {\bfseries 331} (1988)
  31}.

\bibitem{Barrow:2019gzc}
J.D.~Barrow, \emph{{Finite Action Principle Revisited}},
  \href{https://doi.org/10.1103/PhysRevD.101.023527}{\emph{Phys. Rev. D}
  {\bfseries 101} (2020) 023527}
  [\href{https://arxiv.org/abs/1912.12926}{{\ttfamily arXiv:1912.12926}}].

\bibitem{DiTucci:2019bui}
A.~Di~Tucci, J.L.~Lehners and L.~Sberna, \emph{{No-boundary prescriptions in
  Lorentzian quantum cosmology}},
  \href{https://doi.org/10.1103/PhysRevD.100.123543}{\emph{Phys. Rev. D}
  {\bfseries 100} (2019) 123543}
  [\href{https://arxiv.org/abs/1911.06701}{{\ttfamily arXiv:1911.06701}}].

\bibitem{DiTucci:2019dji}
A.~Di~Tucci and J.L.~Lehners, \emph{{No-Boundary Proposal as a Path Integral
  with Robin Boundary Conditions}},
  \href{https://doi.org/10.1103/PhysRevLett.122.201302}{\emph{Phys. Rev. Lett.}
  {\bfseries 122} (2019) 201302}
  [\href{https://arxiv.org/abs/1903.06757}{{\ttfamily arXiv:1903.06757}}].

\bibitem{Sakakihara:2020rdy}
Y.~Sakakihara, D.~Yoshida, K.~Takahashi and J.~Quintin, \emph{{Theories with
  limited extrinsic curvature and a nonsingular anisotropic universe}},
  \href{https://doi.org/10.1103/PhysRevD.102.084004}{\emph{Phys. Rev. D}
  {\bfseries 102} (2020) 084004}
  [\href{https://arxiv.org/abs/2005.10844}{{\ttfamily arXiv:2005.10844}}].

\bibitem{Bernardo:2020nol}
H.~Bernardo, R.~Brandenberger and G.~Franzmann, \emph{{String Cosmology
  backgrounds from Classical String Geometry}},
  \href{https://doi.org/10.1103/PhysRevD.103.043540}{\emph{Phys. Rev. D}
  {\bfseries 103} (2021) 043540}
  [\href{https://arxiv.org/abs/2005.08324}{{\ttfamily arXiv:2005.08324}}].

\bibitem{Misner:1969hg}
C.W.~Misner, \emph{{Mixmaster universe}},
  \href{https://doi.org/10.1103/PhysRevLett.22.1071}{\emph{Phys. Rev. Lett.}
  {\bfseries 22} (1969) 1071}.

\bibitem{Belinsky:1970ew}
V.A.~Belinsky, I.M.~Khalatnikov and E.M.~Lifshitz, \emph{{Oscillatory approach
  to a singular point in the relativistic cosmology}},
  \href{https://doi.org/10.1080/00018737000101171}{\emph{Adv. Phys.} {\bfseries
  19} (1970) 525}.

\bibitem{Ooguri:2006in}
H.~Ooguri and C.~Vafa, \emph{{On the Geometry of the String Landscape and the
  Swampland}},
  \href{https://doi.org/10.1016/j.nuclphysb.2006.10.033}{\emph{Nucl. Phys. B}
  {\bfseries 766} (2007) 21}
  [\href{https://arxiv.org/abs/hep-th/0605264}{{\ttfamily hep-th/0605264}}].

\bibitem{Schutz:1970my}
B.F.~Schutz, \emph{{Perfect Fluids in General Relativity: Velocity Potentials
  and a Variational Principle}},
  \href{https://doi.org/10.1103/PhysRevD.2.2762}{\emph{Phys. Rev. D} {\bfseries
  2} (1970) 2762}.

\bibitem{Brown:1992kc}
J.~Brown, \emph{{Action functionals for relativistic perfect fluids}},
  \href{https://doi.org/10.1088/0264-9381/10/8/017}{\emph{Class. Quant. Grav.}
  {\bfseries 10} (1993) 1579}
  [\href{https://arxiv.org/abs/gr-qc/9304026}{{\ttfamily gr-qc/9304026}}].

\bibitem{DiTucci:2020weq}
A.~Di~Tucci, M.P.~Heller and J.L.~Lehners, \emph{{Lessons for quantum cosmology
  from anti\textendash{}de Sitter black holes}},
  \href{https://doi.org/10.1103/PhysRevD.102.086011}{\emph{Phys. Rev. D}
  {\bfseries 102} (2020) 086011}
  [\href{https://arxiv.org/abs/2007.04872}{{\ttfamily arXiv:2007.04872}}].

\bibitem{Dafermos:2012np}
M.~Dafermos, \emph{{Black holes without spacelike singularities}},
  \href{https://doi.org/10.1007/s00220-014-2063-4}{\emph{Commun. Math. Phys.}
  {\bfseries 332} (2014) 729}
  [\href{https://arxiv.org/abs/1201.1797}{{\ttfamily arXiv:1201.1797}}].

\bibitem{Borissova:2020knn}
J.N.~Borissova and A.~Eichhorn, \emph{{Towards black-hole
  singularity-resolution in the Lorentzian gravitational path integral}},
  \href{https://doi.org/10.3390/universe7030048}{\emph{Universe} {\bfseries 7}
  (2021) 48} [\href{https://arxiv.org/abs/2012.08570}{{\ttfamily
  arXiv:2012.08570}}].

\bibitem{1979grec.conf..504D}
R.H.~{Dicke} and P.J.E.~{Peebles}, \emph{{The big bang cosmology - enigmas and
  nostrums.}},  in \emph{General Relativity: An Einstein centenary survey}
  (S.W.~{Hawking} and W.~{Israel}, eds.), p.~504, Jan., 1979.

\bibitem{Guth:1980zm}
A.H.~Guth, \emph{{The Inflationary Universe: A Possible Solution to the Horizon
  and Flatness Problems}},
  \href{https://doi.org/10.1103/PhysRevD.23.347}{\emph{Phys. Rev. D} {\bfseries
  23} (1981) 347}.

\bibitem{Linde:1981mu}
A.D.~Linde, \emph{{A New Inflationary Universe Scenario: A Possible Solution of
  the Horizon, Flatness, Homogeneity, Isotropy and Primordial Monopole
  Problems}}, \href{https://doi.org/10.1016/0370-2693(82)91219-9}{\emph{Phys.
  Lett. B} {\bfseries 108} (1982) 389}.

\bibitem{Albrecht:1982wi}
A.~Albrecht and P.J.~Steinhardt, \emph{{Cosmology for Grand Unified Theories
  with Radiatively Induced Symmetry Breaking}},
  \href{https://doi.org/10.1103/PhysRevLett.48.1220}{\emph{Phys. Rev. Lett.}
  {\bfseries 48} (1982) 1220}.

\bibitem{Mukhanov:1981xt}
V.F.~Mukhanov and G.V.~Chibisov, \emph{{Quantum Fluctuations and a Nonsingular
  Universe}}, {\emph{JETP Lett.} {\bfseries 33} (1981) 532}.

\bibitem{Linde:1983gd}
A.D.~Linde, \emph{{Chaotic Inflation}},
  \href{https://doi.org/10.1016/0370-2693(83)90837-7}{\emph{Phys. Lett. B}
  {\bfseries 129} (1983) 177}.

\bibitem{Spradlin:2001pw}
M.~Spradlin, A.~Strominger and A.~Volovich, \emph{{Les Houches lectures on de
  Sitter space}},  in \emph{{Les Houches Summer School: Session 76: Euro Summer
  School on Unity of Fundamental Physics: Gravity, Gauge Theory and Strings}},
  p.~423, 10, 2001, \href{https://arxiv.org/abs/hep-th/0110007}{{\ttfamily
  hep-th/0110007}}.

\bibitem{Yoshida:2018ndv}
D.~Yoshida and J.~Quintin, \emph{{Maximal extensions and singularities in
  inflationary spacetimes}},
  \href{https://doi.org/10.1088/1361-6382/aacf4b}{\emph{Class. Quant. Grav.}
  {\bfseries 35} (2018) 155019}
  [\href{https://arxiv.org/abs/1803.07085}{{\ttfamily arXiv:1803.07085}}].

\bibitem{Borde:2001nh}
A.~Borde, A.H.~Guth and A.~Vilenkin, \emph{{Inflationary space-times are
  incomplete in past directions}},
  \href{https://doi.org/10.1103/PhysRevLett.90.151301}{\emph{Phys. Rev. Lett.}
  {\bfseries 90} (2003) 151301}
  [\href{https://arxiv.org/abs/gr-qc/0110012}{{\ttfamily gr-qc/0110012}}].

\bibitem{DiTucci:2019xcr}
A.~Di~Tucci, J.~Feldbrugge, J.L.~Lehners and N.~Turok, \emph{{Quantum
  Incompleteness of Inflation}},
  \href{https://doi.org/10.1103/PhysRevD.100.063517}{\emph{Phys. Rev. D}
  {\bfseries 100} (2019) 063517}
  [\href{https://arxiv.org/abs/1906.09007}{{\ttfamily arXiv:1906.09007}}].

\bibitem{Hofmann:2019dqu}
S.~Hofmann, M.~Schneider and M.~Urban, \emph{{Quantum complete prelude to
  inflation}}, \href{https://doi.org/10.1103/PhysRevD.99.065012}{\emph{Phys.
  Rev. D} {\bfseries 99} (2019) 065012}
  [\href{https://arxiv.org/abs/1901.04492}{{\ttfamily arXiv:1901.04492}}].

\bibitem{Linde:1986fc}
A.D.~Linde, \emph{{Eternal Chaotic Inflation}},
  \href{https://doi.org/10.1142/S0217732386000129}{\emph{Mod. Phys. Lett. A}
  {\bfseries 1} (1986) 81}.

\bibitem{Guth:2007ng}
A.H.~Guth, \emph{{Eternal inflation and its implications}},
  \href{https://doi.org/10.1088/1751-8113/40/25/S25}{\emph{J. Phys. A}
  {\bfseries 40} (2007) 6811}
  [\href{https://arxiv.org/abs/hep-th/0702178}{{\ttfamily hep-th/0702178}}].

\bibitem{Bousso:2011up}
R.~Bousso and L.~Susskind, \emph{{The Multiverse Interpretation of Quantum
  Mechanics}}, \href{https://doi.org/10.1103/PhysRevD.85.045007}{\emph{Phys.
  Rev. D} {\bfseries 85} (2012) 045007}
  [\href{https://arxiv.org/abs/1105.3796}{{\ttfamily arXiv:1105.3796}}].

\bibitem{Guth:2013sya}
A.H.~Guth, D.I.~Kaiser and Y.~Nomura, \emph{{Inflationary paradigm after Planck
  2013}}, \href{https://doi.org/10.1016/j.physletb.2014.03.020}{\emph{Phys.
  Lett. B} {\bfseries 733} (2014) 112}
  [\href{https://arxiv.org/abs/1312.7619}{{\ttfamily arXiv:1312.7619}}].

\bibitem{Ijjas:2013vea}
A.~Ijjas, P.J.~Steinhardt and A.~Loeb, \emph{{Inflationary paradigm in trouble
  after Planck2013}},
  \href{https://doi.org/10.1016/j.physletb.2013.05.023}{\emph{Phys. Lett. B}
  {\bfseries 723} (2013) 261}
  [\href{https://arxiv.org/abs/1304.2785}{{\ttfamily arXiv:1304.2785}}].

\bibitem{Ijjas:2014nta}
A.~Ijjas, P.J.~Steinhardt and A.~Loeb, \emph{{Inflationary schism}},
  \href{https://doi.org/10.1016/j.physletb.2014.07.012}{\emph{Phys. Lett. B}
  {\bfseries 736} (2014) 142}
  [\href{https://arxiv.org/abs/1402.6980}{{\ttfamily arXiv:1402.6980}}].

\bibitem{Rudelius:2019cfh}
T.~Rudelius, \emph{{Conditions for (No) Eternal Inflation}},
  \href{https://doi.org/10.1088/1475-7516/2019/08/009}{\emph{JCAP} {\bfseries
  08} (2019) 009} [\href{https://arxiv.org/abs/1905.05198}{{\ttfamily
  arXiv:1905.05198}}].

\bibitem{Starobinsky:1986fx}
A.A.~Starobinsky, \emph{{Stochastic de Sitter (inflationary) stage in the early
  Universe}}, \href{https://doi.org/10.1007/3-540-16452-9\_6}{\emph{Lect. Notes
  Phys.} {\bfseries 246} (1986) 107}.

\bibitem{Bramberger:2019zks}
S.F.~Bramberger, A.~Di~Tucci and J.L.~Lehners, \emph{{Homogeneous Transitions
  during Inflation: a Description in Quantum Cosmology}},
  \href{https://doi.org/10.1103/PhysRevD.101.063501}{\emph{Phys. Rev. D}
  {\bfseries 101} (2020) 063501}
  [\href{https://arxiv.org/abs/1907.05782}{{\ttfamily arXiv:1907.05782}}].

\bibitem{Kiefer:1998qe}
C.~Kiefer, D.~Polarski and A.A.~Starobinsky, \emph{{Quantum to classical
  transition for fluctuations in the early universe}},
  \href{https://doi.org/10.1142/S0218271898000292}{\emph{Int. J. Mod. Phys. D}
  {\bfseries 7} (1998) 455}
  [\href{https://arxiv.org/abs/gr-qc/9802003}{{\ttfamily gr-qc/9802003}}].

\bibitem{Creminelli:2008es}
P.~Creminelli, S.~Dubovsky, A.~Nicolis, L.~Senatore and M.~Zaldarriaga,
  \emph{{The Phase Transition to Slow-roll Eternal Inflation}},
  \href{https://doi.org/10.1088/1126-6708/2008/09/036}{\emph{JHEP} {\bfseries
  09} (2008) 036} [\href{https://arxiv.org/abs/0802.1067}{{\ttfamily
  arXiv:0802.1067}}].

\bibitem{Obied:2018sgi}
G.~Obied, H.~Ooguri, L.~Spodyneiko and C.~Vafa, \emph{{De Sitter Space and the
  Swampland}},  \href{https://arxiv.org/abs/1806.08362}{{\ttfamily
  arXiv:1806.08362}}.

\bibitem{Ooguri:2018wrx}
H.~Ooguri, E.~Palti, G.~Shiu and C.~Vafa, \emph{{Distance and de Sitter
  Conjectures on the Swampland}},
  \href{https://doi.org/10.1016/j.physletb.2018.11.018}{\emph{Phys. Lett. B}
  {\bfseries 788} (2019) 180}
  [\href{https://arxiv.org/abs/1810.05506}{{\ttfamily arXiv:1810.05506}}].

\bibitem{Agrawal:2018own}
P.~Agrawal, G.~Obied, P.J.~Steinhardt and C.~Vafa, \emph{{On the Cosmological
  Implications of the String Swampland}},
  \href{https://doi.org/10.1016/j.physletb.2018.07.040}{\emph{Phys. Lett. B}
  {\bfseries 784} (2018) 271}
  [\href{https://arxiv.org/abs/1806.09718}{{\ttfamily arXiv:1806.09718}}].

\bibitem{Garg:2018reu}
S.K.~Garg and C.~Krishnan, \emph{{Bounds on Slow Roll and the de Sitter
  Swampland}}, \href{https://doi.org/10.1007/JHEP11(2019)075}{\emph{JHEP}
  {\bfseries 11} (2019) 075}
  [\href{https://arxiv.org/abs/1807.05193}{{\ttfamily arXiv:1807.05193}}].

\bibitem{Dvali:2017eba}
G.~Dvali, C.~Gomez and S.~Zell, \emph{{Quantum Break-Time of de Sitter}},
  \href{https://doi.org/10.1088/1475-7516/2017/06/028}{\emph{JCAP} {\bfseries
  06} (2017) 028} [\href{https://arxiv.org/abs/1701.08776}{{\ttfamily
  arXiv:1701.08776}}].

\bibitem{Dvali:2018jhn}
G.~Dvali, C.~Gomez and S.~Zell, \emph{{Quantum Breaking Bound on de Sitter and
  Swampland}}, \href{https://doi.org/10.1002/prop.201800094}{\emph{Fortsch.
  Phys.} {\bfseries 67} (2019) 1800094}
  [\href{https://arxiv.org/abs/1810.11002}{{\ttfamily arXiv:1810.11002}}].

\bibitem{Brandenberger:2016vhg}
R.~Brandenberger and P.~Peter, \emph{{Bouncing Cosmologies: Progress and
  Problems}}, \href{https://doi.org/10.1007/s10701-016-0057-0}{\emph{Found.
  Phys.} {\bfseries 47} (2017) 797}
  [\href{https://arxiv.org/abs/1603.05834}{{\ttfamily arXiv:1603.05834}}].

\bibitem{Wands:1998yp}
D.~Wands, \emph{{Duality invariance of cosmological perturbation spectra}},
  \href{https://doi.org/10.1103/PhysRevD.60.023507}{\emph{Phys. Rev. D}
  {\bfseries 60} (1999) 023507}
  [\href{https://arxiv.org/abs/gr-qc/9809062}{{\ttfamily gr-qc/9809062}}].

\bibitem{Finelli:2001sr}
F.~Finelli and R.~Brandenberger, \emph{{On the generation of a scale invariant
  spectrum of adiabatic fluctuations in cosmological models with a contracting
  phase}}, \href{https://doi.org/10.1103/PhysRevD.65.103522}{\emph{Phys. Rev.
  D} {\bfseries 65} (2002) 103522}
  [\href{https://arxiv.org/abs/hep-th/0112249}{{\ttfamily hep-th/0112249}}].

\bibitem{Brandenberger:2012zb}
R.H.~Brandenberger, \emph{{The Matter Bounce Alternative to Inflationary
  Cosmology}},  \href{https://arxiv.org/abs/1206.4196}{{\ttfamily
  arXiv:1206.4196}}.

\bibitem{Khoury:2001wf}
J.~Khoury, B.A.~Ovrut, P.J.~Steinhardt and N.~Turok, \emph{{The Ekpyrotic
  universe: Colliding branes and the origin of the hot big bang}},
  \href{https://doi.org/10.1103/PhysRevD.64.123522}{\emph{Phys. Rev. D}
  {\bfseries 64} (2001) 123522}
  [\href{https://arxiv.org/abs/hep-th/0103239}{{\ttfamily hep-th/0103239}}].

\bibitem{Lehners:2007ac}
J.L.~Lehners, P.~McFadden, N.~Turok and P.J.~Steinhardt, \emph{{Generating
  ekpyrotic curvature perturbations before the big bang}},
  \href{https://doi.org/10.1103/PhysRevD.76.103501}{\emph{Phys. Rev. D}
  {\bfseries 76} (2007) 103501}
  [\href{https://arxiv.org/abs/hep-th/0702153}{{\ttfamily hep-th/0702153}}].

\bibitem{Buchbinder:2007ad}
E.I.~Buchbinder, J.~Khoury and B.A.~Ovrut, \emph{{New Ekpyrotic cosmology}},
  \href{https://doi.org/10.1103/PhysRevD.76.123503}{\emph{Phys. Rev. D}
  {\bfseries 76} (2007) 123503}
  [\href{https://arxiv.org/abs/hep-th/0702154}{{\ttfamily hep-th/0702154}}].

\bibitem{Lehners:2008vx}
J.L.~Lehners, \emph{{Ekpyrotic and Cyclic Cosmology}},
  \href{https://doi.org/10.1016/j.physrep.2008.06.001}{\emph{Phys. Rept.}
  {\bfseries 465} (2008) 223}
  [\href{https://arxiv.org/abs/0806.1245}{{\ttfamily arXiv:0806.1245}}].

\bibitem{Barrow:1995cfa}
J.D.~Barrow and M.P.~D\k{a}browski, \emph{{Oscillating Universes}},
  \href{https://doi.org/10.1093/mnras/275.3.850}{\emph{Mon. Not. Roy. Astron.
  Soc.} {\bfseries 275} (1995) 850}.

\bibitem{Steinhardt:2001st}
P.J.~Steinhardt and N.~Turok, \emph{{Cosmic evolution in a cyclic universe}},
  \href{https://doi.org/10.1103/PhysRevD.65.126003}{\emph{Phys. Rev. D}
  {\bfseries 65} (2002) 126003}
  [\href{https://arxiv.org/abs/hep-th/0111098}{{\ttfamily hep-th/0111098}}].

\bibitem{Steinhardt:2004gk}
P.J.~Steinhardt and N.~Turok, \emph{{The Cyclic model simplified}},
  \href{https://doi.org/10.1016/j.newar.2005.01.003}{\emph{New Astron. Rev.}
  {\bfseries 49} (2005) 43}
  [\href{https://arxiv.org/abs/astro-ph/0404480}{{\ttfamily
  astro-ph/0404480}}].

\bibitem{Lehners:2008qe}
J.L.~Lehners and P.J.~Steinhardt, \emph{{Dark Energy and the Return of the
  Phoenix Universe}},
  \href{https://doi.org/10.1103/PhysRevD.79.063503}{\emph{Phys. Rev. D}
  {\bfseries 79} (2009) 063503}
  [\href{https://arxiv.org/abs/0812.3388}{{\ttfamily arXiv:0812.3388}}].

\bibitem{Barrow:2017yqt}
J.D.~Barrow and C.~Ganguly, \emph{{Cyclic Mixmaster Universes}},
  \href{https://doi.org/10.1103/PhysRevD.95.083515}{\emph{Phys. Rev. D}
  {\bfseries 95} (2017) 083515}
  [\href{https://arxiv.org/abs/1703.05969}{{\ttfamily arXiv:1703.05969}}].

\bibitem{Barrow:2017zar}
J.D.~Barrow and C.~Ganguly, \emph{{The Shape of Bouncing Universes}},
  \href{https://doi.org/10.1142/S0218271817430167}{\emph{Int. J. Mod. Phys. D}
  {\bfseries 26} (2017) 1743016}
  [\href{https://arxiv.org/abs/1705.06647}{{\ttfamily arXiv:1705.06647}}].

\bibitem{Ijjas:2019pyf}
A.~Ijjas and P.J.~Steinhardt, \emph{{A new kind of cyclic universe}},
  \href{https://doi.org/10.1016/j.physletb.2019.06.056}{\emph{Phys. Lett. B}
  {\bfseries 795} (2019) 666}
  [\href{https://arxiv.org/abs/1904.08022}{{\ttfamily arXiv:1904.08022}}].

\bibitem{Brandenberger:1988aj}
R.H.~Brandenberger and C.~Vafa, \emph{{Superstrings in the Early Universe}},
  \href{https://doi.org/10.1016/0550-3213(89)90037-0}{\emph{Nucl. Phys. B}
  {\bfseries 316} (1989) 391}.

\bibitem{Tseytlin:1991xk}
A.A.~Tseytlin and C.~Vafa, \emph{{Elements of string cosmology}},
  \href{https://doi.org/10.1016/0550-3213(92)90327-8}{\emph{Nucl. Phys. B}
  {\bfseries 372} (1992) 443}
  [\href{https://arxiv.org/abs/hep-th/9109048}{{\ttfamily hep-th/9109048}}].

\bibitem{Easther:2004sd}
R.~Easther, B.R.~Greene, M.G.~Jackson and D.N.~Kabat, \emph{{String windings in
  the early universe}},
  \href{https://doi.org/10.1088/1475-7516/2005/02/009}{\emph{JCAP} {\bfseries
  02} (2005) 009} [\href{https://arxiv.org/abs/hep-th/0409121}{{\ttfamily
  hep-th/0409121}}].

\bibitem{Battefeld:2005av}
T.~Battefeld and S.~Watson, \emph{{String gas cosmology}},
  \href{https://doi.org/10.1103/RevModPhys.78.435}{\emph{Rev. Mod. Phys.}
  {\bfseries 78} (2006) 435}
  [\href{https://arxiv.org/abs/hep-th/0510022}{{\ttfamily hep-th/0510022}}].

\bibitem{Brandenberger:2008nx}
R.H.~Brandenberger, \emph{{String Gas Cosmology}},  p.~193, 8, 2008,
  \href{https://arxiv.org/abs/0808.0746}{{\ttfamily arXiv:0808.0746}}.

\bibitem{Gasperini:2007zz}
M.~Gasperini, \emph{{Elements of string cosmology}}, Cambridge University Press
  (10, 2007).

\bibitem{Creminelli:2010ba}
P.~Creminelli, A.~Nicolis and E.~Trincherini, \emph{{Galilean Genesis: An
  Alternative to inflation}},
  \href{https://doi.org/10.1088/1475-7516/2010/11/021}{\emph{JCAP} {\bfseries
  11} (2010) 021} [\href{https://arxiv.org/abs/1007.0027}{{\ttfamily
  arXiv:1007.0027}}].

\bibitem{Creminelli:2012my}
P.~Creminelli, K.~Hinterbichler, J.~Khoury, A.~Nicolis and E.~Trincherini,
  \emph{{Subluminal Galilean Genesis}},
  \href{https://doi.org/10.1007/JHEP02(2013)006}{\emph{JHEP} {\bfseries 02}
  (2013) 006} [\href{https://arxiv.org/abs/1209.3768}{{\ttfamily
  arXiv:1209.3768}}].

\bibitem{Nishi:2015pta}
S.~Nishi and T.~Kobayashi, \emph{{Generalized Galilean Genesis}},
  \href{https://doi.org/10.1088/1475-7516/2015/03/057}{\emph{JCAP} {\bfseries
  03} (2015) 057} [\href{https://arxiv.org/abs/1501.02553}{{\ttfamily
  arXiv:1501.02553}}].

\bibitem{Kobayashi:2015gga}
T.~Kobayashi, M.~Yamaguchi and J.~Yokoyama, \emph{{Galilean Creation of the
  Inflationary Universe}},
  \href{https://doi.org/10.1088/1475-7516/2015/07/017}{\emph{JCAP} {\bfseries
  07} (2015) 017} [\href{https://arxiv.org/abs/1504.05710}{{\ttfamily
  arXiv:1504.05710}}].

\bibitem{Nishi:2016ljg}
S.~Nishi and T.~Kobayashi, \emph{{Scale-invariant perturbations from
  null-energy-condition violation: A new variant of Galilean genesis}},
  \href{https://doi.org/10.1103/PhysRevD.95.064001}{\emph{Phys. Rev. D}
  {\bfseries 95} (2017) 064001}
  [\href{https://arxiv.org/abs/1611.01906}{{\ttfamily arXiv:1611.01906}}].

\bibitem{Yoshida:2017swb}
D.~Yoshida, J.~Quintin, M.~Yamaguchi and R.H.~Brandenberger,
  \emph{{Cosmological perturbations and stability of nonsingular cosmologies
  with limiting curvature}},
  \href{https://doi.org/10.1103/PhysRevD.96.043502}{\emph{Phys. Rev. D}
  {\bfseries 96} (2017) 043502}
  [\href{https://arxiv.org/abs/1704.04184}{{\ttfamily arXiv:1704.04184}}].

\bibitem{Mironov:2019qjt}
S.~Mironov, V.~Rubakov and V.~Volkova, \emph{{Genesis with general relativity
  asymptotics in beyond Horndeski theory}},
  \href{https://doi.org/10.1103/PhysRevD.100.083521}{\emph{Phys. Rev. D}
  {\bfseries 100} (2019) 083521}
  [\href{https://arxiv.org/abs/1905.06249}{{\ttfamily arXiv:1905.06249}}].

\bibitem{Volkova:2019jlj}
V.~Volkova, S.~Mironov and V.~Rubakov, \emph{{Cosmological Scenarios with
  Bounce and Genesis in Horndeski Theory and Beyond}},
  \href{https://doi.org/10.1134/S1063776119100236}{\emph{J. Exp. Theor. Phys.}
  {\bfseries 129} (2019) 553}.

\bibitem{Ageeva:2020gti}
Y.~Ageeva, O.~Evseev, O.~Melichev and V.~Rubakov, \emph{{Towards evading the
  strong coupling problem in Horndeski Genesis}},
  \href{https://doi.org/10.1103/PhysRevD.102.023519}{\emph{Phys. Rev. D}
  {\bfseries 102} (2020) 023519}
  [\href{https://arxiv.org/abs/2003.01202}{{\ttfamily arXiv:2003.01202}}].

\bibitem{Ilyas:2020zcb}
A.~Ilyas, M.~Zhu, Y.~Zheng and Y.F.~Cai, \emph{{Emergent Universe and Genesis
  from the DHOST Cosmology}},
  \href{https://doi.org/10.1007/JHEP01(2021)141}{\emph{JHEP} {\bfseries 01}
  (2021) 141} [\href{https://arxiv.org/abs/2009.10351}{{\ttfamily
  arXiv:2009.10351}}].

\bibitem{Hawking:1981gb}
S.~Hawking, \emph{{The Boundary Conditions of the Universe}}, {\emph{Pontif.
  Acad. Sci. Scr. Varia} {\bfseries 48} (1982) 563}.

\bibitem{Hartle:1983ai}
J.B.~Hartle and S.W.~Hawking, \emph{Wave function of the universe},
  \href{https://doi.org/10.1103/PhysRevD.28.2960}{\emph{Phys. Rev. D}
  {\bfseries 28} (1983) 2960}.

\bibitem{Lyons:1992ua}
G.W.~Lyons, \emph{{Complex solutions for the scalar field model of the
  universe}}, \href{https://doi.org/10.1103/PhysRevD.46.1546}{\emph{Phys. Rev.
  D} {\bfseries 46} (1992) 1546}.

\bibitem{Battarra:2014xoa}
L.~Battarra and J.L.~Lehners, \emph{{On the Creation of the Universe via
  Ekpyrotic Instantons}},
  \href{https://doi.org/10.1016/j.physletb.2015.01.028}{\emph{Phys. Lett. B}
  {\bfseries 742} (2015) 167}
  [\href{https://arxiv.org/abs/1406.5896}{{\ttfamily arXiv:1406.5896}}].

\bibitem{Halliwell:1984eu}
J.J.~Halliwell and S.W.~Hawking, \emph{Origin of structure in the universe},
  \href{https://doi.org/10.1103/PhysRevD.31.1777}{\emph{Phys. Rev. D}
  {\bfseries 31} (1985) 1777}.

\bibitem{Feldbrugge:2017kzv}
J.~Feldbrugge, J.L.~Lehners and N.~Turok, \emph{{Lorentzian Quantum
  Cosmology}}, \href{https://doi.org/10.1103/PhysRevD.95.103508}{\emph{Phys.
  Rev. D} {\bfseries 95} (2017) 103508}
  [\href{https://arxiv.org/abs/1703.02076}{{\ttfamily arXiv:1703.02076}}].

\bibitem{DiazDorronsoro:2017hti}
J.~Diaz~Dorronsoro, J.J.~Halliwell, J.B.~Hartle, T.~Hertog and O.~Janssen,
  \emph{{Real no-boundary wave function in Lorentzian quantum cosmology}},
  \href{https://doi.org/10.1103/PhysRevD.96.043505}{\emph{Phys. Rev. D}
  {\bfseries 96} (2017) 043505}
  [\href{https://arxiv.org/abs/1705.05340}{{\ttfamily arXiv:1705.05340}}].

\bibitem{Hartle:2008ng}
J.B.~Hartle, S.~Hawking and T.~Hertog, \emph{{The Classical Universes of the
  No-Boundary Quantum State}},
  \href{https://doi.org/10.1103/PhysRevD.77.123537}{\emph{Phys. Rev. D}
  {\bfseries 77} (2008) 123537}
  [\href{https://arxiv.org/abs/0803.1663}{{\ttfamily arXiv:0803.1663}}].

\bibitem{Vilenkin:1982de}
A.~Vilenkin, \emph{{Creation of Universes from Nothing}},
  \href{https://doi.org/10.1016/0370-2693(82)90866-8}{\emph{Phys. Lett. B}
  {\bfseries 117} (1982) 25}.

\bibitem{Feldbrugge:2017fcc}
J.~Feldbrugge, J.L.~Lehners and N.~Turok, \emph{{No smooth beginning for
  spacetime}},
  \href{https://doi.org/10.1103/PhysRevLett.119.171301}{\emph{Phys. Rev. Lett.}
  {\bfseries 119} (2017) 171301}
  [\href{https://arxiv.org/abs/1705.00192}{{\ttfamily arXiv:1705.00192}}].

\bibitem{Feldbrugge:2017mbc}
J.~Feldbrugge, J.L.~Lehners and N.~Turok, \emph{{No rescue for the no boundary
  proposal: Pointers to the future of quantum cosmology}},
  \href{https://doi.org/10.1103/PhysRevD.97.023509}{\emph{Phys. Rev. D}
  {\bfseries 97} (2018) 023509}
  [\href{https://arxiv.org/abs/1708.05104}{{\ttfamily arXiv:1708.05104}}].

\bibitem{Louko:1988bk}
J.~Louko, \emph{{Canonizing the Hartle-hawking Proposal}},
  \href{https://doi.org/10.1016/0370-2693(88)90008-1}{\emph{Phys. Lett. B}
  {\bfseries 202} (1988) 201}.

\bibitem{Vilenkin:1988yd}
A.~Vilenkin, \emph{{The Interpretation of the Wave Function of the Universe}},
  \href{https://doi.org/10.1103/PhysRevD.39.1116}{\emph{Phys. Rev. D}
  {\bfseries 39} (1989) 1116}.

\bibitem{Lehners:2015sia}
J.L.~Lehners, \emph{{Classical Inflationary and Ekpyrotic Universes in the
  No-Boundary Wavefunction}},
  \href{https://doi.org/10.1103/PhysRevD.91.083525}{\emph{Phys. Rev. D}
  {\bfseries 91} (2015) 083525}
  [\href{https://arxiv.org/abs/1502.00629}{{\ttfamily arXiv:1502.00629}}].

\bibitem{Lehners:2015efa}
J.L.~Lehners, \emph{{New Ekpyrotic Quantum Cosmology}},
  \href{https://doi.org/10.1016/j.physletb.2015.09.032}{\emph{Phys. Lett. B}
  {\bfseries 750} (2015) 242}
  [\href{https://arxiv.org/abs/1504.02467}{{\ttfamily arXiv:1504.02467}}].

\bibitem{Jonas:2020pos}
C.~Jonas and J.L.~Lehners, \emph{{No-boundary solutions are robust to quantum
  gravity corrections}},
  \href{https://doi.org/10.1103/PhysRevD.102.123539}{\emph{Phys. Rev. D}
  {\bfseries 102} (2020) 123539}
  [\href{https://arxiv.org/abs/2008.04134}{{\ttfamily arXiv:2008.04134}}].

\bibitem{Cano:2020oaa}
P.A.~Cano, K.~Fransen and T.~Hertog, \emph{{Novel higher-curvature variations
  of $R^2$ inflation}},  \href{https://arxiv.org/abs/2011.13933}{{\ttfamily
  arXiv:2011.13933}}.

\bibitem{Narain:2021bff}
G.~Narain, \emph{{On Gauss-bonnet gravity and boundary conditions in Lorentzian
  path-integral quantization}},
  \href{https://arxiv.org/abs/2101.04644}{{\ttfamily arXiv:2101.04644}}.

\bibitem{Goroff:1985th}
M.H.~Goroff and A.~Sagnotti, \emph{{The Ultraviolet Behavior of Einstein
  Gravity}}, \href{https://doi.org/10.1016/0550-3213(86)90193-8}{\emph{Nucl.
  Phys. B} {\bfseries 266} (1986) 709}.

\bibitem{Stelle:1976gc}
K.S.~Stelle, \emph{{Renormalization of Higher Derivative Quantum Gravity}},
  \href{https://doi.org/10.1103/PhysRevD.16.953}{\emph{Phys. Rev. D} {\bfseries
  16} (1977) 953}.

\bibitem{Stelle:1977ry}
K.S.~Stelle, \emph{{Classical Gravity with Higher Derivatives}},
  \href{https://doi.org/10.1007/BF00760427}{\emph{Gen. Rel. Grav.} {\bfseries
  9} (1978) 353}.

\bibitem{Salvio:2019ewf}
A.~Salvio, \emph{{Metastability in Quadratic Gravity}},
  \href{https://doi.org/10.1103/PhysRevD.99.103507}{\emph{Phys. Rev. D}
  {\bfseries 99} (2019) 103507}
  [\href{https://arxiv.org/abs/1902.09557}{{\ttfamily arXiv:1902.09557}}].

\bibitem{Donoghue:2019fcb}
J.F.~Donoghue and G.~Menezes, \emph{{Unitarity, stability and loops of unstable
  ghosts}}, \href{https://doi.org/10.1103/PhysRevD.100.105006}{\emph{Phys. Rev.
  D} {\bfseries 100} (2019) 105006}
  [\href{https://arxiv.org/abs/1908.02416}{{\ttfamily arXiv:1908.02416}}].

\bibitem{Akrami:2018odb}
{\scshape Planck} collaboration, Y.~Akrami et~al., \emph{{Planck 2018 results.
  X. Constraints on inflation}},
  \href{https://doi.org/10.1051/0004-6361/201833887}{\emph{Astron. Astrophys.}
  {\bfseries 641} (2020) A10}
  [\href{https://arxiv.org/abs/1807.06211}{{\ttfamily arXiv:1807.06211}}].

\bibitem{Metsaev:1987zx}
R.R.~Metsaev and A.A.~Tseytlin, \emph{{Order alpha-prime (Two Loop) Equivalence
  of the String Equations of Motion and the Sigma Model Weyl Invariance
  Conditions: Dependence on the Dilaton and the Antisymmetric Tensor}},
  \href{https://doi.org/10.1016/0550-3213(87)90077-0}{\emph{Nucl. Phys. B}
  {\bfseries 293} (1987) 385}.

\bibitem{Green:2010wi}
M.B.~Green, J.G.~Russo and P.~Vanhove, \emph{{Automorphic properties of low
  energy string amplitudes in various dimensions}},
  \href{https://doi.org/10.1103/PhysRevD.81.086008}{\emph{Phys. Rev. D}
  {\bfseries 81} (2010) 086008}
  [\href{https://arxiv.org/abs/1001.2535}{{\ttfamily arXiv:1001.2535}}].

\bibitem{Deruelle:2009zk}
N.~Deruelle, M.~Sasaki, Y.~Sendouda and D.~Yamauchi, \emph{{Hamiltonian
  formulation of f(Riemann) theories of gravity}},
  \href{https://doi.org/10.1143/PTP.123.169}{\emph{Prog. Theor. Phys.}
  {\bfseries 123} (2010) 169}
  [\href{https://arxiv.org/abs/0908.0679}{{\ttfamily arXiv:0908.0679}}].

\bibitem{Dyer:2008hb}
E.~Dyer and K.~Hinterbichler, \emph{{Boundary Terms, Variational Principles and
  Higher Derivative Modified Gravity}},
  \href{https://doi.org/10.1103/PhysRevD.79.024028}{\emph{Phys. Rev. D}
  {\bfseries 79} (2009) 024028}
  [\href{https://arxiv.org/abs/0809.4033}{{\ttfamily arXiv:0809.4033}}].

\bibitem{Barrow:1988xh}
J.D.~Barrow and S.~Cotsakis, \emph{{Inflation and the Conformal Structure of
  Higher Order Gravity Theories}},
  \href{https://doi.org/10.1016/0370-2693(88)90110-4}{\emph{Phys. Lett. B}
  {\bfseries 214} (1988) 515}.

\bibitem{Hawking:1984ph}
S.~Hawking and J.~Luttrell, \emph{Higher derivatives in quantum cosmology: (i).
  the isotropic case},
  \href{https://doi.org/https://doi.org/10.1016/0550-3213(84)90380-8}{\emph{Nuclear
  Physics B} {\bfseries 247} (1984) 250 }.

\bibitem{Hohm:2010jc}
O.~Hohm and E.~Tonni, \emph{{A boundary stress tensor for higher-derivative
  gravity in AdS and Lifshitz backgrounds}},
  \href{https://doi.org/10.1007/JHEP04(2010)093}{\emph{JHEP} {\bfseries 04}
  (2010) 093} [\href{https://arxiv.org/abs/1001.3598}{{\ttfamily
  arXiv:1001.3598}}].

\bibitem{Myers:1987yn}
R.C.~Myers, \emph{{Higher Derivative Gravity, Surface Terms and String
  Theory}}, \href{https://doi.org/10.1103/PhysRevD.36.392}{\emph{Phys. Rev. D}
  {\bfseries 36} (1987) 392}.

\bibitem{Deruelle:2017xel}
N.~Deruelle, N.~Merino and R.~Olea, \emph{{Einstein-Gauss-Bonnet theory of
  gravity: The Gauss-Bonnet-Katz boundary term}},
  \href{https://doi.org/10.1103/PhysRevD.97.104009}{\emph{Phys. Rev. D}
  {\bfseries 97} (2018) 104009}
  [\href{https://arxiv.org/abs/1709.06478}{{\ttfamily arXiv:1709.06478}}].

\bibitem{Barrow:2006xb}
J.D.~Barrow and S.~Hervik, \emph{{On the evolution of universes in quadratic
  theories of gravity}},
  \href{https://doi.org/10.1103/PhysRevD.74.124017}{\emph{Phys. Rev. D}
  {\bfseries 74} (2006) 124017}
  [\href{https://arxiv.org/abs/gr-qc/0610013}{{\ttfamily gr-qc/0610013}}].

\bibitem{Barrow:2005qv}
J.D.~Barrow and S.~Hervik, \emph{{Anisotropically inflating universes}},
  \href{https://doi.org/10.1103/PhysRevD.73.023007}{\emph{Phys. Rev. D}
  {\bfseries 73} (2006) 023007}
  [\href{https://arxiv.org/abs/gr-qc/0511127}{{\ttfamily gr-qc/0511127}}].

\bibitem{Barrow:2009gx}
J.D.~Barrow and S.~Hervik, \emph{{Simple Types of Anisotropic Inflation}},
  \href{https://doi.org/10.1103/PhysRevD.81.023513}{\emph{Phys. Rev. D}
  {\bfseries 81} (2010) 023513}
  [\href{https://arxiv.org/abs/0911.3805}{{\ttfamily arXiv:0911.3805}}].

\bibitem{Middleton:2010bv}
J.~Middleton, \emph{{On The Existence Of Anisotropic Cosmological Models In
  Higher-Order Theories Of Gravity}},
  \href{https://doi.org/10.1088/0264-9381/27/22/225013}{\emph{Class. Quant.
  Grav.} {\bfseries 27} (2010) 225013}
  [\href{https://arxiv.org/abs/1007.4669}{{\ttfamily arXiv:1007.4669}}].

\bibitem{Muller:2017nxg}
D.~M\"uller, A.~Ricciardone, A.A.~Starobinsky and A.~Toporensky,
  \emph{{Anisotropic cosmological solutions in $R + R^2$ gravity}},
  \href{https://doi.org/10.1140/epjc/s10052-018-5778-0}{\emph{Eur. Phys. J. C}
  {\bfseries 78} (2018) 311}
  [\href{https://arxiv.org/abs/1710.08753}{{\ttfamily arXiv:1710.08753}}].

\bibitem{Mukhanov:1991zn}
V.F.~Mukhanov and R.H.~Brandenberger, \emph{{A Nonsingular universe}},
  \href{https://doi.org/10.1103/PhysRevLett.68.1969}{\emph{Phys. Rev. Lett.}
  {\bfseries 68} (1992) 1969}.

\bibitem{Brandenberger:1993ef}
R.H.~Brandenberger, V.F.~Mukhanov and A.~Sornborger, \emph{{A Cosmological
  theory without singularities}},
  \href{https://doi.org/10.1103/PhysRevD.48.1629}{\emph{Phys. Rev. D}
  {\bfseries 48} (1993) 1629}
  [\href{https://arxiv.org/abs/gr-qc/9303001}{{\ttfamily gr-qc/9303001}}].

\bibitem{Fleury:2016htl}
P.~Fleury, F.~Nugier and G.~Fanizza, \emph{{Geodesic-light-cone coordinates and
  the Bianchi I spacetime}},
  \href{https://doi.org/10.1088/1475-7516/2016/06/008}{\emph{JCAP} {\bfseries
  06} (2016) 008} [\href{https://arxiv.org/abs/1602.04461}{{\ttfamily
  arXiv:1602.04461}}].

\bibitem{Fleury:2014rea}
P.~Fleury, C.~Pitrou and J.P.~Uzan, \emph{{Light propagation in a homogeneous
  and anisotropic universe}},
  \href{https://doi.org/10.1103/PhysRevD.91.043511}{\emph{Phys. Rev. D}
  {\bfseries 91} (2015) 043511}
  [\href{https://arxiv.org/abs/1410.8473}{{\ttfamily arXiv:1410.8473}}].

\bibitem{ArmendarizPicon:2000dh}
C.~Armendariz-Picon, V.F.~Mukhanov and P.J.~Steinhardt, \emph{{A Dynamical
  solution to the problem of a small cosmological constant and late time cosmic
  acceleration}},
  \href{https://doi.org/10.1103/PhysRevLett.85.4438}{\emph{Phys. Rev. Lett.}
  {\bfseries 85} (2000) 4438}
  [\href{https://arxiv.org/abs/astro-ph/0004134}{{\ttfamily
  astro-ph/0004134}}].

\bibitem{ArmendarizPicon:2000ah}
C.~Armendariz-Picon, V.F.~Mukhanov and P.J.~Steinhardt, \emph{{Essentials of k
  essence}}, \href{https://doi.org/10.1103/PhysRevD.63.103510}{\emph{Phys. Rev.
  D} {\bfseries 63} (2001) 103510}
  [\href{https://arxiv.org/abs/astro-ph/0006373}{{\ttfamily
  astro-ph/0006373}}].

\bibitem{ArkaniHamed:2003uy}
N.~Arkani-Hamed, H.C.~Cheng, M.A.~Luty and S.~Mukohyama, \emph{{Ghost
  condensation and a consistent infrared modification of gravity}},
  \href{https://doi.org/10.1088/1126-6708/2004/05/074}{\emph{JHEP} {\bfseries
  05} (2004) 074} [\href{https://arxiv.org/abs/hep-th/0312099}{{\ttfamily
  hep-th/0312099}}].

\bibitem{Lin:2010pf}
C.~Lin, R.H.~Brandenberger and L.~Perreault~Levasseur, \emph{{A Matter Bounce
  By Means of Ghost Condensation}},
  \href{https://doi.org/10.1088/1475-7516/2011/04/019}{\emph{JCAP} {\bfseries
  04} (2011) 019} [\href{https://arxiv.org/abs/1007.2654}{{\ttfamily
  arXiv:1007.2654}}].

\bibitem{deRham:2017aoj}
C.~de~Rham and S.~Melville, \emph{{Unitary null energy condition violation in
  P(X) cosmologies}},
  \href{https://doi.org/10.1103/PhysRevD.95.123523}{\emph{Phys. Rev. D}
  {\bfseries 95} (2017) 123523}
  [\href{https://arxiv.org/abs/1703.00025}{{\ttfamily arXiv:1703.00025}}].

\bibitem{Mironov:2019haz}
S.~Mironov, V.~Rubakov and V.~Volkova, \emph{{Cosmological scenarios with
  bounce and Genesis in Horndeski theory and beyond: An essay in honor of I.M.
  Khalatnikov on the occasion of his 100th birthday}},
  \href{https://arxiv.org/abs/1906.12139}{{\ttfamily arXiv:1906.12139}}.

\bibitem{Ageeva:2018lko}
Y.A.~Ageeva, O.A.~Evseev, O.I.~Melichev and V.A.~Rubakov, \emph{{Horndeski
  Genesis: strong coupling and absence thereof}},
  \href{https://doi.org/10.1051/epjconf/201819107010}{\emph{EPJ Web Conf.}
  {\bfseries 191} (2018) 07010}
  [\href{https://arxiv.org/abs/1810.00465}{{\ttfamily arXiv:1810.00465}}].

\bibitem{Ageeva:2020buc}
Y.~Ageeva, P.~Petrov and V.~Rubakov, \emph{{Horndeski genesis: consistency of
  classical theory}},
  \href{https://doi.org/10.1007/JHEP12(2020)107}{\emph{JHEP} {\bfseries 12}
  (2020) 107} [\href{https://arxiv.org/abs/2009.05071}{{\ttfamily
  arXiv:2009.05071}}].

\bibitem{Hohm:2019jgu}
O.~Hohm and B.~Zwiebach, \emph{{Duality invariant cosmology to all orders in
  $\alpha'$}}, \href{https://doi.org/10.1103/PhysRevD.100.126011}{\emph{Phys.
  Rev. D} {\bfseries 100} (2019) 126011}
  [\href{https://arxiv.org/abs/1905.06963}{{\ttfamily arXiv:1905.06963}}].

\bibitem{MEISSNER199133}
K.~Meissner and G.~Veneziano, \emph{Symmetries of cosmological superstring
  vacua},
  \href{https://doi.org/https://doi.org/10.1016/0370-2693(91)90520-Z}{\emph{Physics
  Letters B} {\bfseries 267} (1991) 33}.

\bibitem{Meissner:1996sa}
K.A.~Meissner, \emph{{Symmetries of higher order string gravity actions}},
  \href{https://doi.org/10.1016/S0370-2693(96)01556-0}{\emph{Phys. Lett. B}
  {\bfseries 392} (1997) 298}
  [\href{https://arxiv.org/abs/hep-th/9610131}{{\ttfamily hep-th/9610131}}].

\bibitem{Hohm:2019ccp}
O.~Hohm and B.~Zwiebach, \emph{{Non-perturbative de Sitter vacua via $\alpha'$
  corrections}}, \href{https://doi.org/10.1142/S0218271819430028}{\emph{Int. J.
  Mod. Phys. D} {\bfseries 28} (2019) 1943002}
  [\href{https://arxiv.org/abs/1905.06583}{{\ttfamily arXiv:1905.06583}}].

\bibitem{Bernardo:2019bkz}
H.~Bernardo, R.~Brandenberger and G.~Franzmann, \emph{{O$(d,d)$ covariant
  string cosmology to all orders in $\alpha^{\prime}$}},
  \href{https://doi.org/10.1007/JHEP02(2020)178}{\emph{JHEP} {\bfseries 02}
  (2020) 178} [\href{https://arxiv.org/abs/1911.00088}{{\ttfamily
  arXiv:1911.00088}}].

\bibitem{Bernardo:2020zlc}
H.~Bernardo and G.~Franzmann, \emph{{$\alpha'$-Cosmology: solutions and
  stability analysis}},
  \href{https://doi.org/10.1007/JHEP05(2020)073}{\emph{JHEP} {\bfseries 05}
  (2020) 073} [\href{https://arxiv.org/abs/2002.09856}{{\ttfamily
  arXiv:2002.09856}}].

\bibitem{Nunez:2020hxx}
C.A.~N\'u\~nez and F.E.~Rost, \emph{{New non-perturbative de Sitter vacua in
  $\alpha'$-complete cosmology}},
  \href{https://doi.org/10.1007/JHEP03(2021)007}{\emph{JHEP} {\bfseries 03}
  (2021) 007} [\href{https://arxiv.org/abs/2011.10091}{{\ttfamily
  arXiv:2011.10091}}].

\bibitem{Penrose:1900mp}
R.~Penrose, \emph{{Singularities and time-asymmetry, in General Relativity}:
  {An Einstein Centenary Survey, Ed. S. Hawking and W. Israel}},  (Cambridge,
  UK), Univ. Pr., 1979.

\bibitem{Bedroya:2019snp}
A.~Bedroya and C.~Vafa, \emph{{Trans-Planckian Censorship and the Swampland}},
  \href{https://doi.org/10.1007/JHEP09(2020)123}{\emph{JHEP} {\bfseries 09}
  (2020) 123} [\href{https://arxiv.org/abs/1909.11063}{{\ttfamily
  arXiv:1909.11063}}].

\bibitem{Bedroya:2020rmd}
A.~Bedroya, \emph{{de Sitter Complementarity, TCC, and the Swampland}},
  \href{https://arxiv.org/abs/2010.09760}{{\ttfamily arXiv:2010.09760}}.

\end{thebibliography}\endgroup

\end{document}